\documentclass[dvipdfm]{pasj00}
\begin{document}
\SetRunningHead{Yoshino et al.}{Soft X-ray Diffuse Emission}
\Received{2008/12/02}
\Accepted{2009/05/02}

\title{Energy Spectra of  the Soft X-ray Diffuse Emission in Fourteen Fields Observed with Suzaku}

\author{
T. \textsc{Yoshino}\altaffilmark{1}\thanks{Present Address is NEC corporation, Nisshin-cho 1-10, Fuchu, 183-8551, Japan.}, 
K. \textsc{Mitsuda}\altaffilmark{1}, 
N. Y. \textsc{Yamasaki}\altaffilmark{1}, 
Y. \textsc{Takei}\altaffilmark{1}, 
T. \textsc{Hagihara}\altaffilmark{1}, 
K.  \textsc{Masui}\altaffilmark{1}\thanks{Present Address is FUJIFILM Advanced Research Laboratories,
Ushijima 577, Kaisei-machi, Ashigara-gun, Kanagawa 258-8577, Japan.}, \\
M. \textsc{Bauer}\altaffilmark{1,2}\thanks{Present Address is Randox Laboratories Ltd.,
55 Diamond Road, Crumlin, Co. Antrim, United Kingdom, BT29 4QY.},
D. \textsc{McCammon}\altaffilmark{3},
R.  \textsc{Fujimoto}\altaffilmark{4},
Q.D.  \textsc{Wang}\altaffilmark{5}, 
and 
Y. \textsc{Yao}\altaffilmark{6,7}
}
\altaffiltext{1}{Institute of Space and Astronautical Science, Japan Aerospace Exploration Agency, 3-1-1 Yoshinodai, Sagamihara, 229-8510, Japan}
\altaffiltext{2}{Max-Planck-Institut f\"ur extraterrestrische Physik, 85748 Garching, Germany}
\altaffiltext{3}{Department of Physics, University of Wisconsin, Madison, 1150 University Avenue, Madison, WI~53706, USA}
\altaffiltext{4}{Department of Physics, Kanazawa University, Kanazawa, 920-1192, Japan}
\altaffiltext{5}{Department of Astronomy, University of Massachusetts Amherst, MA~01003, USA}
\altaffiltext{6}{Massachusetts Institute of Technology (MIT) Kavli Institute for Astrophysics and Space Research, 70 Vassar Street, Cambridge, MA~02139, USA}
\altaffiltext{7}{University of Colorado, CASA, 389 UCB, Boulder, CO~80309, USA}

\KeyWords{Galaxy: disk --- Galaxy: halo ---  X-rays: diffuse background --- X-rays: ISM } 
\maketitle

\begin{abstract}
The soft diffuse X-ray emission of  twelve fields observed with Suzaku are presented together with two additional 
fields from previous analyses.   
All have galactic longitudes $65^\circ   <  \ell  <  295^\circ$ to avoid contributions from the very bright diffuse source that extends at least $30^\circ$  from the Galactic center. 
The surface brightnesses of
the Suzaku nine fields 
for which apparently uncontaminated ROSAT All Sky Survey (RASS) were available
were statistically consistent with the RASS values,  
with an upper limit  for differences of $17 \times 10^{-6}~{\rm c~s}^{-1}~{\rm amin}^{-2}$ 
in {\it R45}-band.  
The O\textsc{vii} and O\textsc{viii} intensities are well correlated to each other, and  O\textsc{vii} emission shows  an intensity floor at  $\sim~2~{\rm photons~s}^{-1}~{\rm cm}^{-2}~{\rm str}^{-1}$ (LU).  The high-latitude O\textsc{viii} emission shows a tight correlation with excess of O\textsc{vii} emission above the floor, with  (O\textsc{viii}~intensity) = 0.5~$\times$~[(O\textsc{vii}~intensity)~--~2 LU], 
suggesting that  temperatures averaged over different  line-of-sight show 
a narrow distribution around $\sim$~0.2 keV.
We consider that  the offset intensity of  O\textsc{vii} arises from the
 Heliospheric solar wind charge exchange and perhaps from the local hot
 bubble,  and that the excess O\textsc{vii} (2-7 LU) is  emission from
 more distant parts of the Galaxy. 
The total bolometric luminosity of this galactic emission is
estimated to be $4 \times 10^{39}~{\rm erg~s}^{-1}$, and its
characteristic temperature may be related to the virial temperature of the Galaxy. 

\end{abstract}

\section{Introduction}
\label{sec:intro}

The soft X-ray diffuse backgound below 1 keV is considered to 
consist of  the contribution of faint extragalactic sources 
and emissions from highly ionized ions,
such as C\textsc{vi}, O\textsc{vii},  O\textsc{viii}, Fe\textsc{xvii}, and Ne\textsc{ix}, 
in solar neighborhoods and in our Galaxy.  The extragalactic contribution, which
we refer to as the Cosmic X-ray Background (CXB) in this paper, 
is estimated to be about 40 \%  of emission in  the
so-called ROSAT {\it R45} band  \citep{McCammon_etal_2002}, which is
approximately $\sim 0.44 - 1$ keV.  The emission from highly ionized 
ions is considered to arise from at least three different origins.
Among them, the solar wind charge exchange (SWCX) induced emission 
from the Heliosphere 
\citep{Cox_1998, Cravens_2000, Lallement_2004}, 
and the thermal emission from the hot gas in local hot bubble 
(LHB) \citep{McCammon_Sanders_1990}  are
hard to be separated from each other using emission spectra above $\sim 0.4$ keV.
In this energy range, the sum of the
two emission components are approximated by 
a thin-thermal emission of  ${\rm k}T \sim 0.1$ keV without absorption \citep{Smith_etal_2007, Henley_etal_2007, Galeazzi_etal_2007, Kuntz_Snowden_2008, Masui_etal_2009}.  
The remaining emission is considered to arise from more distant part of the galaxy; mostly 
above or beyond
the bulk of absorption in the Galactic disk.
\citet{Kuntz_Snowden_2000} called this component 
the ``transabsorption''  emission (TAE) and separated it
in the ROSAT all sky map 
utilizing  the directional dependence of the absorption column density.  They 
found that the emission spectrum can be described by a two-temperature thermal
emission model of temperatures $kT \sim$  0.10 and 0.25 keV.
However, because there is no constraint on distance other than absorption, 
it is hard to constrain the origins conclusively.  

New insight has been obtained from combined analyses of the absorption lines 
observed in the energy
spectra of extragalactic objects and emission lines of the same ion species
observed nearby. 
\citet{Yao_etal_2009} analyzed the absorption spectra 
of LMC X-3 obtained with the transmission grating (HETG) onboard Chandra, 
and the emission spectra from the blank fields about 30' away from 
LMC X-3 observed with the CCD camera (XIS) onboard Suzaku. 
The joint spectral fit of the data shows the hot gas attributed to the TAE component
cannot be  isothermal. Instead, 
a thick Galactic hot gaseous disk whose temperature and density decrease
exponentially  from the Galactic midplane can consistently explain the observations. 
They obtained
scale heights of 1.4 and 2.8 kpc, and  the midplane values of 0.31 keV and
$1.4 \times 10^{-3} ~{\rm cm}^{-3}$ for the temperature and the density, respectively.

A similar result was obtained  for 
the sight line toward Mrk 421, although the the X-ray emission is
mostly based on the ROSAT All Sky Survey data, thus the emission lines are
not resolved spectroscopically \citep{Yao_Wang_2007}.

In this paper, we analyze energy spectra of the soft X-ray diffuse emission of twelve different 
fields observed with Suzaku \citep{Mitsuda_etal_2007} and combine them with Suzaku results
of two other fields in literature.  
By virtue of the good energy response function of the CCD camera \citep{Koyama_etal_2007}
and the high sensitivity when combined with the X-ray mirror \citep{Serlemitsos_etal_2007},
we clearly detected
O\textsc{vii} emission from all the fields,
and O\textsc{viii} emission from most of them.  

In this paper we concentrate on the XIS1 data which has
much larger effective area below 1~keV than other XIS sensors because
it employs backside illuminated CCD.
Throughout this paper, we quote single parameter errors
at the 90~\% confidence level unless otherwise specified.

\section{Observations and data reduction}
\label{sec:obs_reduction}

\subsection{Observations and standard data screening}

\begin{table*}
\begin{footnotesize}
\begin{center}
\caption{Log of observations, ordered by $|b|$ \label{tbl:obs_log}}
\begin{tabular}{rllrrlccc}
\hline\hline
\multicolumn{2}{l}{Data set} & \multicolumn{1}{c}{Obs ID} &
 \multicolumn{1}{c}{Date} & \multicolumn{2}{c}{Exposure (ks)}  & 
\multicolumn{2}{c}{Aim point}\\
ID&Field Name (Short Name)      &                            &
 &       Total       & Cleaned  & ($\ell, b$) & ($E_{\rm Lon}, E_{\rm Lat}$)$^*$ \\\hline
{1}  & GB1428+4217 ({GB}) & 701092010 & Jun 12-13, 2006 & 48.7 &
 34.9 & (75.9, 64.9) & (194.2, 52.7)\\
{2}  &High latitude B ({HL-B}) & 500027020 & Feb 17-20, 2006&
 103.6 & 29.7 & (272.4, -58.3)& (4.4, -61.4)\\
{3}	&Lockman hole 2 ({LH-2}) &	101002010 & May 17-19,
 2006& 80.4  & 40.0 & (149.7, 53.2) & (137.1, 45.1) \\ 
{4}  & Lockman hole 1  ({LH-1})& 100046010 & Nov 14-15, 2005&
 77.0   &  61.7 & (149.0, 53.2) & (137.2, 45.5)\\
{5}	  &Off Filament$^{\rm a}$ ({Off-FIL}) &
 501001010 & Mar 1-2, 2006 & 80.1   & 59.6 &  (278.7, -47.1) & (354.8, -72.6) \\
{6}	  &On Filament$^{\rm a}$ ({On-FIL}) & 501002010
 & Mar 3-6 , 2006 &  101.4  & 59.2  &  (278.7 , -45.3) & (354.1, -74.4) \\
{7}	 & High latitude A ({HL-A}) & 500027010 & Feb 14-15,
 2006&  73.6  &  53.2  & (68.4, 44.4) & (228.8, 63.5)  \\
{8}  & MBM12 off cloud$^{\rm b}$ ({M12off} ) &
 501104010 & Feb 6-8, 2006 &   75.3   & 51.0 & (157.3, -36.8) & (44.5, 2.3) \\
{9}  &LMC X-3 Vicinity$^{\rm c}$ ({LX-3}) & 500031010 & Mar
 17-18, 2006& 	82.0 	 & 56.1 & (273.4, -32.6) & (41.2, -86.2) \\
{10}	 &North Ecliptic Pole 1$^{\rm d}$  ({NEP1}) &
 100018010 & Sep 2-4, 2005&  	106.2    & 58.7 & (95.8, 28.7) & (334.8,
 88.7) \\
{11}	 &North Ecliptic Pole 2  ({NEP2})
 & 500026010 &  Feb 10-12, 2006 & 75.6   	 & 16.5 & (95.8, 28.7) &
 (334.8, 88.7) \\
{12}   &Low latitude 86-21 ({LL21} ) & 502047010 & 	May
 9-10, 2007 &    	81.5 	 & 57.0 & (86.0, -20.8) & (347.6, 38.4) \\
{13}  &Low latitude 97+10 ({LL10}) & 503075010 & Apr 15-16,
 2008 &  79.8  & 40.8 &  (96.6, 10.4) & (0.7, 70.6) \\
\hline
{R1} &MBM12 on cloud$^{\rm b, e}$ ({M12on})   & 500015010 &
 Feb 3-6, 2006 & 102.9 &  68.0& (159.2, -34.5) & (47.2, 2.6) \\
{R2} &Midplane 235$^{\rm e}$ ({MP235}) &  502021010 & Apr
 22-25, 2007& 189.5 & 53.0 & (235.0, 0.0) & (119.5, -40.6) \\
\hline
\multicolumn{7}{l}{
\rlap{\parbox[t]{.9\textwidth}{
Results previously published by 
$^{\rm a}$~{\citet{Henley_etal_2007}},
$^{\rm b}$~{\citet{Smith_etal_2007}},
$^{\rm c}$~{\citet{Yao_etal_2009}},
$^{\rm d}$~{\citet{Fujimoto_etal_2007}},
$^{\rm e}$~{\citet{Masui_etal_2009}}.
}}}\\
\multicolumn{7}{l}{$^{\rm *}$~{Ecliptic coordinate}}
\end{tabular}
\end{center}
\end{footnotesize}
\end{table*}

In Table~\ref{tbl:obs_log}, we show the observations we used in this paper.
The analysis results of the seven data sets,  
Off Filament ({Off-FIL}), 
On Filament ({On-FIL}),
North Ecliptic Pole 1 ({NEP1}), 
MBM 12 off cloud ({M12off}),  
LMC X-3 Vicinity ({LX-3}), 
MBM 12 on cloud ({M12on}), 
and Midplane 235 ({MP235})
have been already
published.  

The data reduction done by \citet{Yao_etal_2009} for  {LX-3}  is 
consistent with that shown in this section. 
Since \citet{Fujimoto_etal_2007} ({NEP1}) and \citet{Smith_etal_2007} ({M12off}) 
used version 0.7 processed data, we re-analyzed the data from the data reduction. 
The second observation of the North Ecliptic Pole direction ({NEP2}) 
was made about a half year after the first NEP observation.  The aim points of
the two observations are identical, although the roll angle was rotated by 
about 180$^\circ$. 

The two fields at the bottom of Table~\ref{tbl:obs_log}, {M12on} and {MP235}, are
special directions compared to other twelve fields.
{M12on} is  towards the molecular cloud MBM12 located at $\sim 100$ pc with a
bright Cataclysmic Variable in the field of view behind the cloud, and {MP235} is
in the galactic plane at $b = 235^\circ$ with
a large ($N_{\rm H} = 9 \times 10^{21}~{\rm cm}^{-2}$) galactic absorption.
They therefore require spectral models different from those of the
other fields in this paper.   The spectral results from these two directions
have been already published \citep{Smith_etal_2007, Masui_etal_2009} and 
were analyzed in the same manner as used in this paper, including the geocoronal SWCX removal.  
We therefore simply
adopt their analysis, although we have updated their fits to {M12on} 
with more recent calibration data.  For these two fields we will simply show the
results in tables and figures and will not show the details of analysis in the text.

In all the observations, the XIS was set to the normal clocking mode and the data format was either
$3 \times 3$ or $5 \times 5$.  The Spaced-raw Charge Injection (SCI) was on for
Low latitude 86-21 ({LL21}) and Low latitude 97+10 ({LL10}).  
We used version 2.0 processed Suzaku data. 
We first cleaned the data using the selection criteria:  elevation from sunlit/dark  earth rim $>$ 20/5 deg, 
cut off rigidity $>$ 8 GV.
We checked the dependency of the 0.4 - 0.7 keV counting rate on the Oxygen column density of the sunlit atmosphere in the line of sight using the MSIS atmosphere model (see \cite{Fujimoto_etal_2007, Smith_etal_2007, Miller_etal_2008}).  We found  that the counting rate was constant as a function of the column density for the cleaned data.   Thus there is no significant neutral O emission from Earth atmosphere in the filtered data.

\subsection{Removal of Point sources}

We then constructed an X-ray image in 0.3 to 2~keV energy range.  We detected
point sources in the $17.'8 \times 17.'8$ field of view  for all the data sets except for {LL10}.
We removed a circular region centered at the
source position from the further analysis.   The radius of the circular region was
determined so that the counts from
the point source outside the circular region becomes  less than 
3 \%  of the diffuse X-ray emission in 0.3 to 1~keV energy range.  
The sum of contribution of point sources is estimated to be
less than 6\% of the diffuse emission for all the fields except for 
{LH-1} and {LL21}.  The contributions of the two fields,
{LH-1} and {LL21}, are respectively 8 and 12 \%.  
For these fields, 
we analyzed the point source spectra and used the mirror point spread function 
to determine their resiual contribution in O\textsc{vii} and O\textsc{viii}.
We extracted sum of point-source spectra 
from the circular source regions and performed model fittings.  
As the background spectra, we used the spectra extracted from source-free
circular regions whose distances from the optical axis of the telescope are
equivalent to the source regions, respectively.  
We estimated the upper limits of O\textsc{vii}
and O\textsc{viii} emission intensities of the source regions with spectral fits,
then their contamination to the SXDB spectra.  The upper limit
of contamination was 0.4 LU for O\textsc{vii} emission for the both fields.
It was 0.2 LU and 0.6 LU for O\textsc{viii} emission for 
{LH-1} and {LL10}, respectively. 
For GB1428+4217 ({GB}) a point source of an intensity of 
$2.6 \times 10^{-13}$ ${\rm erg~s}^{-1}~{\rm cm}^{-2}$ in the 
energy band of  0.3 to 1 keV 
( $22 \times 10^{-13}$ ${\rm erg~s}^{-1}~{\rm cm}^{-2}$ in 1 to 10 keV)
was detected at the center of the field of view.  
We removed a circular region of  a 5 arc minute radius for this observation.  
The contribution of the source in the counting rate of the remaining region is 
estimated to be only  0.2 \%.

\subsection{Removal of geocoronal SWCX}
\label{subsec:remov_geocorona}
The spectrum below 1~keV could be
contaminated by the SWCX induced emission from the geocorona (e.g. 
\citet{Cravens_etal_2001}).
\citet{Fujimoto_etal_2007} reported a flare-like increase of X-ray flux during the Suzaku
observation of  {NEP1}, which was clearly visible in the light curve
of 256 s time bins. In all other observations, the increase is too small to be recognized 
in the light curve of 256 or 512 s time bins, but still it could be statistically 
significant
in the X-ray spectra when they are  integrated over a certain amount of time
 (e.g. $\gtrsim 30$ ks).
Thus as the last stage of the data reduction, we removed time intervals in which the 
data can be contaminated by the SWCX from the geocorona.

We consider  two parameters which are related to Suzaku's
geocoronal SWCX inensity.  The first one is the solar wind flux near the Earth.
We calculated the solar-wind proton flux at the Earth
using ACE SWEPAM data\footnote{The data available at 
http://www.srl.caltech.edu/ACE/ACS/}.  
When the ACE SWEPAM data are not available, 
we used WIND SWE data \footnote{The data available at \\
http://web.mit.edu/afs/athena/org/s/space/www/wind.html} or OMNI data from CDAWeb (Coordinated Data Analysis Web) \footnote{The data available at 
http://cdaweb.gsfc.nasa.gov/cdaweb/sp\_phys/}. 

The second parameter is the Earth-center to magnetopause (ETM) distance,
where the magnetopause is defined as the lowest position along the sight line of Suzaku where 
geomagnetic field is open to interplanetary space. 
The ETM distance is in the range of $\sim 2$ to $\sim 10$ 
Earth radii ($R_{\rm E}$ ), while 
Suzaku is in a low earth orbit of an about 650 km altitude.
Thus it
is always looking at the sky through the magnetopause. 
The probability of a contamination in Suzaku
spectra by the SWCX from the geocorona increases if the shortest 
Earth-center to magnetopause (ETM) distance is 
$\lesssim $ 5
$R_{\rm E}$  \citep{Fujimoto_etal_2007}.  
We calculated the ETM distance
every 256 s for all the observation periods using the  T96 magnetic field model
\citep{Tsyganenko_Synov_2005} .  We obtained the interplanetary plasma parameters
required for the calculations from the CDAWeb.  

When the sun is quiet the solar wind proton flux stays in the range of 
$(1 - 4) \times 10^8~ {\rm protons~s}^{-1} ~{\rm cm}^{-2}$, and shows slow (times scales of 
$\sim 10$ ks) time variations.  On the other hand, during flaring events, it increases and
can go up
to $\sim 5 \times 10^9~ {\rm protons~s}^{-1} ~{\rm cm}^{-2}$, and shows fast ($\lesssim 1$ ks)
time variations.

For the four observations, GB1428+4217  ({GB}), Lockman hole 1 ({LH-1}), 
On Filament ({On-FIL}) and Low latitude 86-21({LL21}), 
the proton flux was always below $4 \times 10^8~ {\rm protons~s}^{-1} ~{\rm cm}^{-2}$.  
Among them, during {GB} and {On-FIL} observations, the proton flux showed slow variation and the data sets can be divided into two subsets of $(1-2) \times 10^9~ {\rm protons~s}^{-1} ~{\rm cm}^{-2}$ and $(2-3) \times 10^9~ {\rm protons~s}^{-1} ~{\rm cm}^{-2}$ for {GB},
and of $(2-3) \times 10^9~ {\rm protons~s}^{-1} ~{\rm cm}^{-2}$ and $(3-4) \times 10^9~ {\rm protons~s}^{-1} ~{\rm cm}^{-2}$ for {On-FIL}, respectively.
Both during the  {GB} and {On-FIL}  observations, the ETM distance varied in 2 to 10 $R_{\rm E}$ range.  We created spectra for the subsets and performed spectral fits to determine the O\textsc{vii} and O\textsc{viii} emission intensities.  The fitting model used in this analysis
will be described in detail in the next section (\ref{subsec:det_line_intensities}).
We found the difference in the O\textsc{vii} line intensities was less than 2 $\sigma$
for both fields.  This corresponds to 
1.7 LU and 1.3 LU for  {GB} and {On-FIL}, respectively.
For O\textsc{viii}, the difference was within 1 $\sigma$, which is 0.5 and 0.6 LU
 for   {GB} and {On-FIL}, respectively.
 
\begin{figure}
\vspace{5cm}
\begin{center}
\FigureFile(0.8\columnwidth, ){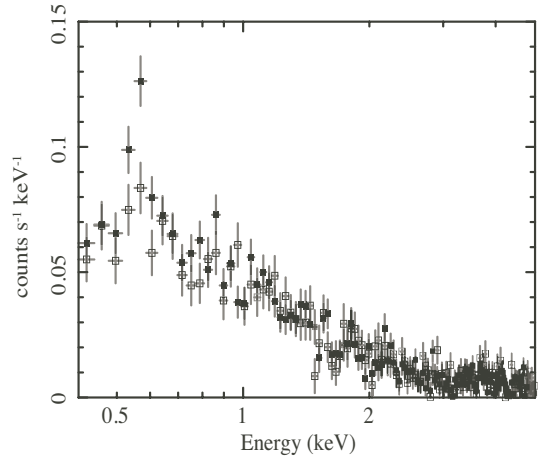}
\end{center}
\caption{
Energy spectra of High Latitude B ({HL-B})  for the time intervals with solar wind proton flux
higher (filled squares) or lower (open squares)  than $ 4 \times 10^8~
 {\rm protons~s}^{-1} ~{\rm cm}^{-2}$.
The vertical error bars are the $1-\sigma$ statistical errors.  
The non X-ray background was subtracted.
There are large ($> 3~\sigma$)
discrepancies between the O\textsc{vii} emission ($\sim 0.56$ keV) 
intensities of two spectra, which indicates significant
geocoronal SWCX contamination in the higher proton-flux data.
\label{fig:HLB_compared}}
\end{figure}

Among all other observations, in part  of  {HL-B}, {HL-B}, and  {LH-2} observations, 
the proton flux
stayed below  $4 \times 10^8~ {\rm protons~s}^{-1} ~{\rm cm}^{-2}$, and 
showed variations of
more than $1 \times 10^8~ {\rm protons~s}^{-1} ~{\rm cm}^{-2}$ was observed.  
We thus also subdivided the data sets according to the proton flux.  
We found that the variations of O\textsc{vii} and O\textsc{viii} intensities
were within 2 $\sigma$ level for all those data sets. 
We thus conclude that when the proton flux is 
$< 4 \times 10^8~ {\rm protons~s}^{-1} ~{\rm cm}^{-2}$
the contamination of geocoronal SWCX is relataively small, although there still could remain
possibility of 
contamination of $\sim 1.5$ LU and $\sim 0.5$ LU levels for O\textsc{vii} and O\textsc{viii},
respectively.  
We thus decided to use all the data for the four observations, 
 ({GB}, {LH-1}, {On-FIL}), and {LL21},

For the seven observations,
High latitude B  ({HL-B}),  Lockman hole 2 ({LH-2}), 
Off Filament ({Off-FIL}),
LMC X-3 Vicinity ({LX-3}),  North Ecliptic Pole ({NEP1}),  the second
North Ecliptic Pole observation 
({NEP2}) and
Low latitude 97+10 ({LL10}), 
the proton flux shows shows flare-like increase and exceeds
$4 \times 10^8~ {\rm protons~s}^{-1} ~{\rm cm}^{-2}$  in about a half of the obsevation.
We thus 
subdivided those data into two subsets according to the proton flux, and determined the
O\textsc{vii} and O\textsc{viii} intensities.   
We found significant (more than $2~\sigma$ in O\textsc{vii} or O\textsc{viii}) difference between the two spectra for four cases:
{HL-B}, {LH-2}, {NEP1} and {NEP2}. 
An example of spectral comparisons is shown in  
Figure~\ref{fig:HLB_compared}. 
For these four data sets,  we decided to use only the time intervals 
in which the proton flux  was lower than  $4 \times 10^8~ {\rm protons~s}^{-1} ~{\rm cm}^{-2}$.
These results are consistent with the geocoronal SWCX contamination reported for
Suzaku observations of 4U1830-303 vicinities \citep{Mitsuda_etal_2008}.

\newcommand{\mfw}{0.08\textwidth}
\begin{table*}
\begin{center}
\caption{Summary of Geocoronal SWCX removal process\label{tbl:geocorona_log}}
\begin{tabular}{lllllll}
\hline\hline
\parbox[t]{0.15\textwidth}{ID}  &\parbox[t]{\mfw}{(1)} &\parbox[t]{\mfw}{(2)}  
    & \parbox[t]{\mfw}{(3)}   & \parbox[t]{\mfw}{(4)}  & \parbox[t]{\mfw}{(5)} & \parbox[t]{\mfw}{(6)}\\\hline
{1}  ({GB})              & \includegraphics[angle=0,width=2.5mm]{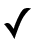}\\
{2} ({HL-B})           &               & \includegraphics[angle=0,width=2.5mm]{check.eps}\\
{3} ({LH-2})  &                & \includegraphics[angle=0,width=2.5mm]{check.eps}\\
{4}  ({LH-1}) &\includegraphics[angle=0,width=2.5mm]{check.eps} \\
{5}	({Off-FIL})       &               &                 &              & \includegraphics[angle=0,width=2.5mm]{check.eps}\\
{6}	 ({On-FIL})      & \includegraphics[angle=0,width=2.5mm]{check.eps}\\
{7}  ({HL-A})          &                  &               &               &                 &\includegraphics[angle=0,width=2.5mm]{check.eps}\\
{8}  ({M12off} )    &                &                 &               &                 & \includegraphics[angle=0,width=2.5mm]{check.eps} \\
{9}   ({LX-3})               &               &                & \includegraphics[angle=0,width=2.5mm]{check.eps}\\
{10}	 ({NEP1})       &                & \includegraphics[angle=0,width=2.5mm]{check.eps}\\
{11} ({NEP2})  &               & \includegraphics[angle=0,width=2.5mm]{check.eps}\\
{12} ({LL21} )   & \includegraphics[angle=0,width=2.5mm]{check.eps}\\
{13}  ({LL10})     &                &                 &             & \includegraphics[angle=0,width=2.5mm]{check.eps}\\\hline
{R1}  ({M12on})     &                &                 &             &                    &                 & \includegraphics[angle=0,width=2.5mm]{check.eps}\\
{R2} ({MP235})   &                    & \includegraphics[angle=0,width=2.5mm]{check.eps}\\
\hline
\multicolumn{6}{l}{
\rlap{\parbox[t]{.77\textwidth}{
(1) The solar wind flux was alway $< 4 \times 10^8~ {\rm protons~s}^{-1}~{\rm cm}^{-2}$.  All time intervals were adopted.
}}}\\
\multicolumn{6}{l}{
\rlap{\parbox[t]{.77\textwidth}{
(2) Excess in the  high-solar-wind-flux spectrum.  Time intervals with high solar wind flux were discarded.
}}}\\
\multicolumn{6}{l}{
\rlap{\parbox[t]{.77\textwidth}{
(3) The geomagnetic field was open to the anti-sun side, when the ETM distance was low ($< 5 R_{\rm E}$).  All time intervals were adopted.
}}}\\
\multicolumn{6}{l}{
\rlap{\parbox[t]{.77\textwidth}{
(4) The geomagnetic field was occasionally open to the sun side when the  ETM distance was low.  Time intervals with high solar wind flux were discarded.
}}}\\
\multicolumn{6}{l}{
\rlap{\parbox[t]{.77\textwidth}{
(5) Proton flux stayed in  $3 - 6 \times 10^8~ {\rm protons~s}^{-1}~{\rm cm}^{-2}$. 
 Time intervals with $ {\rm ETM ~distance} < 5 R_{\rm E}$ were discarded.
}}}\\
\multicolumn{6}{l}{
\rlap{\parbox[t]{.77\textwidth}{
(6) Proton flux stayed in  $3 - 7 \times 10^8~ {\rm protons~s}^{-1}~{\rm cm}^{-2}$. 
ETM distance was always  $ > 10 R_{\rm E}$.  All time intervals were used.
}}}\\

\end{tabular}
\end{center}
\end{table*}

In order to find out why no difference was found in the remaining 
three observations, we checked the ETM distance of these observations.
As already reported in \citet{Yao_etal_2009}, 
the ETM distance became as short as $ 1.4 R_{\rm E}$
during the {LX-3}  observation.  However, 
the magnetic field
was open to anti-sun direction when the ETM distance was  
$ < 5 R_{\rm E}$ during which solar-wind particles cannot penetrate. 
This explains
why there was no significant difference between the high and low 
proton-flux spectra for this data set.
We thus decided to adopt all the data for this observation.
During the {LL10} and {Off-FIL} observation, 
the magnetic field was occasionally open to sun side when the ETM distance dropped 
below $5 R_{\rm E}$.  
We decided to simply discard the time intervals with high proton fluxes for these data
because we still 
have enough statistics.  

There remain two data sets, High latitude A  ({HL-A}) and MBM12 off cloud ({M12off}).
During the {HL-A} observation, 
the proton flux varied
in the range of $3 - 6 \times 10^8 ~ {\rm protons~s}^{-1} ~{\rm cm}^{-2}$ but the
time interval exceeding $4 \times 10^8 ~ {\rm protons~s}^{-1} ~{\rm cm}^{-2}$
was short ($\sim 1/8$ of data).  
During the {M12off} observation, it stayed at 
a level of $\sim 5 \times 10^8 ~ {\rm protons~s}^{-1} ~{\rm cm}^{-2}$.  
Thus for these data sets,  we can not compare spectra with low and high proton fluxes. 
For {HL-A}, we subdivided the data sets according to the ETM distance and created two energy spectra with ETM distances = $5 - 10  R_{\rm E}$ and $> 10  R_{\rm E}$.  
We found that the two spectra with different ETM distances show no significant difference.
Thus we will use data with the ETM distances $>5  R_{\rm E}$.  There was a 
small fraction of the data with the ETM distance $< 5 R_{\rm E}$, which we decided to discard.   
For {M12off}, the ETM distance stayed in $7 - 9 R_{\rm E}$.  Thus further subdivision was not possible, and we decided to use all the data. 

We tried to exclude time intervals for which contamination of time variable
geocoronal SWCX was significant.
However we should keep it in mind that there is still uncertainty of contamination by geocoronal SWCX at  1.5 LU and 0.5 LU levels for  O\textsc{vii} and O\textsc{viii}, respectively.   In addition, the last data set, {M12off}, could  have
larger contamination.
In Table~\ref{tbl:geocorona_log}, we summarize the process we applied to remove time intervals which are suspected to be contaminated by the SWCX from the geocorona.  We also show the process for {M12on} 
and {MP235} fields in the table.

\subsection{Non X-ray background}

The non X-ray background  (NXB) spectrum was constructed from the dark Earth database
using the standard method in which the cut off rigidity distributions of the on-source
and the  background data were made identical \citep{Tawa_etal_2008}.   We found 
5 to 10 \% discrepancies in the counting rates above 10~keV 
between  the NXB and the observation
data, suggesting background uncertainty of this level.  
We also analyzed front-iluminated CCD data (XIS0, XIS2, and XIS3. but XIS2 was
not available after November 9, 2006) to determine the intensity of the
Cosmic X-ray Background (CXB) above 2 keV for each sensors.  
We determined the CXB intensity
by the spectral fits.  
The model function will be described  in section 
\ref{sec:analysis_results}.
The intensities determined by different sensors were  consistent
with each other within the 90\% statistical errors 
for all observations except for {NEP2}.
For this observation, the intensity from XIS1 were not consistent with those
from other three sensors.  
It was necessary to adjust the NXB level
by about 30\% for XIS1 to obtain consistent result.  
This suggests that not only the NXB level
but also its spectral shape may vary from the one extracted from the
standard NXB data base.  The CXB intensity of {NEP2} after the NXB correction
was consistent with the CXB intensity estimated from {NEP1} data.
Since in the energy range below 1~keV, the non X-ray background
is only about 10 \% of the diffuse X-ray background, the spectral
parameters derived in section \ref{sec:analysis_results} are not
sensitive to this level of background 
uncertainty.

In Suzaku NXB non-subracted energy spectra, there are three
 instrumental lines at 1.486,
1.740,  and 2.123 keV 
\citep{Tawa_etal_2008}.
Since the line intensity and the line spread function of the detector vary with time, 
the residuals of those lines appears in the NXB subtracted spectra as lines or line-like structures in  the 1.3 to 2.1 keV range. 
In order to study those features in our spectra, 
we first fitted all the NXB subtracted spectra 
in the energy range 1.1 to 5~keV.  
We employed a power-law function absorbed with a Galactic absorption for the
continuum and narrow Gaussian functions for the NXB lines.  
We found that we need to include a line at 1.828 keV for {HL-B},
and a line at 2.157 keV for {LL10}  to obtain an acceptable fit.
For other data sets, such a line was not required.
In the further spectral fits, we include  these Gaussian functions with the parameters fixed
at the best fit values for these two data sets.  The results of the further analysis do 
not depend on inclusion or non-inclusion of these lines in the models of spectral fits. 

\subsection{Arf and rmf files}

In order to perform spectral fitting, we
generated an efficiency file (arf file) for a spatially uniform emission, using
the xissimarfgen software version 2008-04-05 
 \citep{Ishisaki_etal_2007},
assuming a 20$'$-radius flat field as the input emission of the generator.  
A pulse height re-distribution matrix (rmf file) was created by 
the script xisrmfgen version 2007-05-14.  The calibration files,
ae\_xi1\_quanteff\_20070402.fits, ae\_xi1\_rmfparam\_20080311.fits, ae\_xi1\_makepi\_20080131.fits, and ae\_xi1\_contami\_20071224.fits were used to generate the arf and rmf files.

\subsection{Contamination of the optical blocking filter}

The degradation of low energy efficiency due to the contamination on the XIS
optical blocking filter was included in the arf file. The gradient of the 
contaminant thickness over  the optical blocking filter is also taken into account.   
Systematic errors 
in the contaminant thickness are estimated to be about 10 \%
(The Suzaku Technical Description\footnote{available from http://www.astro.isas.jaxa.jp/suzaku/doc/suzaku\_td/}, see also \citet{Fujimoto_etal_2007} for early data and \citet{Yamasaki_etal_2009} for recent data).
We performed all the analysis described in the next section not only for the nominal contaminant thicknesses but also assuming  10\% thicker and thinner contaminants.  
Because the contaminant thickness is gradually increasing with time, 
we created the arf files of 10\%  thicker or thiner contaminants by shifting the observation dates from the real dates, except for  the 10 \% thicker cases of 
{LL21} and {LL10}.  
For the two cases, 
 the +10 \% thickness is larger than that of the most recent date in the latest calibration database.  
We thus added an absorption model in the model spectra to represent the extra contamination.  
For all data sets, we found that the best-fit parameters changed by only small amount when we varied the assumed contamination thickness by 
$\pm 10$ \%.  Among the parameters, the O\textsc{vii} emission intensity is most sensitive to the contaminant thickness.  The  O\textsc{vii} emission intensity varied at most by 0.8 LU
 (LU = ${\rm photons~s}^{-1} {\rm cm}^{-2} {\rm str}^{-1}$), while the statistical errors
 were 0.5 to 1. 5 LU.   Thus we will show only the results with nominal contamination thickness.

\newcommand{\figw}{0.4}
\begin{figure*}
\newcounter{keepfignumhs}
\setcounter{keepfignumhs}{\value{figure}}
\begin{scriptsize}
\begin{tabular}{lll}
\begin{minipage}{\figw\textwidth}
({1}) GB1428+4217\vspace{-0.055\textwidth}\\
\FigureFile(0.99\textwidth, ){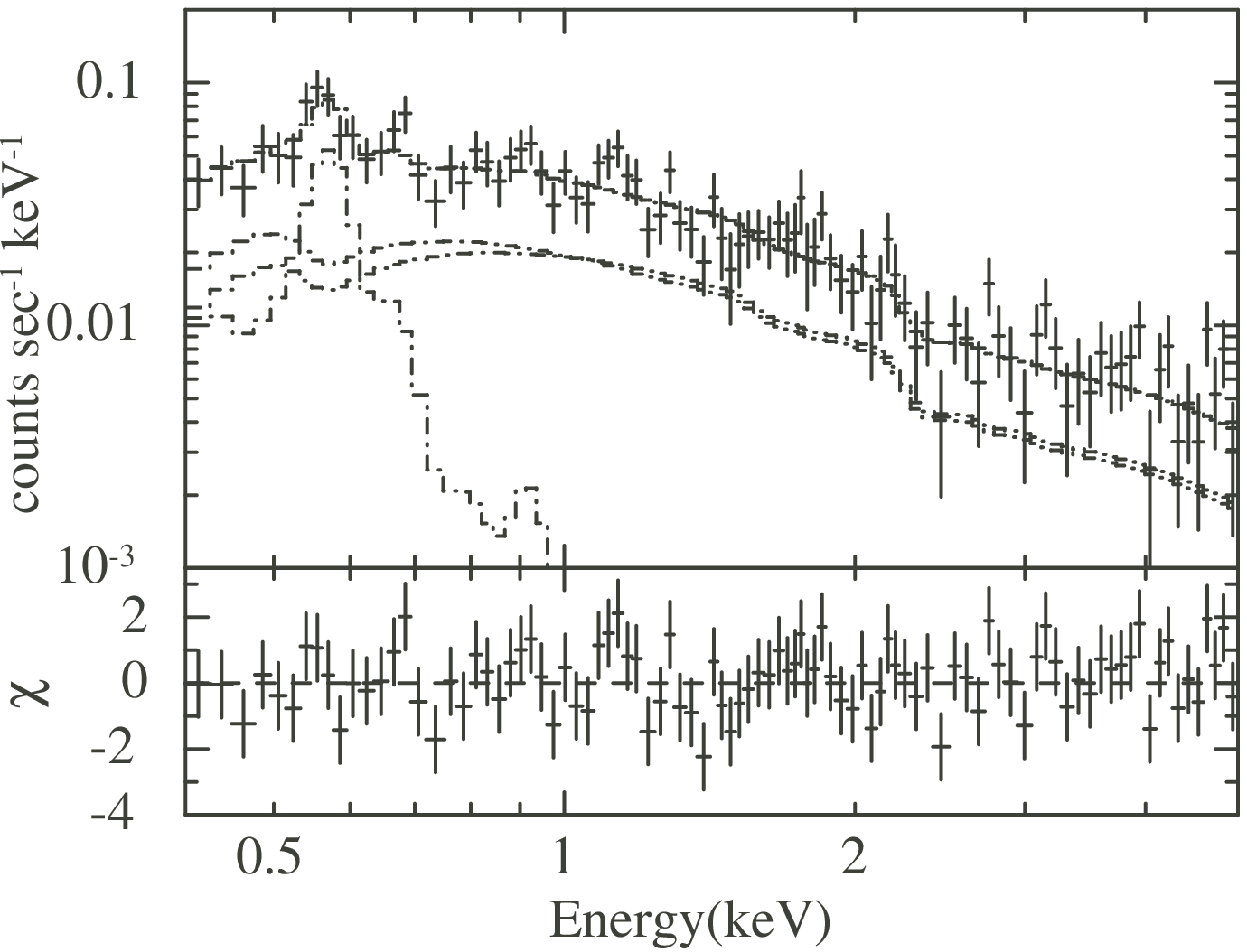}
\end{minipage}
 &
\begin{minipage}{\figw\textwidth}
({2}) High latitude B\vspace{-0.055\textwidth}\\
\FigureFile(0.99\textwidth, ){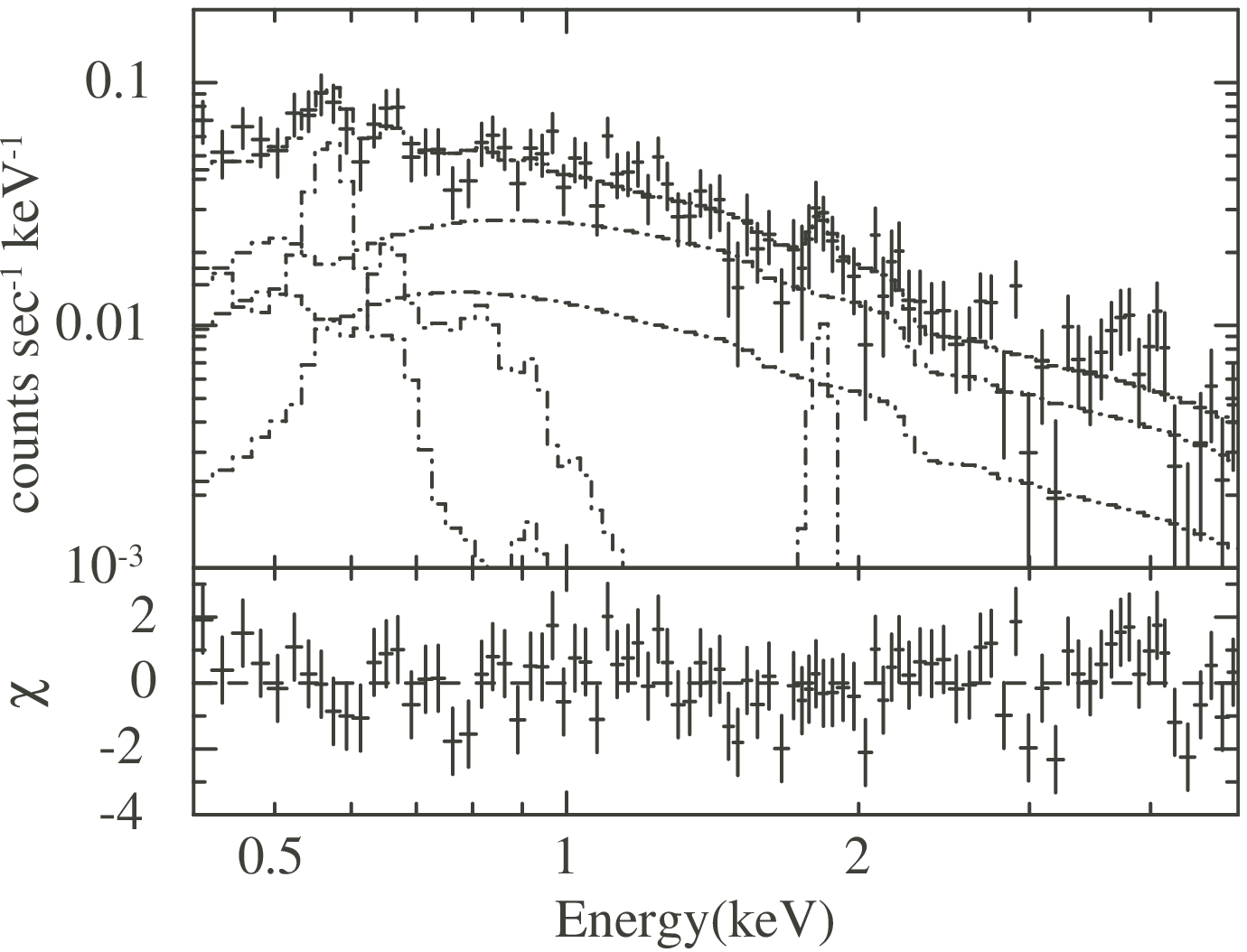}
\end{minipage}
\\
\begin{minipage}{\figw\textwidth}
({3}) Lockman hole 2\vspace{-0.055\textwidth}\\
\FigureFile(0.99\textwidth, ){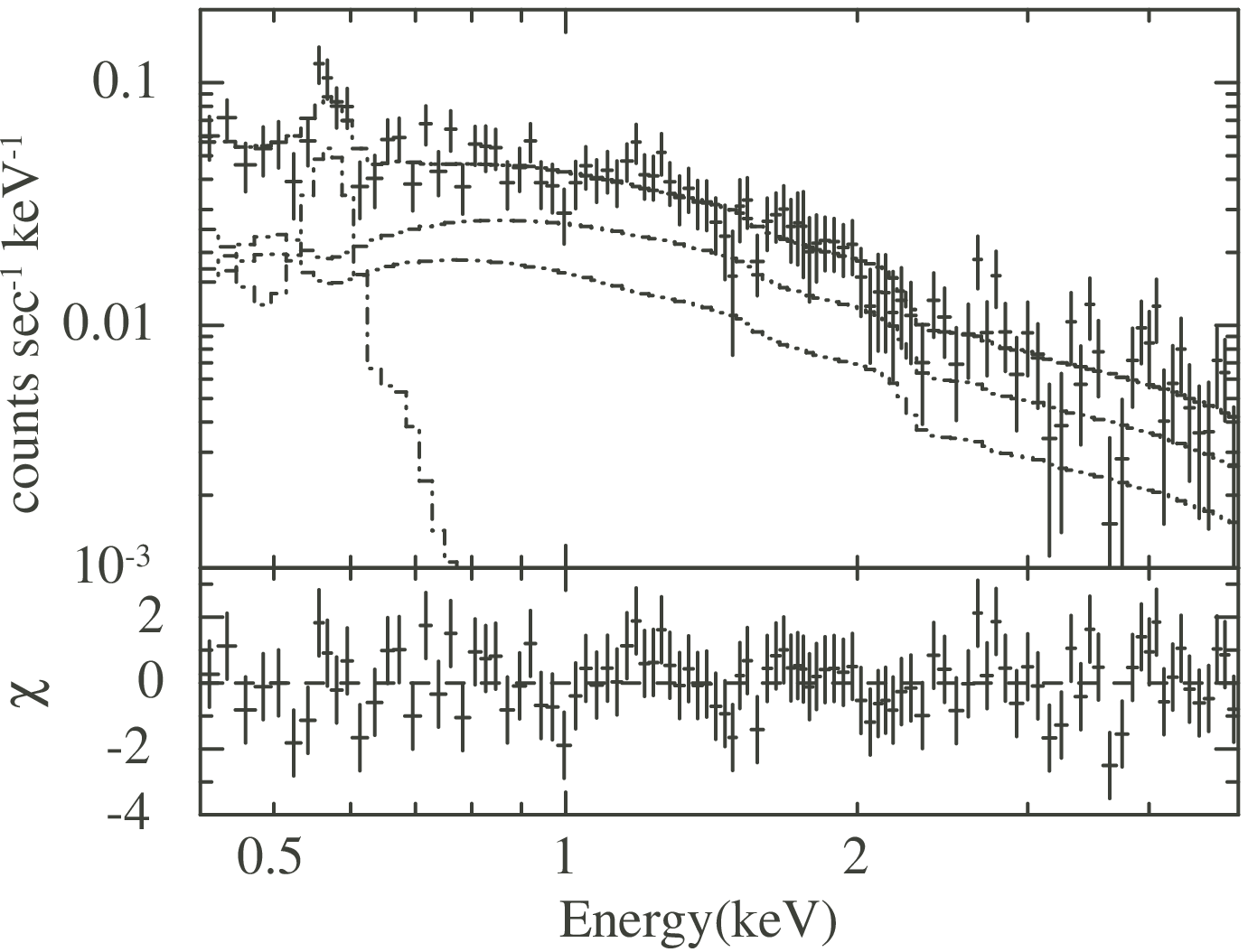}
\end{minipage}
&
\begin{minipage}{\figw\textwidth}
({4})  Lockman hole 1\vspace{-0.055\textwidth}\\
\FigureFile(0.99\textwidth, ){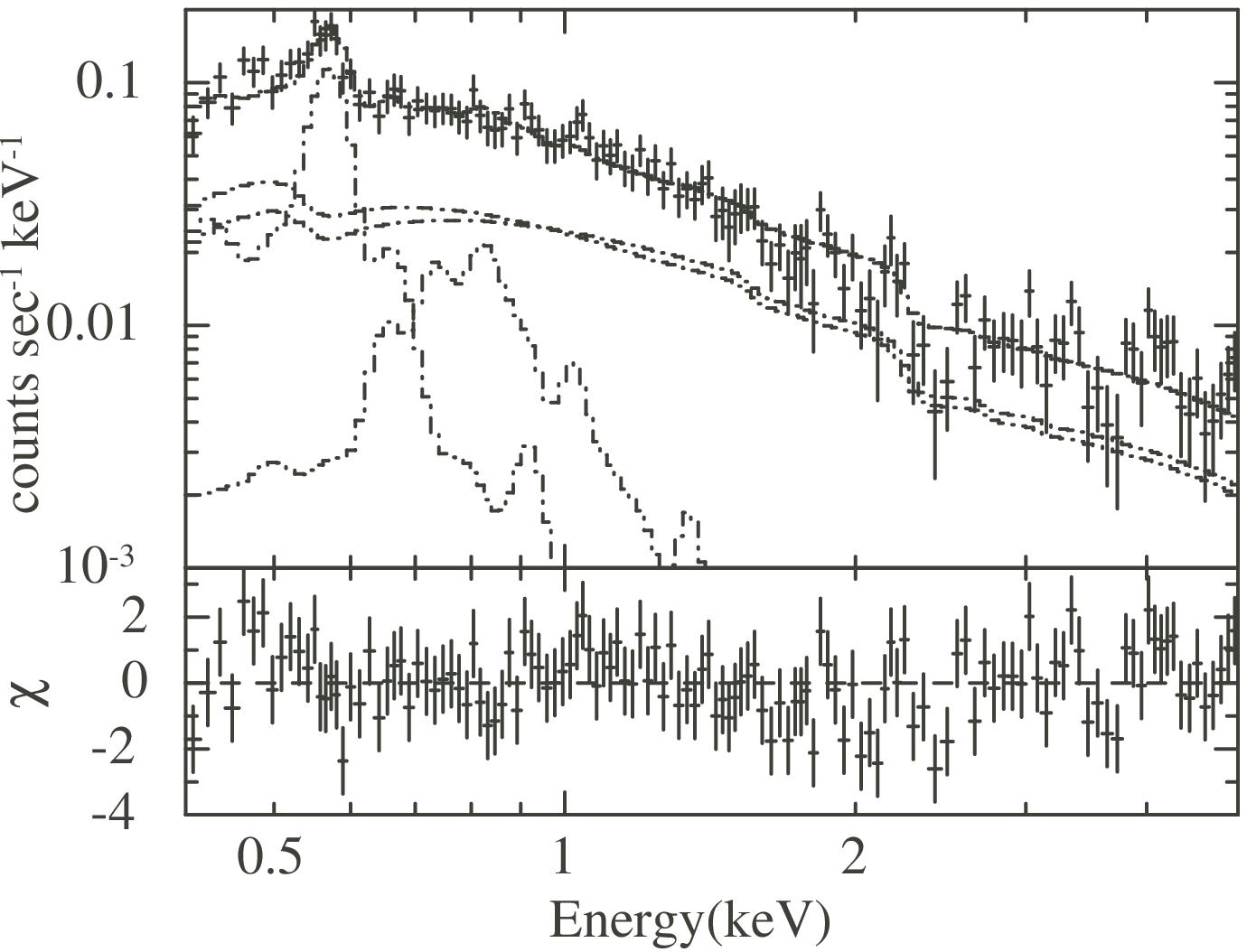}
\end{minipage}
\\
\begin{minipage}{\figw\textwidth}
({5}) Off Filament\vspace{-0.055\textwidth}\\
\FigureFile(0.99\textwidth, ){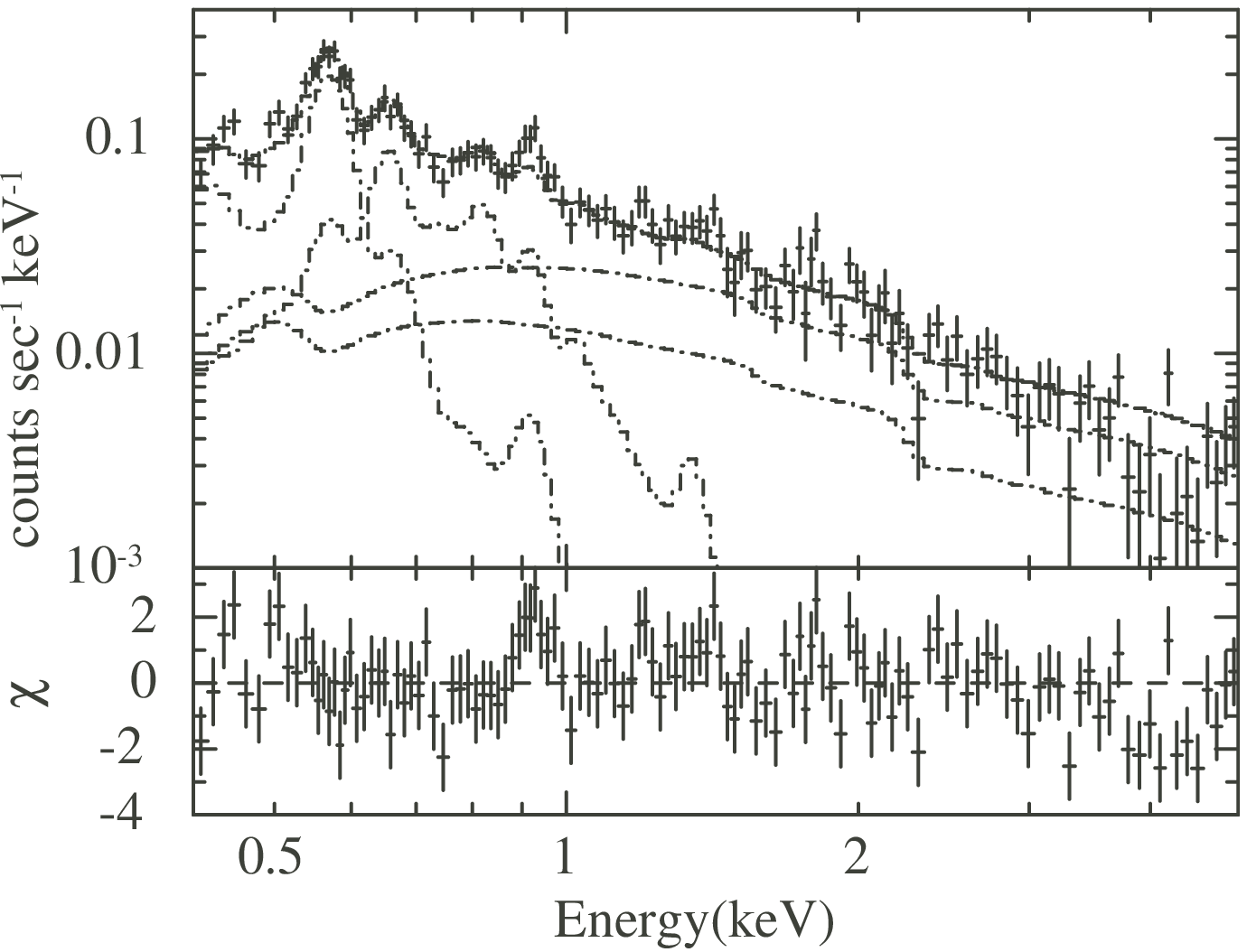}
\end{minipage}
&
\begin{minipage}{\figw\textwidth}
({6}) On Filament\vspace{-0.055\textwidth}\\
\FigureFile(0.99\textwidth, ){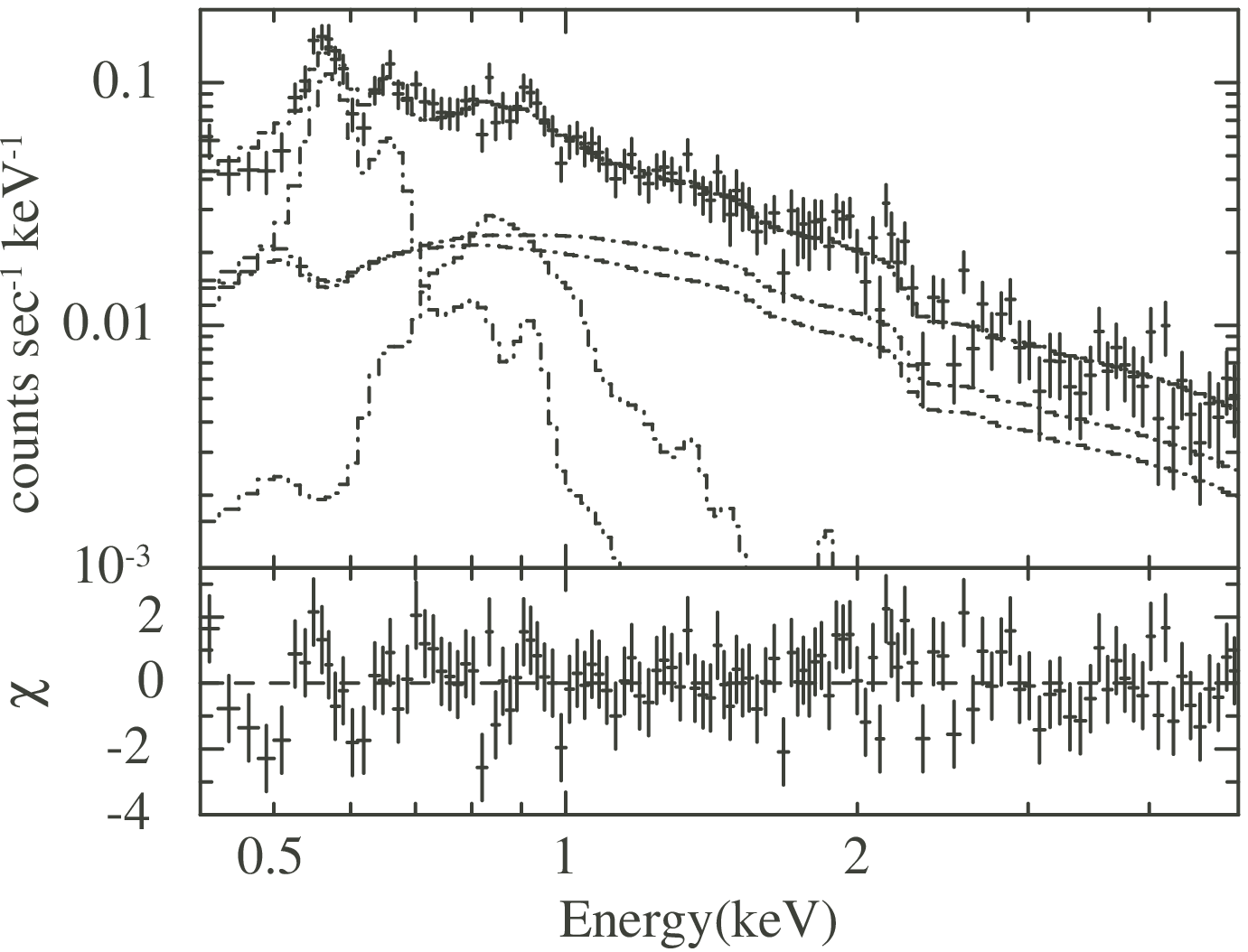}
\end{minipage}
\end{tabular}
\end{scriptsize}
\caption{
Observed spectra (crosses) with NXB subtracted, best-fit
 model~1 and its components (step functions)  convolved with the
 instrument response function and residuals of the fit (bottom panels).
 The model function consists of three spectral components, CXB,
 TAE and SWCX+LHB. The TAE and SWCX+LHB were respresented by thin
 thermal emission model of solar abundance. The CXB component was represented by a double broken power-law model. Thus there are two curves for this component.  The vertical error bars of data points correspond to the  $1~\sigma$ statistical errors.  Some of the spectra show
 a large excess over the model at Ne K emission ($\sim 0.9$ keV). The hard spectral component which is dominant above $\sim$ 2 keV for {M12on} represents the Cataclysmic Variable behind the MBM-12 molecular cloud.  The observed spectrum of {MP235} is found in Figure~\ref{fig:spectrum_fixedfit}.
\label{fig:spectrum}}
\end{figure*}

\addtocounter{figure}{-1}

\begin{figure*}
\begin{scriptsize}
\begin{tabular}{lll}
\begin{minipage}{\figw\textwidth}
({7}) High latitude A\vspace{-0.055\textwidth}\\
\FigureFile(0.99\textwidth, ){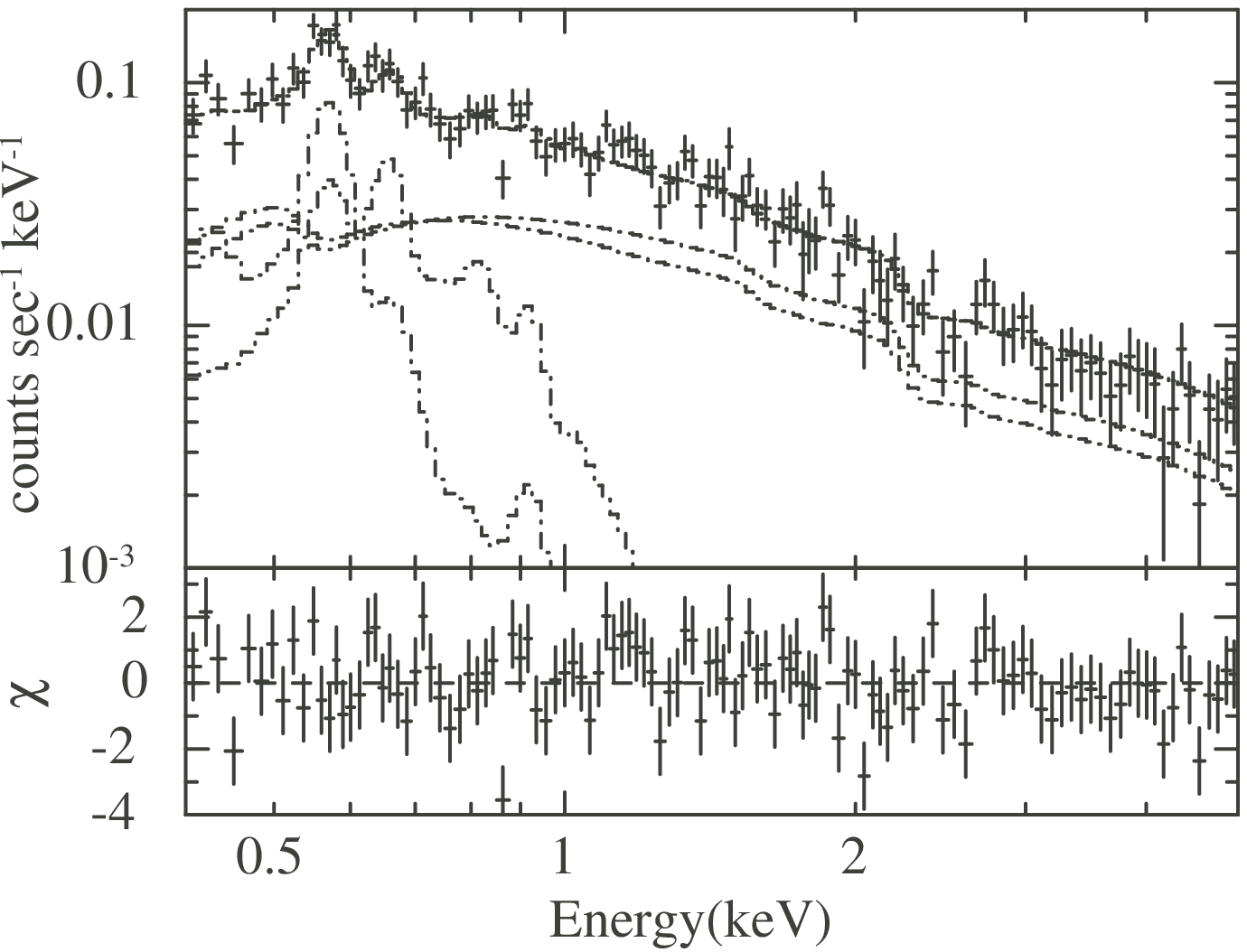}
\end{minipage}
&
\begin{minipage}{\figw\textwidth}
({8}) MBM12 off cloud\vspace{-0.055\textwidth}\\
\FigureFile(0.99\textwidth, ){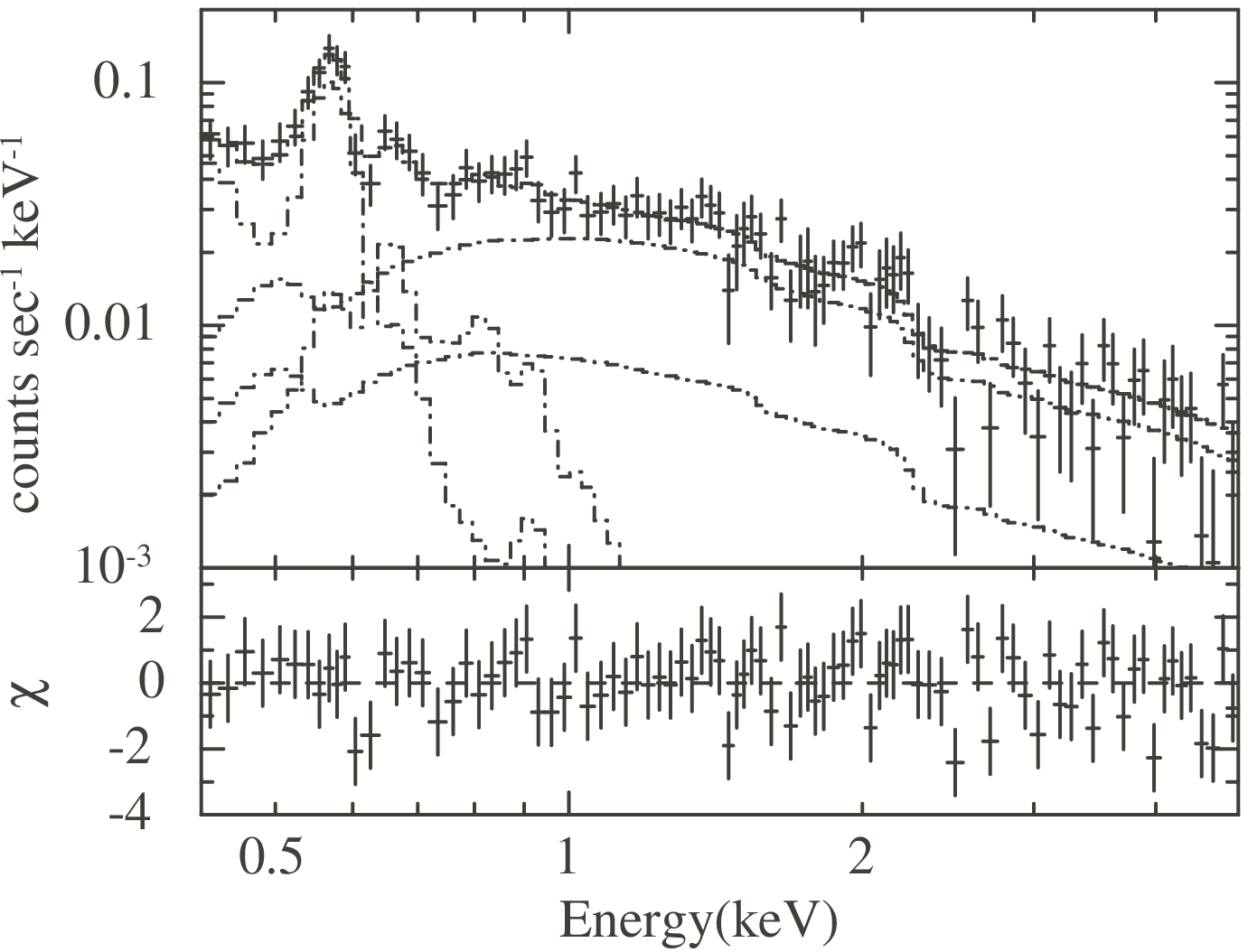}
\end{minipage}
\\
\begin{minipage}{\figw\textwidth}
({9}) LMC X-3 Vicinity\vspace{-0.055\textwidth}\\
\FigureFile(0.99\textwidth, ){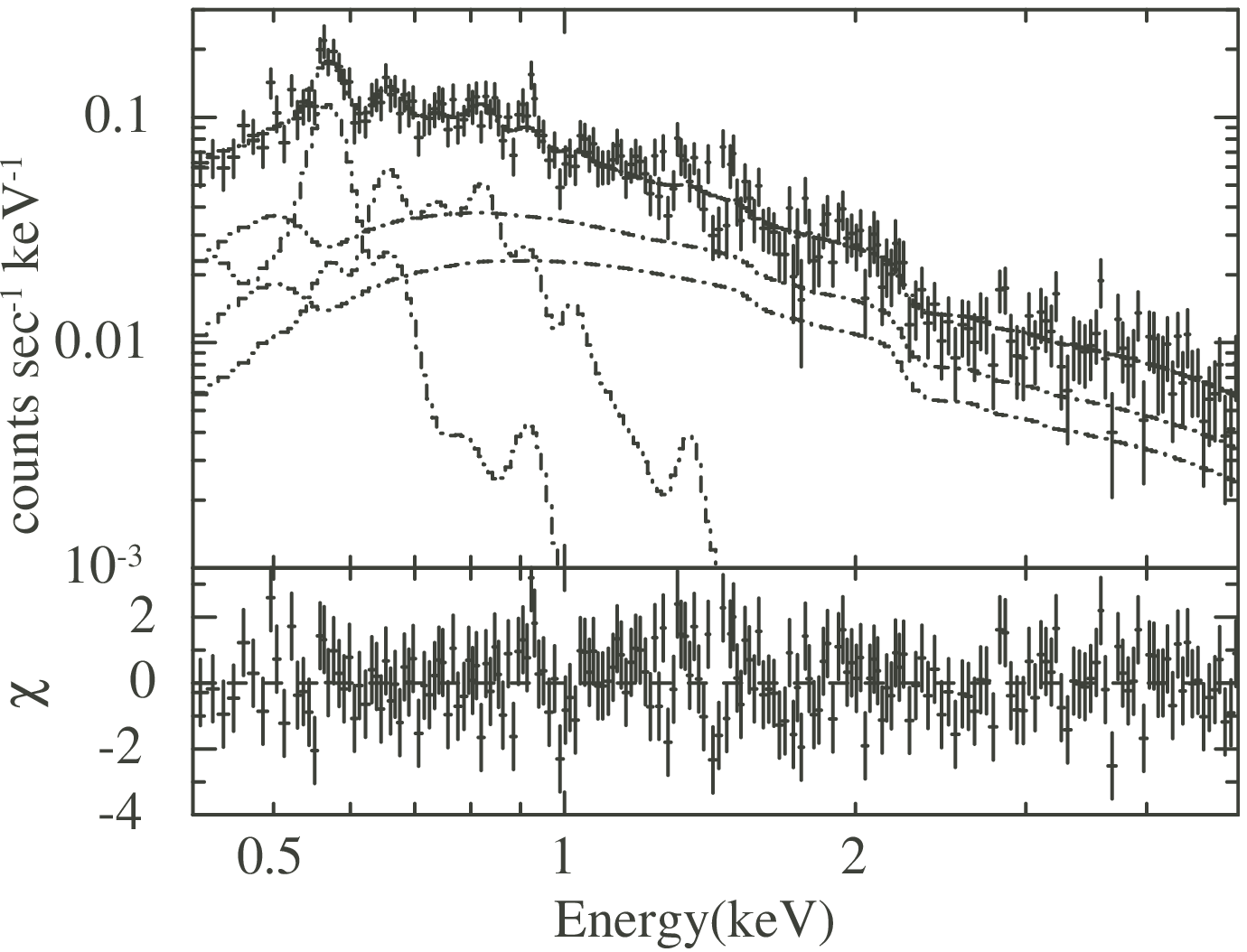}
\end{minipage}
&
\begin{minipage}{\figw\textwidth}
({10})  North Ecliptic Pole 1\vspace{-0.055\textwidth}\\
\FigureFile(0.99\textwidth, ){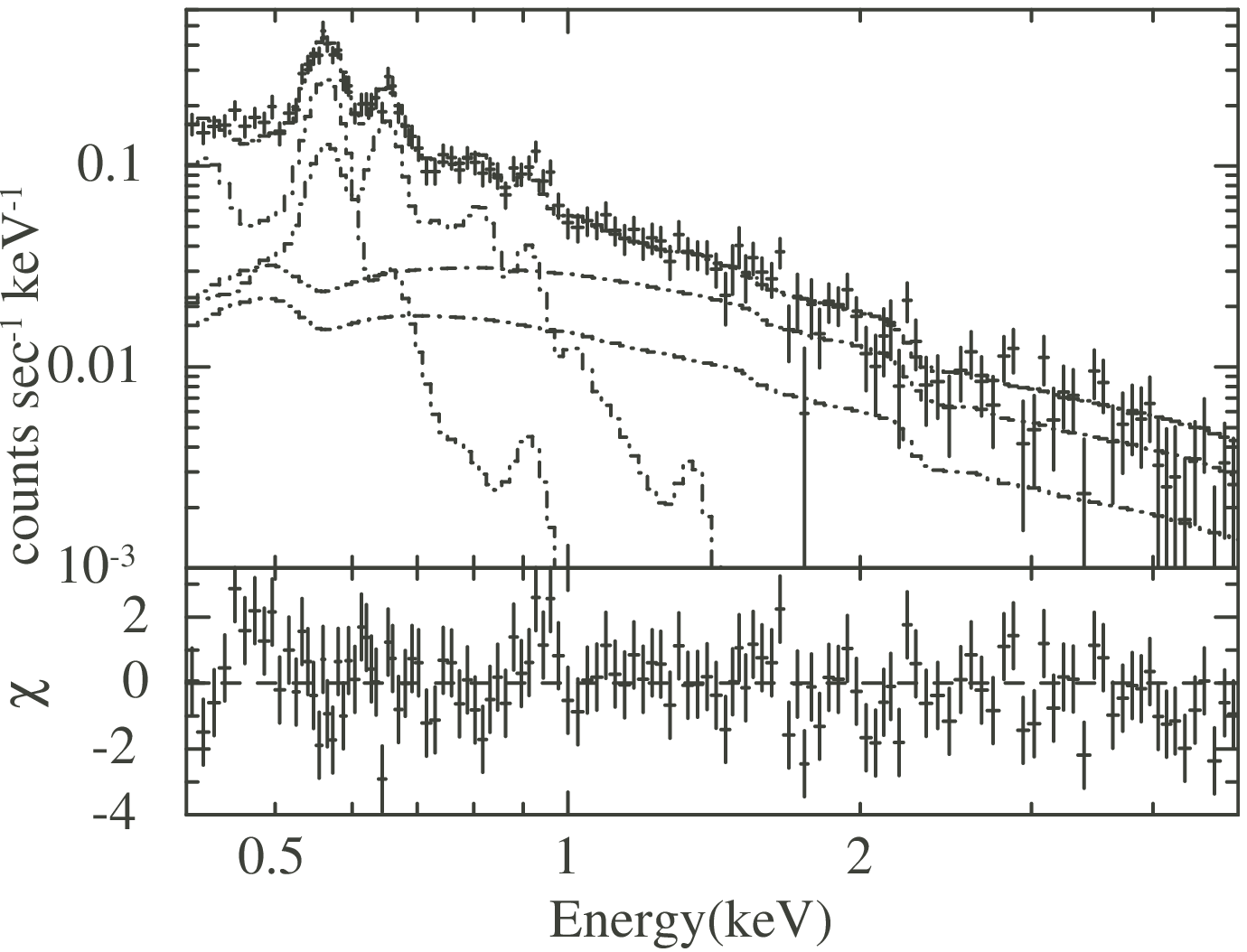}
\end{minipage}
\\
\begin{minipage}{\figw\textwidth}
({11})  North Ecliptic Pole 2\vspace{-0.055\textwidth}\\
\FigureFile(0.99\textwidth, ){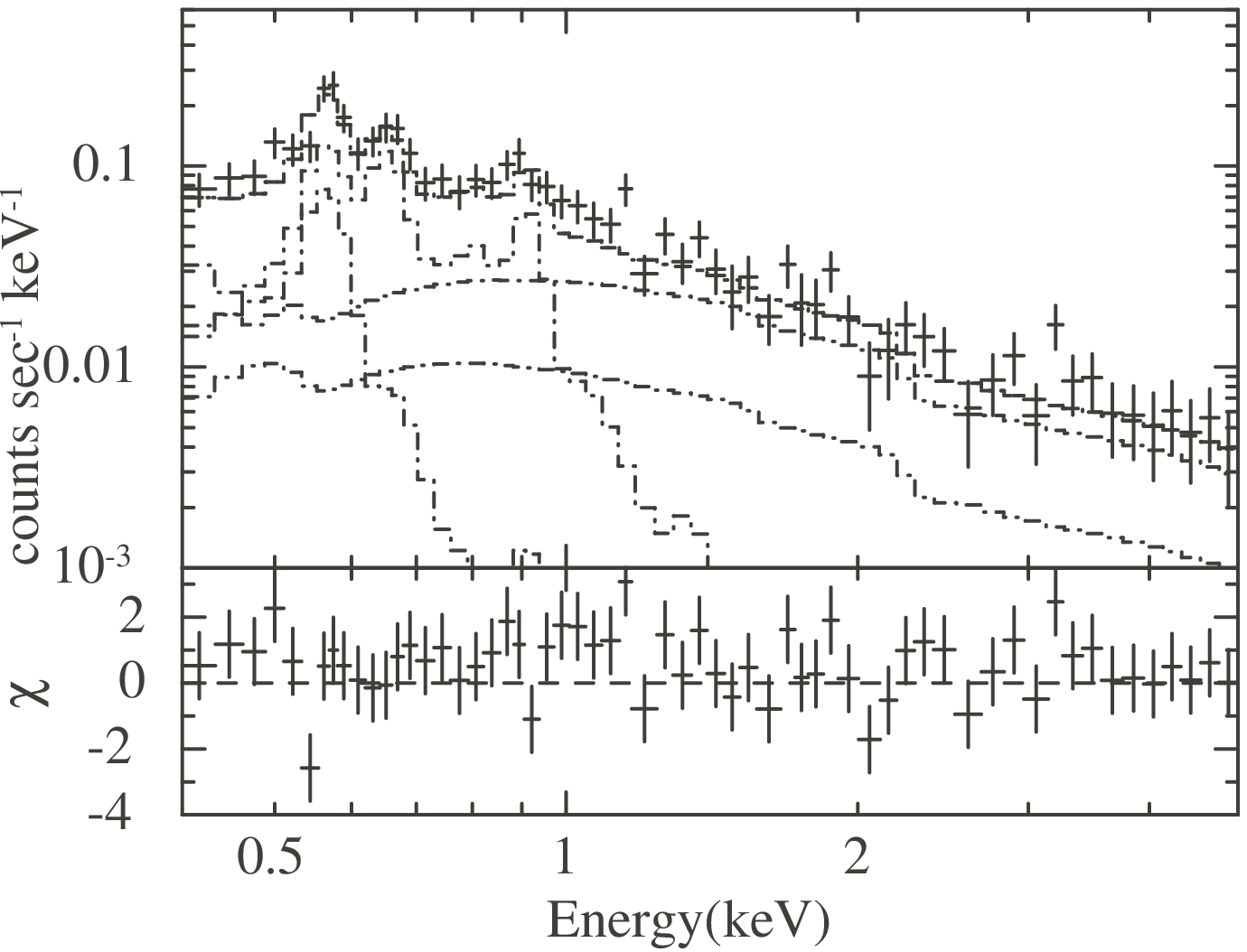}
\end{minipage}
&
\begin{minipage}{\figw\textwidth}
({12}) Low latitude 86-21\vspace{-0.055\textwidth}\\
\FigureFile(0.99\textwidth, ){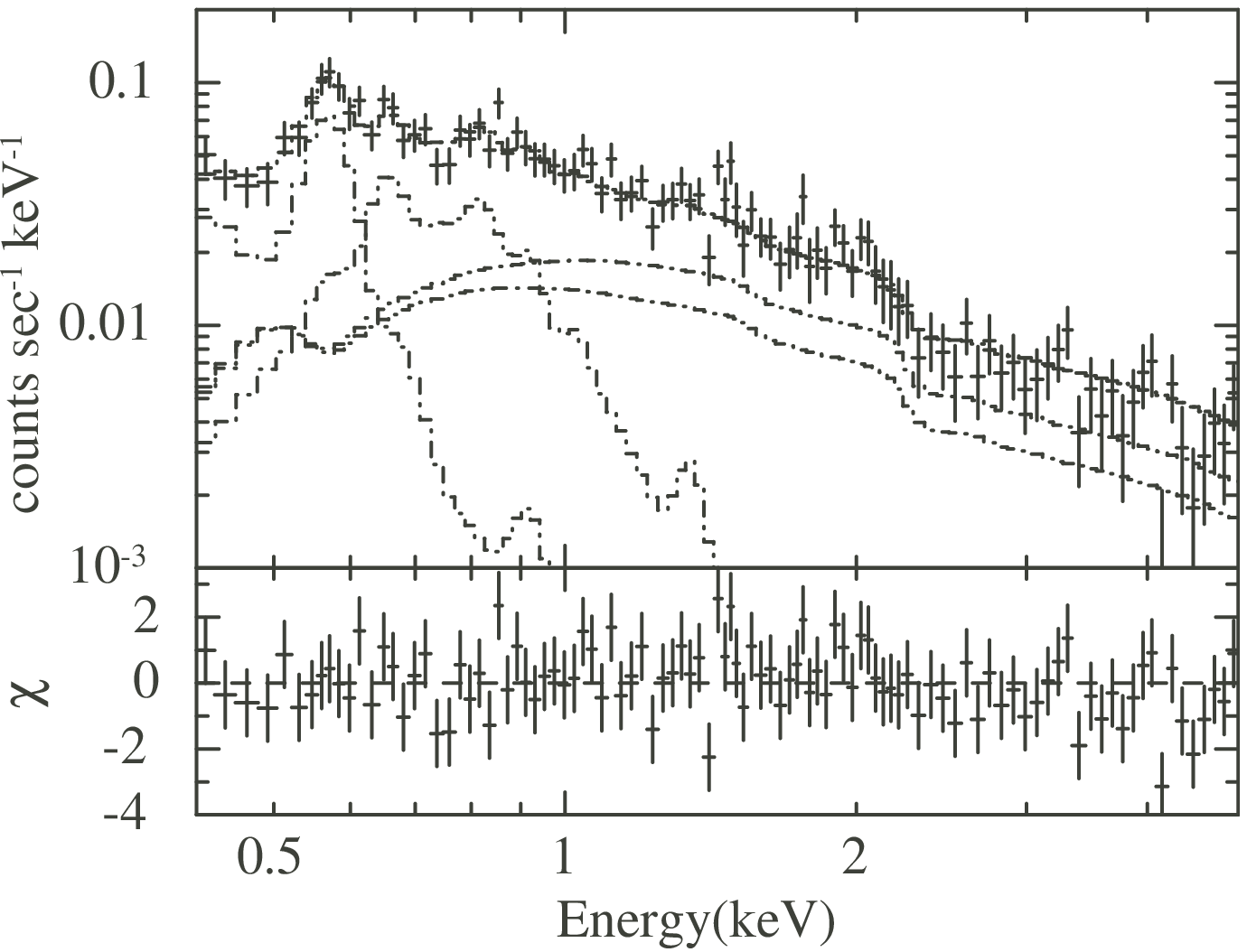}
\end{minipage}
\\
\begin{minipage}{\figw\textwidth}
({13}) Low latitude 97+10\vspace{-0.055\textwidth}\\
\FigureFile(0.99\textwidth, ){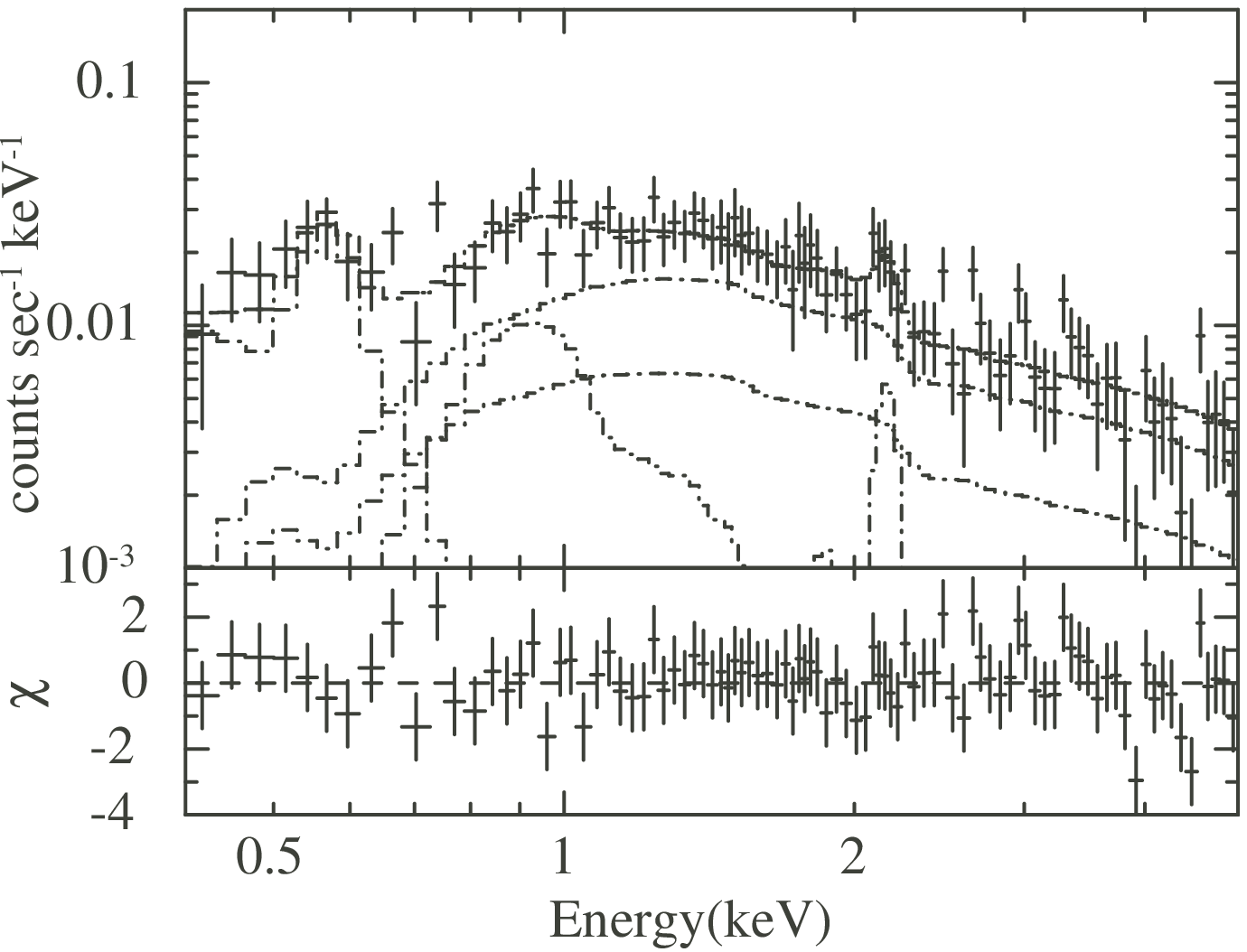}
\end{minipage}
&
\begin{minipage}{\figw\textwidth}
({R1}) MBM12 on cloud\vspace{-0.055\textwidth}\\
\FigureFile(0.99\textwidth, ){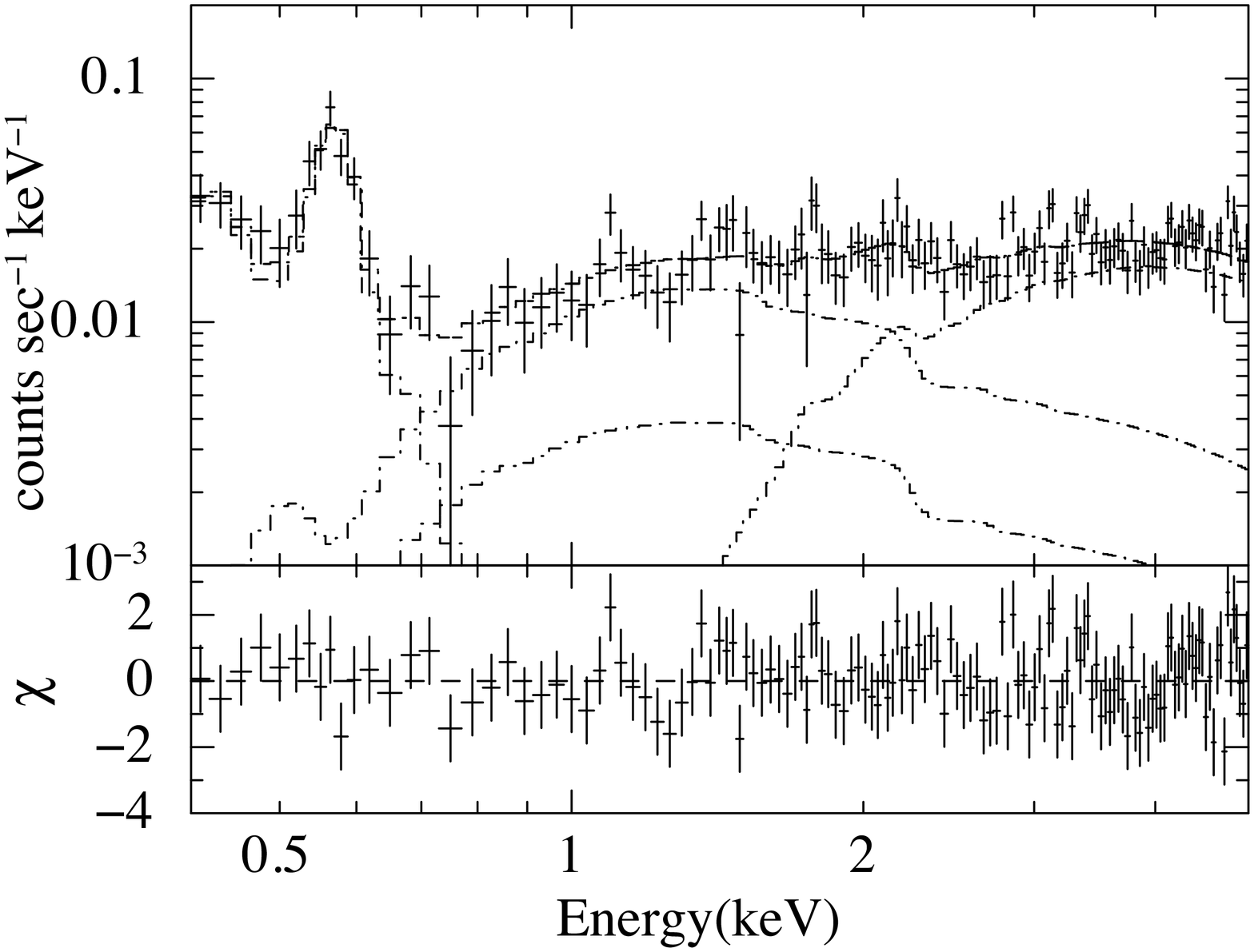}
\end{minipage}
\end{tabular}
\end{scriptsize}
\caption{
Continued.
}
\end{figure*}

\section{Analysis and results}
\label{sec:analysis_results}

\subsection{Determination of O\textsc{vii} and O\textsc{viii} emission intensities}
\label{subsec:det_line_intensities}

In Figure~\ref{fig:spectrum}, the energy spectra of thirteen data sets are shown, together with the reference data set, {M12on}. (The spectrum of {MP235} is shown in  Figure~\ref{fig:spectrum_fixedfit}).
We can clearly see the O\textsc{vii} ($\sim 0.56$ keV) emission lines in all of the spectra.

We first  fitted the spectra in 0.4-5.0 keV with a model function, consisting of the CXB component represented by a power-law function or broken power-law functions absorbed with Galactic neutral medium, and two thin thermal-emission components which we consider to represent the Heliospheric SWCX induced emission plus the thermal emission from the local hot bubble, and the thermal emission from more distant part of the Galaxy.   
As the thin thermal emission model, we used APEC  
(http://hea-www.harvard.edu/APEC/).  
The first APEC component representing the sum of 
the SWCX  from the Heliosphere and  the  emission from the local hot bubble  (hereafter SWCX+LHB)
is unaffected by the Galactic absorption.  
The contribution of the LHB is considered to be at most one third of this component and could be  as small as $\sim 1/20$ (see the discussion by \citet{Smith_etal_2007}). 
The SWCX induced X-ray emission is a non-thermal process.  Thus there is no logical reason that the spectrum of this component can be approximated by thermal emission.  In particular, fine structure of lines and 
line intensity ratios of 
different transitions must be very different from those of thermal emission
 (e.g. \citet{Kharchenko_etal_2003}).   
However,  the spectrum of this component has been successfully represented by  thin thermal models.  Since Suzaku does not have enough energy resolution to resolve fine structures of lines, and since this component mainly contributes to the Suzaku spectra of the energy range below $\sim 0.6$ keV, the same approximation can be applied to the present data.  

Galactic emission beyond the LHB may arise inside  the bulk of Galactic absorption. 
However, we assume that the second thermal component 
is subject to the total  Galactic absorption; i.e. the transabsorption emission (TAE).  
This assumption is consistent with the discussion in the next subsection
(\ref{subsec:ovii_oviii_cor}).
We fixed the column densities of absorption,  $N_{\rm H}$,  for the CXB and the TAE components to the total H column density determined by 21cm observations \citep{Dickey_Lockman_1990} except for On Filament ({On-FIL})  and Off Filament 
({Off-FIL}) data. 
For these two data sets, \citet{Henley_etal_2007} and \citet{Henley_Shelton_2008} used the column densities estimated from the 
100-$\mu$m intensity from the all-sky IRAS maps using the conversion 
relation for the southern Galactic hemisphere by Snowden et al. (2000). 
We thus tried spectral fits using two different $N_{\rm H}$ values.

The average AGN spectrum below $\sim$ 1~keV becomes steeper than above $\sim$ 1~keV, 
and the average photon index was determined to be  1.96  below $\sim$ 1~keV by  \citet{Hasinger_etal_1993}.  
We thus tried two different models for the CXB: a power-law model with a photon index fixed to 1.4,  and  a double broken power-law  model  used by  \citet{Smith_etal_2007}.  
The latter model consists of two broken power-law functions with photon indices of 1.54 and 1.96 below 1.2~keV and a photon index of 1.4 above the energy.  
We  fixed the normalization of the broken power law component with a low-energy photon index of 1.54   to 5.7  photons s$^{-1}$ keV$^{-1}$ str$^{-1}$ at 1keV and  set the normalization of the other broken power-law component free.

First we fixed the elemental abundances of both thin thermal components to the solar value  \citep{AG_1989}, and set 
the temperature and normalization free.  We call these models with the double broken power-law CXB model and with the simple power-law CXB model respectively model 1 and model 1'.

\begin{table*}
\begin{center}
\caption{Results of three-component (CXB, TAE, SWCX+LHB) spectral fits with the double broken power-law CXB (model 1) 
\label{tbl:specfit_bknpow}}
\begin{tabular}{lrcccccc}
\hline\hline
 &  $N_{\rm H} $$^{\rm a}$  & CXB$^{\rm b}$  & \multicolumn{2}{c}{TAE}  &  \multicolumn{2}{c}{SWCX+LHB} & $\chi^2$/dof\\
ID            & $(10^{20} {\rm cm}^{-2})$ &  Norm$^{\rm c}$      &   $kT$ (keV)   &      Norm$^{\rm d}$                  & $kT$ (keV)     &  Norm$^{\rm d}$            \\\hline
{1}  (GB) & 1.40 & $5.8 \pm${\scriptsize 0.6} & 0.264$^{\rm e}$   & $< 1.7$$^{\rm e}$ &
 0.142$_{ -0.020 }^{+ 0.049  }$  & $6.2 \pm${\scriptsize 1.4} & 91.25/92 \\ 
{2}  (HL-B) & 3.36 & $2.7 \pm${\scriptsize 0.6} &  0.272$_{ -0.076 }^{+ 0.081  }$ &1.6$_{  -0.8 }^{+   1.6  }$  & 0.120$^{\rm f}$ & 6.8$_{  -3.6 }^{+   1.9  }$ & 94.45/87  \\ 
{3} (LH-2)  & 0.56 &  $3.6 \pm${\scriptsize 0.5} & 0.264$^{\rm e}$  & $ < 1.0$$^{\rm e}$ &  $0.099 \pm${\scriptsize 0.027} & 16.3$_{  -3.9 }^{+   5.4  }$  & 94.74/99  \\ 
{4} (LH-1) & 0.56 & 5.6$_{  -0.5 }^{+   0.4  }$ & 0.476$_{ -0.149 }^{+ 0.279  }$   & 1.3$_{  -0.5 }^{+   0.8  }$ & 0.137$_{ -0.017 }^{+ 0.011  }$ & 8.3$_{  -1.4 }^{+   3.7  }$  & 154.08/127  \\  
{5} (Off-FIL) & 4.19 & $3.0 \pm${\scriptsize 0.4} & 0.271$_{ -0.023 }^{+ 0.018  }$ & 7.2$_{  -1.0 }^{+   1.7  }$ & 0.118$_{ -0.008 }^{+ 0.006  }$  &  25.5$_{  -5.5 }^{+   5.5  }$ & 173.17/122 \\ 
{5}' (Off-FIL) &1.90 & $2.9 \pm${\scriptsize 0.4} & 0.273$_{ -0.024 }^{+ 0.011  }$ & 5.8$_{  -0.9 }^{+   1.4  }$ & 0.122$_{ -0.009 }^{+ 0.015  }$ & 21.9$_{  -5.9 }^{+   5.4  }$ & 169.90/122 \\ 
{6} (On-FIL) & 4.61 & $4.8 \pm${\scriptsize 0.5} & 0.644$_{ -0.075 }^{+ 0.081  }$   &  1.9$_{  -0.2 }^{+   0.5  }$ & 0.182$_{ -0.013 }^{+ 0.012  }$ & $7.0 \pm${\scriptsize 0.7} & 128.57/117 \\
{6}'  (On-FIL) & 9.60 & 5.1$_{  -0.6 }^{+   0.5  }$ & 0.627$_{ -0.063 }^{+ 0.072  }$  & 2.7$_{  -0.5 }^{+   0.6  }$ &  $0.183 \pm${\scriptsize 0.010} &  8.1$_{  -0.7 }^{+   0.6  }$ & 127.35/117 \\ 
{7} (HL-A) & 1.02 &$4.9 \pm${\scriptsize 0.5} & 0.237$_{ -0.039 }^{+ 0.045  }$ & 3.5$_{  -1.1 }^{+   1.9  }$ &0.120$^{\rm f}$ & 10.0$_{  -4.5 }^{+   3.2  }$  & 141.01/116 \\
{8} (M12off)  & 8.74& $1.8 \pm${\scriptsize 0.5}  & 0.245$_{ -0.118 }^{+ 0.104  }$ & 2.4$_{  -1.4 }^{+   3.4  }$ & 0.102$_{ -0.080 }^{+ 0.016  }$ &  21.4$_{  -6.5 }^{+  21.1  }$ & 86.27/89   \\ 
{9}  (LX-3) & 4.67  & 8.7$\pm${\scriptsize 0.5}  & 0.310$_{ -0.029 }^{+ 0.087  }$  &  6.0$_{  -2.5 }^{+   1.3  }$  & 0.140$_{ -0.019 }^{+ 0.031  }$ & 9.9$_{  -1.9 }^{+   4.1  }$ & 211.71/196 \\ 
{10} (NEP1) &4.40 & 3.0$_{  -0.4 }^{+   0.5  }$ & 0.244$_{ -0.017 }^{+ 0.023  }$ &  9.5$_{  -1.9 }^{+   1.7  }$ & 0.113$_{ -0.011 }^{+ 0.009  }$ & 23.8$_{  -4.1 }^{+   6.3  }$ & 158.44/120 \\  
{11} (NEP2)  &4.40 & $2.8 \pm${\scriptsize 0.8} & 0.271$_{ -0.039 }^{+
 0.040  }$  & 7.7$_{  -2.3 }^{+   3.1  }$ & 0.116$_{ -0.025 }^{+ 0.020
 }$  & 18.9$_{  -8.5 }^{+  17.3  }$ & 60.85/55  \\ 
{12}  (LL21) & 7.24 & $4.3 \pm${\scriptsize 0.5} & 0.284$_{ -0.035 }^{+ 0.042  }$  &6.8$_{  -1.7 }^{+   2.1  }$  & 0.109$_{ -0.020 }^{+ 0.019  }$ &  24.5$_{  -9.6 }^{+  27.0  }$ & 105.03/97   \\ 
{13} (LL10) & 27.10 & 2.6$_{  -0.6 }^{+   0.7  }$& 0.738$_{ -0.184 }^{+ 0.243  }$  & 1.5$_{  -0.7 }^{+   0.6  }$ & 0.112$_{ -0.032 }^{+ 0.049  }$ & 8.1$_{  -5.3 }^{+  26.2  }$ &  84.28/ 89 \\
\hline
{R1} {M12on}$^{\rm g}$ & 40.00 & $1.7 \pm${\scriptsize 0.6}&- &- & 0.095$_{-0.007 }^{+0.014}$&18.8$_{-8.0}^{+8.9}$&141.62/136\\
{R2} {MP235}& 90.00 & $6.0 \pm${\scriptsize 1.0} &( 0.765$_{-0.041}^{+0.039}$&$3.73 \pm${\scriptsize 0.38} )$^{\rm h}$  &0.108$_{-0.035}^{+0.048}$ &12.4$_{-7.9 }^{+64}$&75.80/84\\
\hline
\multicolumn{8}{l}{
\rlap{\parbox[t]{.9\textwidth}{
$^{\rm a}$~{
The absorption column densities for the CXB and TAE components were fixed to the tabulated values. They were estimated from \citet{Dickey_Lockman_1990}, except for {5}'  and {6}', for which the values were estimated from the 100 $\mu$m intensity of the direction \citep{Henley_Shelton_2008}. }
}}}\\
\multicolumn{8}{l}{
\rlap{\parbox[t]{.9\textwidth}{
$^{\rm b}$~{Two broken power-law model was adopted.  The  photon
 indices below 1.2 keV were fixed to 1.52 and 1.96.  The normalization of the former index component is fixed to 5.7  ${\rm phtons~s}^{-1}{\rm cm}^{-2}{\rm~keV}^{-1}{\rm str}^{-1}$@1keV.}
}}}\\
\multicolumn{8}{l}{
\rlap{\parbox[t]{.9\textwidth}{
$^{\rm c}$~{The normalization of the broken power-law component with a photon index of 1.96 below 1.2 keV.}
}}}\\
\multicolumn{8}{l}{
\rlap{\parbox[t]{.9\textwidth}{
$^{\rm d}$~{The emission measure integrated over the line of sight, $(1/4\pi)\int n_{\rm e} n_{\rm H} ds$ in the unit of $10^{14}{\rm cm}^{-5}~{\rm str}^{-1}$.}
}}}\\
\multicolumn{8}{l}{
\rlap{\parbox[t]{.9\textwidth}{
$^{\rm e}$~{The TAE component was not significant.  The upper limit of the normalization was determined for the average TAE temperature of the TAE component of other observations.}
}}}\\
\multicolumn{8}{l}{
\rlap{\parbox[t]{.9\textwidth}{
$^{\rm f}$~{Because of the strong coupling between the TAE and LHB+SWCX components, the temperatures of the two components were not well determined.  We thus fixed the temperature of the SWCX+LHB component to the average value of other obervations.}
}}}\\
\multicolumn{8}{l}{
\rlap{\parbox[t]{.9\textwidth}{
$^{\rm g}$~{A spectral component for the bright point source represented by an absorbed bremsstrahlung was included in the fits but is not tabulated. The model parameters are $kT = 200 $ keV(fixed), the normalization factor = $2.66 \pm 0.14 \times 10^{-12} {\rm ~cm^{-5} ~str^{-1}} $ (when a point source was replaced with a 20' radius flat emission), and $N_{\rm H} =5.5 \times 10^{22} {\rm cm}^{-2}$ (fixed).}
}}}\\
\multicolumn{8}{l}{
\rlap{\parbox[t]{.9\textwidth}{
$^{\rm h}$~{This component represents the "narrow bump" component \citep{Masui_etal_2009} rather than the TAE.  No absorption was included in this component.  This component emits O\textsc{viii} emissions and  the O abundance was set free (the best-fit value is $3.1_{-1.4}^{+1.3}$ solar.}
}}}\\

\end{tabular}
\end{center}
\end{table*}

\begin{table*}
\begin{center}
\caption{Results of three-component (CXB, TAE, SWCX+LHB) spectral fits with the power-law CXB model (model 1')
\label{tbl:specfit_pow}}
\begin{tabular}{lrcccccc}
\hline\hline
 &  $N_{\rm H} $$^{\rm a}$  & CXB$^{\rm b}$  & \multicolumn{2}{c}{TAE}  &  \multicolumn{2}{c}{SWCX+LHB} & $\chi^2$/dof\\
ID            & $(10^{20} {\rm cm}^{-2})$ &  Norm$^{\rm c}$      &   $kT$ (keV)   &      Norm$^{\rm d}${}                  & $kT$ (keV)     &  Norm$^{\rm d}$            \\\hline
{1}  (GB) & 1.40 & $10.9 \pm${\scriptsize 0.6} & 0.203$_{ -0.050 }^{+ 0.086  }$ &  3.3$_{  -1.4 }^{+   4.1  }$  & 0.112$^{\rm e}$  &  8.1$_{  -4.7 }^{+   6.3  }$   & 89.86/91  \\ 
{2} (HL-B) &3.36  & $8.0 \pm${\scriptsize 0.5}  &   0.273$_{ -0.043 }^{+ 0.066  }$ & 2.3$_{  -0.9 }^{+   1.0  }$ & 0.112$^{\rm e}$   & 9.5$_{  -2.9 }^{+   2.6  }$  & 96.37/87  \\ 
{3} (LH-2)  & 0.56 & 8.7$_{  -0.6 }^{+   0.5  }$ & 0.335$_{ -0.148 }^{+ 0.160  }$  &  1.0$_{  -0.6 }^{+   1.7  }$ &  0.087$_{ -0.030 }^{+ 0.021  }$ & 36.4$_{ -23.9 }^{+ 128.5  }$  & 89.76/97  \\ 
{4} (LH-1) & 0.56 & 10.6$_{  -0.5 }^{+   0.4  }$ &0.354$_{ -0.038 }^{+ 0.069  }$ & $2.5 \pm${\scriptsize   0.6} &0.116$_{ -0.008 }^{+ 0.013  }$  & 15.2$_{  -4.1 }^{+   4.2  }$ & 172.90/127  \\ 
{5}	 (OFF-FIL) &4.19 & $8.2 \pm${\scriptsize 0.4}  & 0.273$_{ -0.019 }^{+ 0.017  }$ & 7.9$_{  -1.0 }^{+   1.5  }$  & 0.115$_{ -0.006 }^{+ 0.009  }$ & 28.9$_{  -5.8 }^{+   5.4  }$ & 178.04/122 \\
{5}'	(OFF-FIL) & 1.90 &  $8.2 \pm${\scriptsize 0.4} & 0.276$_{ -0.021 }^{+ 0.012  }$ &  6.5$_{  -0.9 }^{+   1.3  }$ & 0.118$_{ -0.008 }^{+ 0.011  }$ & 25.7$_{  -5.7 }^{+   5.8  }$ & 174.37/122  \\
{6}  (ON-FIL) &4.61 & 10.0$_{  -0.4 }^{+   0.5  }$ & 0.626$_{ -0.065 }^{+ 0.073  }$ &$2.1 \pm${\scriptsize 0.5}   & $0.182 \pm${\scriptsize 0.011} &  7.8$_{  -0.6 }^{+   0.7  }$ & 124.99/ 117   \\
{6}'  (ON-FIL) & 9.60 & $10.2 \pm${\scriptsize 0.5} & 0.618$_{ -0.056 }^{+ 0.062  }$  & 3.0$_{  -0.6 }^{+   0.5  }$& 0.183$_{ -0.009 }^{+ 0.010  }$  & 8.8$_{  -0.6 }^{+   0.7  }$ & 131.00/117 \\ 
{7} (HL-A)  & 1.02 & $10.1 \pm${\scriptsize 0.4} &  0.242$_{ -0.049 }^{+ 0.036  }$ & 4.9$_{  -1.3 }^{+   3.0  }$  & 0.093$_{ -0.046 }^{+ 0.020  }$ & 27.8$_{ -13.3 }^{+  56.1  }$ &  142.17/115 \\ 
{8} (M12off)  & 8.74 & $7.2 \pm${\scriptsize 0.4} & 0.260$_{ -0.069 }^{+ 0.071  }$   & 2.7$_{  -1.0 }^{+   2.8  }$ & 0.101$_{ -0.012 }^{+ 0.015  }$  & 23.9$_{ -10.6 }^{+  10.6  }$ & 86.99/89   \\ 
{9} (LX-3) & 4.67 &  $13.4 \pm${\scriptsize 0.5} & 0.308$_{ -0.023 }^{+ 0.032  }$   & 7.6$_{  -1.5 }^{+   1.2  }$ & 0.125$_{ -0.013 }^{+ 0.021  }$  & 14.7$_{  -4.1 }^{+   5.8  }$  & 214.27/196   \\
{10} (NEP1)  &  4.40 & $8.1 \pm${\scriptsize 0.4} & 0.250$_{ -0.016 }^{+ 0.023  }$ &  9.9$_{  -1.8 }^{+   1.7  }$  & 0.112$_{ -0.011 }^{+ 0.008  }$  &  26.8$_{  -4.1 }^{+   7.7  }$  & 171.74/120  \\
{11} (NEP2)  &  4.40 & $8.1 \pm${\scriptsize 0.7}  &  $0.275 \pm${\scriptsize 0.036} & 8.3$_{  -1.9 }^{+   2.9  }$  & 0.113$_{ -0.023 }^{+
 0.024  }$  & 21.8$_{  -9.0 }^{+  19.2  }$ & 63.15/55  \\
{12} (LL21)  & 7.24& $9.5 \pm${\scriptsize 0.5}& 0.287$_{ -0.031 }^{+ 0.033  }$ &  7.6$_{  -1.5 }^{+   1.7  }$ & 0.101$_{ -0.012 }^{+ 0.018  }$ & 33.5$_{ -14.5 }^{+  25.7  }$  & 104.43 /97   \\
{13} (LL10) &27.10 & 7.8$_{  -0.6 }^{+   0.5  }$& 0.721$_{ -0.166 }^{+ 0.190  }$&  1.9$_{  -0.6 }^{+   0.8  }$ & 0.113$_{ -0.030 }^{+ 0.052  }$ & 8.0$_{  -5.3 }^{+  19.9  }$ & 84.70/89   \\ 
\hline
{R1} {M12on}$^{\rm f}$ & 40.00 & 7.2$_{ -0.5 }^{+ 0.6 }$ & - & - & 0.096$_{ -0.007 }^{+ 0.014 }$ & 18.5$_{ -7.8 }^{+ 8.9 }$ & 141.34/136 \\
{R2} {MP235} & 90.0 & $11.1 \pm${\scriptsize 0.9} & (0.766$_{ -0.039 }^{+ 0.041 }$ & 3.75$_{ -0.37 }^{+ 0.40 }$)$^{\rm g}$ & 0.105$_{ -0.032 }^{+ 0.051 }$ & 14.1$_{ -9.6 }^{+ 64 }$ & 76.54/84 \\
\hline
\multicolumn{8}{l}{
\rlap{\parbox[t]{.92\textwidth}{
$^{\rm a}$~{The absorption column densities for the CXB and TAE components were fixed to the tabulated values. They were estimated from \citet{Dickey_Lockman_1990}, except for {5}'  and {6}'.  The column densities were estimated from the 100 $\mu$m intensity of the direction.  }
}}}\\
\multicolumn{8}{l}{
\rlap{\parbox[t]{.92\textwidth}{
$^{\rm b}$~{A power-law model with a photon index of 1.4 was adopted.}
}}}\\
\multicolumn{8}{l}{
\rlap{\parbox[t]{.92\textwidth}{
$^{\rm c}$~{The normalization of the power-law function with  the unit of ${\rm phtons~s}^{-1}{\rm cm}^{-2}{\rm~keV}^{-1}{\rm str}^{-1}$@1keV.}
}}}\\
\multicolumn{8}{l}{
\rlap{\parbox[t]{.92\textwidth}{
$^{\rm d}$~{The emission measure integrated over the line of sight,  $(1/4\pi)\int n_{\rm e} n_{\rm H} ds$ in the unit of $10^{14}{\rm cm}^{-5}~{\rm str}^{-1}$.}
}}}\\
\multicolumn{8}{l}{
\rlap{\parbox[t]{.92\textwidth}{
$^{\rm e}$~{Because of the strong coupling between the TAE and LHB+SWCX components, the temperatures of the two components were not well determined.  We thus fixed the temperature of the SWCX+LHB component to the average values of other obervations.}
}}}\\
\multicolumn{8}{l}{
\rlap{\parbox[t]{.92\textwidth}{
$^{\rm f}$~{
The model parameters for the point source were  $kT = 200 $ keV(fixed), the normalization factor = $2.65 \pm 0.14 \times 10^{-12} {\rm ~cm^{-5} ~str^{-1}} $ (for a 20' radius flat emission), and $N_{\rm H} =5.5 \times 10^{22} {\rm cm}^{-2}$ (fixed). See table note (g) of Table~\ref{tbl:specfit_bknpow}.}
}}}\\
\multicolumn{8}{l}{
\rlap{\parbox[t]{.92\textwidth}{
$^{\rm g}$~{"Narrow bump" component. The best-fit O abundance was $3.1_{-1.2}^{+1.3}$. See table note (h) of  Table~\ref{tbl:specfit_bknpow}.}
}}}\\

\end{tabular}
\end{center}
\end{table*}

In Tables \ref{tbl:specfit_bknpow} and \ref{tbl:specfit_pow}, we summerize
 the results of the spectral fits.
The best fit model functions convolved with the instrument response functions  are plotted in  
Figure~\ref{fig:spectrum} for fits with model 1.
For {HL-B} and {HL-A} with model 1, and for {GB} and {HL-B} for model 1', 
the temperatures of the two thermal components could not be constrained well because of strong coupling between the two components.  
In these cases, we fixed the temperature of the SWCX+LHB components to the average values of other spectra.  For {GB} and {LH-2} with model 1, the existence of the TAE component was not significant.  Thus for these two cases we fixed the temperature of the TAE component to the average value of other spectra other than {LL10} (0.264 keV), and estimated the upper limit of the normalization parameter.
Although some different  fitting procedures were necessary  between models 1 and 1' 
for a few data sets as described above,
the best-fit parameter values of the TAE and SWCX+LHB components determined
using different CXB models are consistent within the statistical errors.
We performed all the fits shown in this paper using two different CXB models.  
However,  since all of those results are
consistent with each other within the statistical errors, we will only show the results with the double broken power-law model for the CXB  in the further analysis.

For the {On-FIL} and {Off-FIL}
results in Tables \ref{tbl:specfit_bknpow} and \ref{tbl:specfit_pow}, we find the best-fit parameters to be consistent with each other within the statistical errors for both absorption values.   We performed all the fittings shown later for these data sets with both column densities but continued to find consistent results, and so will show only the results using the absorption column densities from \citet{Dickey_Lockman_1990}.

The resultant $\chi^2$ values were  generally good  (reduced $\chi^2$ is 
0.95 -- 1.42 for 87 -- 196 degrees of freedom.)
The best-fit values of the temperatures of the two thin-thermal models  are contained in a relatively narrow ranges.  They are  from 0.10 to 0.14  keV for  the LHB + SWCX component
and  from 0.22  and 0.48  keV for the TAE, except for the TAE temperatures of 
 {On-FIL} ($kT = 0.644 $ keV) and  {LL10} ($kT = 0.738 $ keV).

\begin{table*}
\begin{center}
\caption{Results of spectral fits to determine O\textsc{vii} and
 O\textsc{viii} line intensities with broken-power-law CXB model (model 1)\label{tbl:specfit_lines_bknpow}}
\begin{tabular}{lcccccccc}
\hline\hline
 &   CXB$^{\rm a}$  & TAE$^{\rm a,b}$ &  SWCX+LHB$^{\rm b}$  & \multicolumn{2}{c}{O\textsc{vii}} & \multicolumn{2}{c}{O\textsc{viii}$^{\rm f}$} & $\chi^2$/dof\\
ID     &  Norm$^{\rm c}$ & Norm$^{\rm d}$  &  Norm$^{\rm d}$  &
Centroid &  Norm$^{\rm e}$  & Centroid &  Norm$^{\rm e}$     \\\hline
{1}  (GB)  & $5.7 \pm${\scriptsize 0.6}  & - &  4.5$_{  -4.5 }^{+   5.6  }$ &  0.560$_{ -0.008 }^{+ 0.010  }$ &3.6$_{  -1.0 }^{+   0.9  }$    &  0.664$_{ -0.018 }^{+ 0.017  }$ & 0.7$\pm${\scriptsize 0.5}  & 83.19/87 \\
{2}  (HL-B) &  $2.4 \pm${\scriptsize 0.6} & 1.9$_{  -1.5 }^{+   1.6  }$
 &  $14.9 \pm${\scriptsize 5.4}  & 0.560$_{ -0.010 }^{+ 0.009  }$ &
 $2.3 \pm${\scriptsize 0.7}  & 0.654$_{ -0.014 }^{+ 0.010  }$  & 0.9$\pm${\scriptsize 0.4}  & 80.06/80\\
{3} (LH-2)  	& $3.6 \pm${\scriptsize 0.5} & - & 18.3$_{ -10.1 }^{+   9.1  }$ & 0.568$_{ -0.006 }^{+ 0.008  }$  & $2.5 \pm${\scriptsize 0.7} & 0.654$^{\rm g}$ & $< 0.4$ & 93.58/97\\
{4}  (LH-1) &  $5.6 \pm${\scriptsize 0.5} & $1.2 \pm${\scriptsize 0.4 }
 &  11.8$_{  -3.3 }^{+   3.2  }$   & 0.562$_{ -0.002 }^{+ 0.003  }$ &
 $4.1 \pm${\scriptsize 0.5} & 0.659$_{ -0.017 }^{+ 0.015  }$ & 0.5$\pm${\scriptsize 0.3} & 124.53/ 92 	\\
{5} (OFF-FIL) & $3.0 \pm${\scriptsize 0.4  } & $9.1 \pm${\scriptsize 1.3} & 31.5$_{  -4.7 }^{+   4.5  }$ &  0.566$_{ -0.004 }^{+ 0.002
 }$  &$ 9.5 \pm${\scriptsize 0.7 } & 0.653$_{ -0.005 }^{+ 0.004  }$ &
 2.5$\pm${\scriptsize 0.4}  & 154.66/118 \\
{6} (ON-FIL) &  $5.0 \pm${\scriptsize 0.5 }  & $2.1 \pm${\scriptsize  0.5}  & 4.7$_{  -3.7 }^{+   3.5  }$ & 0.563$_{ -0.005 }^{+ 0.004  }$
 & 5.2$_{  -0.6 }^{+   1.0  }$ & 0.660$_{ -0.006 }^{+ 0.008  }$ &
 1.6$\pm${\scriptsize 0.3} & 113.18/113 \\
{7}  (HL-A) & $5.0 \pm${\scriptsize  0.5}  &  $4.9 \pm${\scriptsize 2.1  }
 & $13.7 \pm${\scriptsize 4.7}  & $0.564 \pm${\scriptsize  0.004} &
 $4.5 \pm${\scriptsize 0.6} & 0.647$_{ -0.011 }^{+ 0.007  }$ & 1.6$\pm${\scriptsize  0.4} & 134.71/111  \\
{8}  (M12off) 	& 1.9$_{  -0.5 }^{+   0.4  }$& 2.9$_{  -1.9 }^{+   2.1  }$& 24.3$_{  -5.3 }^{+   5.2  }$  & $0.564 \pm${\scriptsize  0.004 } & $4.4 \pm${\scriptsize 0.5}  & 0.662$_{ -0.010 }^{+ 0.011  }$  & 0.6$\pm${\scriptsize 0.3 } & 79.05/85\\
{9}  (LX-3)  & $8.8 \pm${\scriptsize 0.5} &  7.3$_{  -1.3 }^{+   1.2  }$ &  11.9$_{  -4.2 }^{+   4.1  }$ & $0.570 \pm${\scriptsize 0.004} & $6.0 \pm${\scriptsize 0.7} & $0.654 \pm${\scriptsize  0.006} & 1.8$\pm${\scriptsize 0.4} &  206.76/192 \\
{10} (NEP1) &3.1$_{  -0.5 }^{+   0.3  }$ &13.4$_{  -1.7 }^{+   2.0  }$
 & 29.1$_{  -3.4 }^{+   3.5  }$ & 0.567$_{ -0.001 }^{+ 0.003  }$ &
 $8.9 \pm${\scriptsize 0.5 } & 0.653$_{ -0.002 }^{+ 0.004  }$ & 2.8$_{  -0.3 }^{+   0.3  }$ & 165.23/116  \\ 
{11} (NEP2) & 3.4$_{  -0.7 }^{+   0.8  }$ & 8.2$_{  -2.7 }^{+   2.6  }$
 & 25.1$_{  -7.2 }^{+   7.3  }$   &  0.568$_{ -0.004 }^{+ 0.006  }$ &
 $7.0 \pm${\scriptsize 1.1} & 0.656$_{ -0.008 }^{+ 0.006  }$ &  2.6$\pm
 0.7$ & 56.20/51   \\
{12}  (LL21)  &	$4.6 \pm${\scriptsize 0.5} &  8.4$_{  -1.8 }^{+   1.7  }$   &  27.3$_{  -7.5 }^{+   7.4  }$  &0.568$_{ -0.005 }^{+ 0.006  }$ & 6.4$_{  -0.9 }^{+   0.8  }$ & $0.652 \pm${\scriptsize 0.010} & 1.8$\pm${\scriptsize 0.5} & 108.14 /93\\
{13}  (LL10) &  $2.6 \pm${\scriptsize 0.6} & $1.7 \pm${\scriptsize 0.7}
& $10.5 \pm${\scriptsize 5.6}  & 0.554$_{ -0.018 }^{+ 0.019  }$   &
1.7$_{  -0.7 }^{+   1.3  }$  &  0.642$_{ -0.034 }^{+ 0.022  }$ &
0.4$_{  -0.3 }^{+   0.3  }$ & 80.63/87  \\
\hline
{R1} {M12on}$^{\rm h}$ & 1.7$_{ -0.5 }^{+ 0.5 }$ & - & 20.5$_{ -3.9 }^{+ 4.2 }$ & 0.564$_{ -0.006 }^{+ 0.005 }$ & $2.9\pm0.5$ & 0.654$^{\rm g}$ & $< 0.4$ & 141.69/135 \\
{R2} {MP235} & 6.25$_{ -1.0 }^{+ 0.9 }$ & (4.0$_{ -0.3 }^{+ 0.4 })$$^{\rm i}$ & 19.5$_{ -10.5 }^{+ 7.8 }$ & 0.578$_{ -0.011 }^{+ 0.005 }$ & 2.1$_{ -0.6 }^{+ 0.8 }$ & 0.673$_{ -0.024 }^{+ 0.007 }$ & $0.9 \pm${\scriptsize 0.3} & 77.18/83 \\
\hline
\multicolumn{8}{l}{
\rlap{\parbox[t]{.95\textwidth}{
$^{\rm a}$~{The absorption column densities for the CXB and TAE components were fixed to the values tabulated in Table~\ref{tbl:specfit_bknpow}. }
}}}\\
\multicolumn{8}{l}{
\rlap{\parbox[t]{.95\textwidth}{
$^{\rm b}$~{The temperature of TAE and SWCX+LHB  components were fixed to the best fit values tabulated in Table~\ref{tbl:specfit_bknpow}.  The O abundance of the both components was set to 0 in the fits.}
}}}\\
\multicolumn{8}{l}{
\rlap{\parbox[t]{.95\textwidth}{
$^{\rm c}$~{The unit is ${\rm phtons~s}^{-1}{\rm cm}^{-2}{\rm~keV}^{-1}{\rm str}^{-1}$@1keV. }
}}}\\
\multicolumn{8}{l}{
\rlap{\parbox[t]{.95\textwidth}{
$^{\rm d}$~{The emission measure integrated over the line of sight, $(1/4\pi)\int n_{\rm e} n_{\rm H} ds$ in the unit of $10^{14}{\rm cm}^{-5}~{\rm str}^{-1}$.}
}}}\\
\multicolumn{8}{l}{
\rlap{\parbox[t]{.95\textwidth}{
$^{\rm e}$~{The unit is  LU =${\rm phtons~s}^{-1}~{\rm cm}^{-2}~{\rm str}^{-1}$.}
}}}\\
\multicolumn{8}{l}{
\rlap{\parbox[t]{.95\textwidth}{
$^{\rm f}$~{Contribution of O\textsc{vii} ${\rm K}_\beta$ emission 
in the Gaussian fitting function is subtracted using 
${\rm K}_\alpha$ intensity (see text).}
}}}\\
\multicolumn{8}{l}{
\rlap{\parbox[t]{.95\textwidth}{
$^{\rm g}$~{The centroid energy was fixed to obtain the upper limit of the intensity.}
}}}\\
\multicolumn{8}{l}{
\rlap{\parbox[t]{.9\textwidth}{
$^{\rm h}$~{An additional spectral component was included to represent the point source. See table note (g) of Table~\ref{tbl:specfit_bknpow}.}
}}}\\
\multicolumn{8}{l}{
\rlap{\parbox[t]{.9\textwidth}{
$^{\rm i}$~{"Narrow bump" component. The O abundance of this component was also set to 0.  See table note (h) of  Table~\ref{tbl:specfit_bknpow}.}
}}}\\

\end{tabular}
\end{center}
\end{table*}

We then estimated  the surface brightness of O\textsc{vii} ${\rm K}_\alpha$ 
and O\textsc{viii} ${\rm K}_\alpha$ emission.  We set the Oxygen abundances of the two APEC models to 0 and added two Gaussian functions that represent O\textsc{vii} ${\rm K}_\alpha$\ and O\textsc{viii} ${\rm K}_\alpha$ emission.  
We set the intrinsic width of the Gaussian functions to a value small compared to the detector energy resolution. We did not include absorption in the Gaussian components.   We set the centroid energies  and the normalization parameters of the Gaussians and the normalization parameters of the CXB and APEC components free.  
We set the intensities of the TAE component to 0 for {GB} and {LH-1}, because we had obtained only upper limits for this component.     The surface brightness of a  line is calculated from the best-fit value of the Gaussian normalization.

The O\textsc{vii} ${\rm K}_\beta$ emission is at 666 eV which will therefore blend with the
O\textsc{viii} ${\rm K}_\alpha$  emission at 654 eV.   We estimated the O\textsc{vii} ${\rm K}_\beta$
intensity from the best-fit value of 
Gaussian normalization of  O\textsc{vii} ${\rm K}_\alpha$ and subtracted it from the Gaussian
normalization of  O\textsc{viii} ${\rm K}_\alpha$.  
The ratio, $u$,  of O\textsc{vii} ${\rm K}_\beta$ to O\textsc{vii} ${\rm K}_\alpha$ intensities is  
a slow function of the plasma temperature
for thermal emission.  
As we will show later, 
the temperature of the TAE component determined from O emission is in the range of $kT =$  0.19 to 0.23 keV.  Then $u = 0.056$.  
If the emission is due to SWCX, $u = 0.083$ \citep{Kharchenko_etal_2003}.   
We estimated the O\textsc{vii} ${\rm K}_\beta$  intensity assuming that
2 LU of O\textsc{vii}${\rm K}_\alpha$ is from SWCX and the rest from thermal emission 
of $kT =$ 0.19 to 0.23 keV (see below).  
The difference of the estimation from the totally SWCX or totally thermal cases
is at most 0.2 LU, 
which is smaller than the statistical errors of O\textsc{viii} intensity.

In Table~\ref{tbl:specfit_lines_bknpow} 
we summarize  the results of the spectral fits.   The centroid energy of O\textsc{viii} had to be fixed to the theoretical value for {LH-2}, because we obtained only an upper limit for this data set.   
Except for this case, the centroid energies are consistently determined to be equal to their expected values within the systematic errors of energy scale calibration (5 eV, \cite{Koyama_etal_2007}, also see the Suzaku Technical Descricription).

The O\textsc{vii} emission intensities of {NEP1} and {NEP2} are not 
consistent with each other within the 90 \% statistical errors, while
the O\textsc{viii} emission intensities are consistent.  The difference between
the best fit values of 
O\textsc{vii} is
1.9 LU.  In the previous section, we discussed that time variable geocoronal
SWCX of 1.5 - 2 LU level can remain in our spectra.  
Thus, the difference can be partly due to contamination of the geocoronal SWCX.  
Temporal variations of Heliospheric SWCX could also contribute to the difference. 

\subsection{Correlation between O\textsc{vii} and O\textsc{viii} intensities}
\label{subsec:ovii_oviii_cor}

\begin{figure*}
\begin{center}
\FigureFile(0.8\textwidth, ){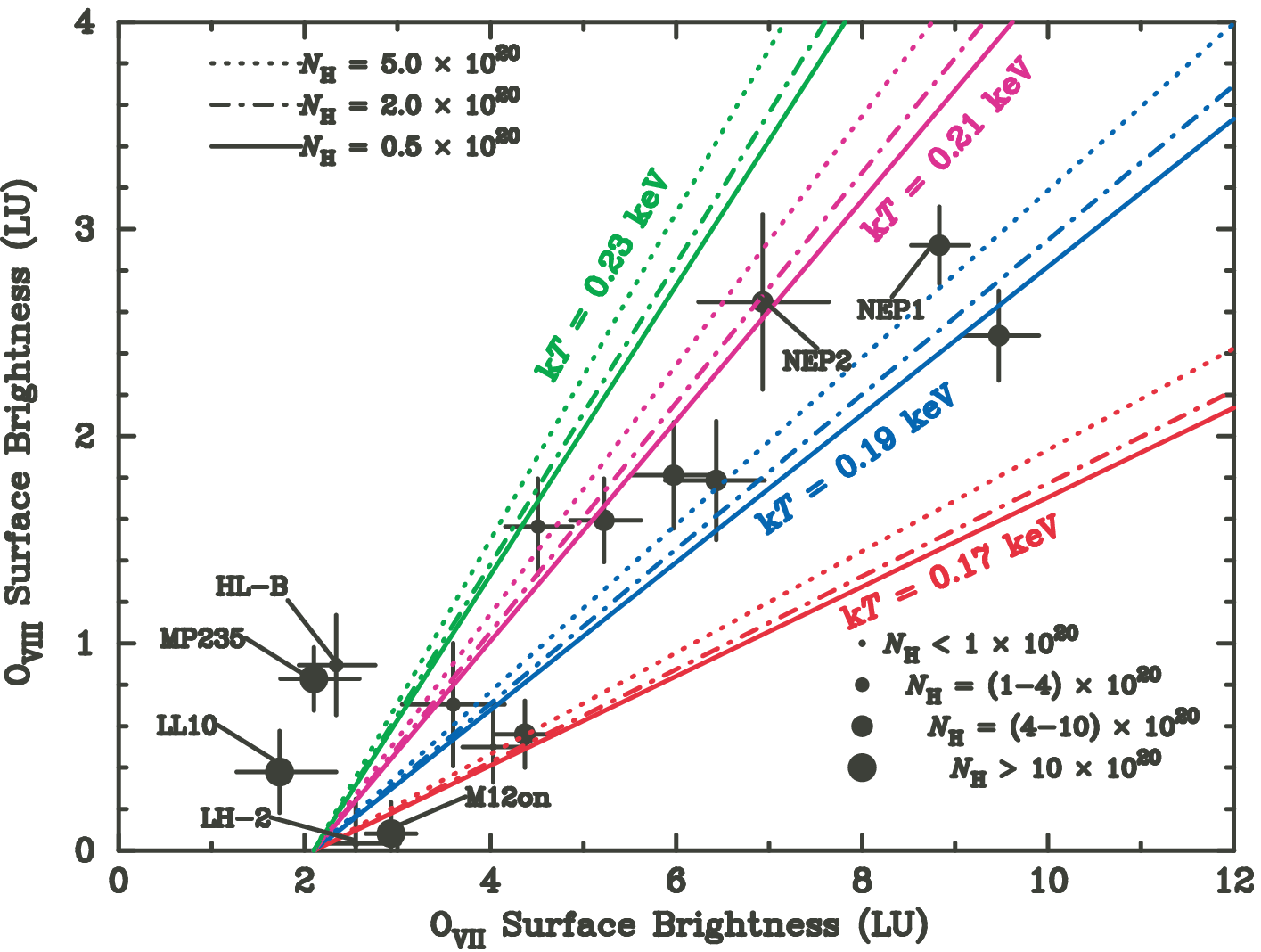}
\end{center}
\caption{Relation between O\textsc{vii} and O\textsc{viii} surface brightnesses for the 14 sky fields observed with Suzaku.  The horizontal and vertical bars of data points show the $1~\sigma$ errors
of the estimation.  The contribution of  O\textsc{vii} ${\rm K}_\beta$ emission in the Gaussian function for 
O\textsc{viii} ${\rm K}_\alpha$ is subtracted (see Table \ref{tbl:specfit_lines_bknpow} and text). 
The diagonal lines show the relation between  O\textsc{vii} and O\textsc{viii}, assuming an offset O\textsc{vii} emission of  2.1 LU and  emission from a hot plasma of the temperature and the absorption column density shown in the figure.  The Galactic absorption column density of the observation fields are indicated by the maker size of the data points.  The short names of five data points on the intensity floor of O\textsc{vii} emission are also shown. 
\label{fig:oviii-ovii_Bkn}}
\end{figure*}

In Figure~\ref{fig:oviii-ovii_Bkn}, we plot the derived O\textsc{viii} emission 
intensity as a function of 
the O\textsc{vii} intensity, and include the data points for  {M12on} and 
{MP235} from \citet{Masui_etal_2009}.  
In the figure we notice two remarkable characteristics.
First,  there is a floor in the O\textsc{vii} intensity at $\sim 2$ LU. 
 Second, all the data points
except for three ({HL-B}, {MP235}, and {LL10}) approximately follow the relation
 (O\textsc{viii} intensity) =  0.5 $\times$ [ (O\textsc{vii} intensity) - 2 LU].
\citet{Masui_etal_2009} proposed that the O\textsc{viii} emission line of the midplane field, {MP235}, is associated with a $\sim 0.8$ keV component that may arise from faint young dM stars in the Galactic disk.  As we discuss later, the low latitude field at $b=10^\circ$, {LL10}, is also considered to contain emission of similar origin.  Such high temperature emission produces little O\textsc{vii}.
We thus consider that the floor of O\textsc{vii} emission and the strong correlation 
between additional O\textsc{vii} and O\textsc{viii} intensities suggest that  the  O\textsc{vii} emission 
consists of two emission components of  different origins; 
an emission component which emits  O\textsc{vii} of $\sim 2$  LU and no or weak  O\textsc{viii} emission (the floor component), and the other component which emits O\textsc{vii} of $\sim 0$ to $\sim 7$ LU and
the  O\textsc{viii} intensity is about a half of O\textsc{vii} (the linear component).

Taking into account possible systematic error of O\textsc{vii} intensity
due geocoronal SWCX ($\sim 1.5$ LU), {M12on} and  {LH-2} 
are also on the floor.
The O\textsc{vii} emission  of the {M12on} field must arise from the near side of the MBM-12 molecular cloud whose distance is 60 - 300 pc.  
Assuming the midplane space-averaged neutral density \citep{Ferriere_2001}, 
the absorption length of O\textsc{vii} emission is estimated to be 300 pc
for {MP235}.  The total Galactic absorption column density of {LL10}
($|b| = 10^\circ$)  is 
$2.7 \times 10^{21} {\rm cm}^{-2}$, through which only
10 \% of O\textsc{vii} photons can penetrate.  
These indicate that  the floor component is "local" emission and is
associated with  the SWCX + LHB component of the spectral model 
(see also discussion in
the first paragraph of section \ref{sec:dis:TAE}).  
The emission temperature
of the SWCX + LHB component is $kT \sim 0.1$ keV when it is described
with thermal emission.
This is consistent with
the weak O\textsc{viii} emission of the floor component.

Then  the linear component is likely to arise from more distant part of our
Galaxy.  It is remarkable that  LL21 ($|b| = 20^\circ$) contains significant
linear component  (6.4 - 2 = 4.4 LU), even though the low Galactic latitude. 
The absorption column density of this direction, $7.2 \times 10^{20} ~{\rm cm}^{-2}$,
is about quarter of  {LL10} column density and the transmission of 
the O\textsc{vii} emission is about 54 \%.  This suggests that a large fraction
of the linear component arises from beyond the bulk of Galactic absorption,
although the possibility that the difference between
LL10 and LL21 is mainly due to spatial fluctuation of the emission measure
cannot be ruled out.   
We thus consider that the linear component  is associated with the TAE component.   
This component emits   O\textsc{vii} and O\textsc{viii} with an intensity ratio of about  2 to 1 and shows field-to-field variations of  
$\sim 0$ to $\sim 7$ LU in O\textsc{vii} intensity.
The lines in Figure~\ref{fig:oviii-ovii_Bkn} show O\textsc{vii} - O\textsc{viii} intensity relations for  different temperatures and absorption column densities.  We have assumed 2.1 LU as the O\textsc{vii} floor intensity, using the value for  the midplane field ({MP235}).   The lines suggest that the average temperatures along each line of sight of the plasmas emitting the O\textsc{viii} emission and the  O\textsc{vii} emission above the floor are in a relatively narrow range of $kT \sim$ 0.19 to 0.23 keV if they are arising from collisionally equilibrium plasma in spite of  the large O\textsc{vii} intensity variations  ($\sim 0$ to $\sim 7$ LU).

\begin{figure*}
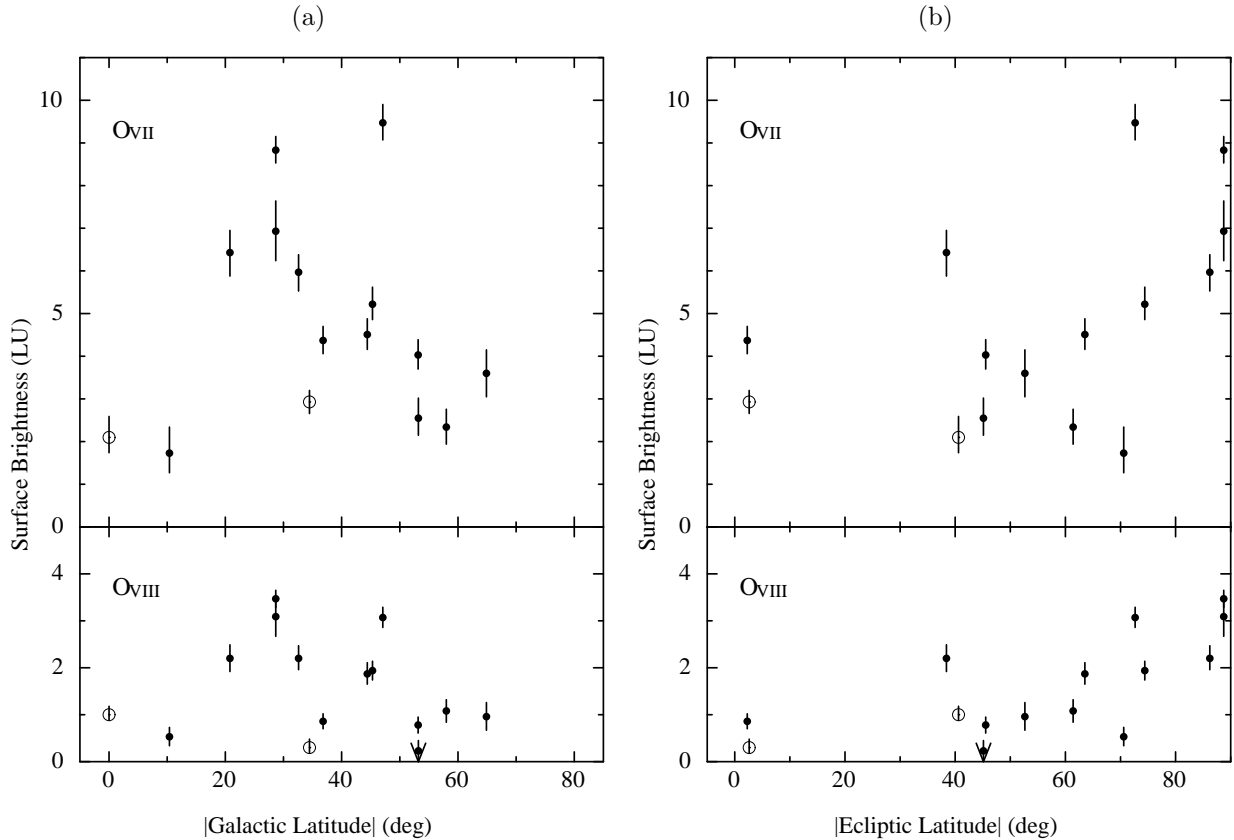

\begin{tabular}{cc}
(a) & (b) \\
 \FigureFile(0.44\textwidth, ){fig4_a.ps} &
 \FigureFile(0.44\textwidth, ){fig4_b.ps}\\
\end{tabular}
\caption{
O\textsc{vii} and O\textsc{viii} emission intensities as  functions of 
absolute Galactic latitude $|b|$ (a), 
and as functions of absolute ecliptic latitude (b).  
The two data points marked with an open circle are O\textsc{vii} intensities of the MBM 12 on cloud direction and  the midplane direction (235, 0).  Both are taken from \citet{Masui_etal_2009}.  
\label{fig:ovii_vs_latitude}}
\end{figure*}

In Figure~\ref{fig:ovii_vs_latitude}, we plotted the O\textsc{vii} and O\textsc{viii} 
emission intensities as functions of absolute Galactic latitude $|b|$ (a), and as a function of absolute ecliptic latitude (b).  The two data points marked with an open circle are {M12on} and MP235.  
The correlations of O\textsc{vii} and O\textsc{viii} emission intensities to  Galactic latitude and 
ecliptic latitude are not very clear.  Near solar minimum, Heliospheric SWCX emission
is expected to be enhanced in low ecliptic latitude region of $\lesssim 20^\circ$
by $\sim 1.5$ LU  \citep{Koutroumpa_etal_2006}.
Such small difference may be masked by the spatial variations of the TAE, 
systematic errors in geocoronal-SWCX removal, and
time variations of Heliospheric SWCX.   
The relevance of Galactic latitude (for TAE) and ecliptic latitude (for
SWCX emission ) are discussed in sections \ref{sec:dis:TAE} 
and \ref{sec:dis:H-SWCX}, respectively. 

\subsection{Spectral models consistently describe O emission}

If the O\textsc{vii} emission originates in the two plasma emission components of the model as discussed above, we would expect approximately constant temperature and normalization for the SWCX+LHB component that provides the floor to the Ovii rate and a constant temperature ($kT \sim 0.2$ keV) for the TAE component with its constant O\textsc{viii}/O\textsc{vii} ratio. 
However, the results of the spectral fits shown in 
Table~\ref{tbl:specfit_bknpow} (and  Table~\ref{tbl:specfit_pow}
do not
show such tendency. The Ovii emission intensities of the SWCX+LHB component estimated using the best-fit parameters in Table~\ref{tbl:specfit_bknpow} vary from 2 to 5.5 LU. The temperature of the TAE component is significantly higher than 0.2 keV and is far from constant.

We consider that this discrepancy is related to the Fe and Ne emission.
When we fixed the abundances (Fe to O, and Ne to O ratios) of the hot gas in the spectral fits, 
the temperature of the TAE component was mostly determined by Fe-L and/or Ne-K emission rather than O emission.  Then the temperature and the intensity of the SWCX+LHB component was optimized to fill the remaining emission.  We notice relatively large residual near Ne K emission energy, in some of the residual plots  in Figure~\ref{fig:spectrum}.
We will now try fixing the temperature and the normalization of the SWCX+LHB component, and allowing  the Ne and Fe abundances of the TAE component to vary to see if we can still get good fits and if the temperatures of the second component will be closer to 0.2 keV.  This is model 2.  
 
As discussed in the previous subsection, the O\textsc{vii} emission from the three fields,
{M12on},  {MP235}, and  {LL10} 
contains only the SWCX+LHB component.
The temperature and normalization (emission measure) derived for that
component (see Tables \ref{tbl:specfit_bknpow} and \ref{tbl:specfit_pow}) are, 
respectively, about 0.11 keV and
$14 \times 10^{14} {\rm cm}^{-5}~ {\rm str}^{-1}$ for all three fields, which is consistent
with the parameters obtained for model fitting of data from the
LMC X-3 vicinity ({LX-3}) by \citet{Yao_etal_2009} 
(0.103  keV and 18.4  $\times 10^{14} {\rm cm}^{-5}~ {\rm str}^{-1}$), which were derived from previous XMM-Newton and Suzaku observations of directions with absorption gas.  
For the model 2 spectral fits, we therefore 
adopted the best fit value of the midplane observation (Table~\ref{tbl:specfit_bknpow}) 
as representative of SWCX+LHB parameters.

\begin{table*}
\caption{Results of three-component (CXB, TAE, SWCX+LHB) spectral fit with SWCX+LHB temperature and normalization fixed, and with double broken power law CXB (model 2)
\label{tbl:specfit_fix_bknpow}}
\begin{center}
\begin{tabular}{lccccccc} \hline \hline
ID & $N_{\rm H} $$^{\rm a}$ & CXB$^{\rm b}$  & \multicolumn{4}{c}{TAE} & $\chi^2$/dof\\ 
     & $10^{20}$cm$^{-2}$ &   Norm$^{\rm c}$ & kT (keV)  &  Ne  & Fe & Norm$^{\rm d}$   &     \\  \hline
{1} (GB) & 1.40 &  5.6$_{  -0.7 }^{+   0.7  }$  & 0.222$_{ -0.068 }^{+ 0.111  }$ & 1$^{\rm e}$  &1$^{\rm e}$   &  1.2$_{  -1.0 }^{+   1.0  }$  &95.12/ 92 \\
{2} (HL-B)  &3.36  & 2.5$_{  -0.6 }^{+   0.5  }$ &   0.296$_{ -0.049 }^{+ 0.093  }$ & 1$^{\rm e}$   & 1$^{\rm e}$   &1.6$_{  -0.7 }^{+   0.9  }$  & 95.14/88 \\
{3}  (LH-2) & 0.56 &  3.5$_{  -0.5 }^{+   0.4  }$  & 0.222&1$^{\rm e}$   &1$^{\rm e}$  & $<   0.7$ & 95.33/100 \\
{4}  (LH-1) & 0.56   & 5.6$_{  -0.5 }^{+   0.5  }$ & 0.237$_{ -0.058 }^{+ 0.262  }$ & 3.79$_{ -2.02 }^{+ 6.43  }$ & 2.49$_{ -1.98 }^{+ 13.60  }$ & 2.0$_{  -0.6 }^{+   0.6  }$ & 162.31/127 \\
{5} (OFF-FIL) & 4.19 &  3.1$_{  -0.4 }^{+   0.4  }$ & 0.181$_{ -0.008 }^{+ 0.008  }$ & 3.53$_{ -0.80 }^{+ 1.01  }$ & 2.99$_{ -1.21 }^{+ 2.75  }$  & 15.3$_{  -1.1 }^{+   1.1  }$  & 159.33/122 \\
{6} (ON-FIL) & 4.61  & 5.0$_{  -0.5 }^{+   0.5  }$  & 0.239$_{ -0.025 }^{+ 0.027  }$  & 3.46$_{ -1.19 }^{+ 1.91  }$ & 1.88$_{ -0.85 }^{+ 1.97  }$ &  5.2$_{  -0.9 }^{+   0.9  }$  &  130.94/117 \\ %
{7} (HL-A) & 1.02  &  4.9$_{  -0.5 }^{+   0.5  }$&  0.243$_{ -0.032 }^{+ 0.031}$& 0.91$_{ -0.78 }^{+ 1.06  }$& 0.51$_{ -0.48 }^{+ 1.16  }$ & 4.3$_{  -0.8 }^{+   0.8  }$  & 138.89/115 \\
{8}  (M12off) & 8.74 & 2.0$_{  -0.4 }^{+   0.4  }$ & 0.184$_{ -0.030 }^{+ 0.023  }$  & 1$^{\rm e}$   &1$^{\rm e}$   & 4.9$_{  -1.2 }^{+   1.2  }$    &89.83/91 \\
{9}  (LX-3) &4.67 & 8.8$_{  -0.5 }^{+   0.5  }$ &  0.213$_{ -0.018 }^{+ 0.026  }$&3.04$_{ -1.168 }^{+ 1.282  }$  & 3.12$_{ -1.590 }^{+ 1.978  }$  & 7.7$_{  -0.5 }^{+   1.2  }$ & 210.94/196 \\
{10} (NEP1)  & 4.40 &  3.2$_{  -0.4 }^{+   0.5  }$   & 0.191$_{ -0.006 }^{+ 0.006  }$ &  2.09$_{ -0.47 }^{+ 0.53  }$ & 1.69$_{ -0.55 }^{+ 0.75  }$  &  15.3$_{  -0.9 }^{+   0.9  }$  &  164.60/120 \\
{11} (NEP2) & 4.40 &3.4$_{  -0.7 }^{+   0.7  }$ &  0.206$_{ -0.019 }^{+
 0.041  }$ & 2.65$_{ -1.34 }^{+ 1.43  }$ & 1.59$_{ -1.177 }^{+ 1.743  }$
 & 10.8$_{  -1.8 }^{+   1.8  }$ & 57.94/55 \\ 
{12}  (LL21) & 7.24& 4.7$_{  -0.5 }^{+   0.5  }$  & 0.193$_{ -0.016 }^{+ 0.019  }$  &  2.00$_{ -0.86 }^{+ 1.15  }$  & 2.88$_{ -1.33 }^{+ 3.53  }$  & 11.1$_{  -1.7 }^{+   1.7  }$ &112.77/97  \\
{13}  (LL10) &27.10 & 2.5$_{  -0.7 }^{+   0.7  }$ & 0.752$_{ -0.177 }^{+ 0.240  }$& 1$^{\rm e}$  & 1$^{\rm e}$  &1.6$_{  -0.7 }^{+   0.7  }$ & 91.44/91  \\ 
\hline
\multicolumn{8}{l}{
\rlap{\parbox[t]{.85\textwidth}{
$^{\rm a}$~{The absorption column densities for the CXB and TAE components were fixed to the tabulated values. }
}}}\\
\multicolumn{8}{l}{
\rlap{\parbox[t]{.85\textwidth}{
$^{\rm b}$~{Two broken power-law model was adopted.  The  photon
 indexes below 1.2 keV were fixed to 1.52 and 1.96 .  The normalization of the former index component is fixed to 5.7, and only the normalization of the other component was allowed to vary.}
}}}\\
\multicolumn{8}{l}{
\rlap{\parbox[t]{.85\textwidth}{
$^{\rm c}$~{The normalization of one of the broken power-law components with  the unit of ${\rm phtons~s}^{-1}{\rm cm}^{-2}{\rm~keV}^{-1}{\rm str}^{-1}$@1keV. }
}}}\\
\multicolumn{8}{l}{
\rlap{\parbox[t]{.85\textwidth}{
$^{\rm d}$~{The emission measure integrated over the line of sight,  $(1/4\pi)\int n_{\rm e} n_{\rm H} ds$ in the unit of $10^{14}{\rm cm}^{-5}~{\rm str}^{-1}$.}
}}}\\
\multicolumn{8}{l}{
\rlap{\parbox[t]{.85\textwidth}{
$^{\rm e}$~{The abundance was not constrained well, thus fixed to the solar value.}
}}}\\
\end{tabular}
\end{center}
\end{table*}

\begin{figure*}
\setcounter{keepfignumhs}{\value{figure}}
\begin{scriptsize}
\begin{tabular}{lll}
\begin{minipage}{\figw\textwidth}
({1}) GB1428+4217\vspace{-0.055\textwidth}\\
\FigureFile(0.99\textwidth, ){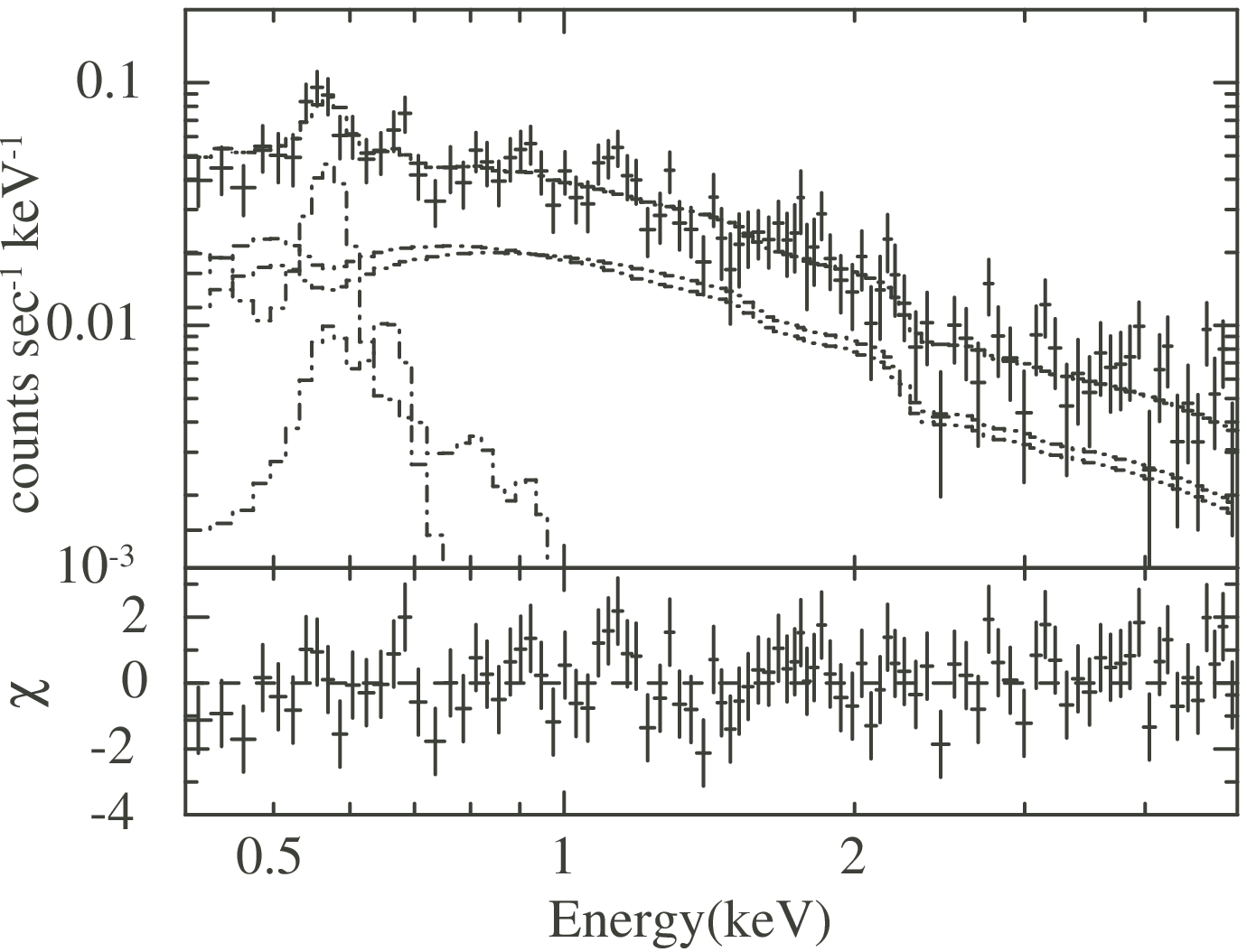}
\end{minipage}
&
\begin{minipage}{\figw\textwidth}
({2}) High latitude B\vspace{-0.055\textwidth}\\
\FigureFile(0.99\textwidth, ){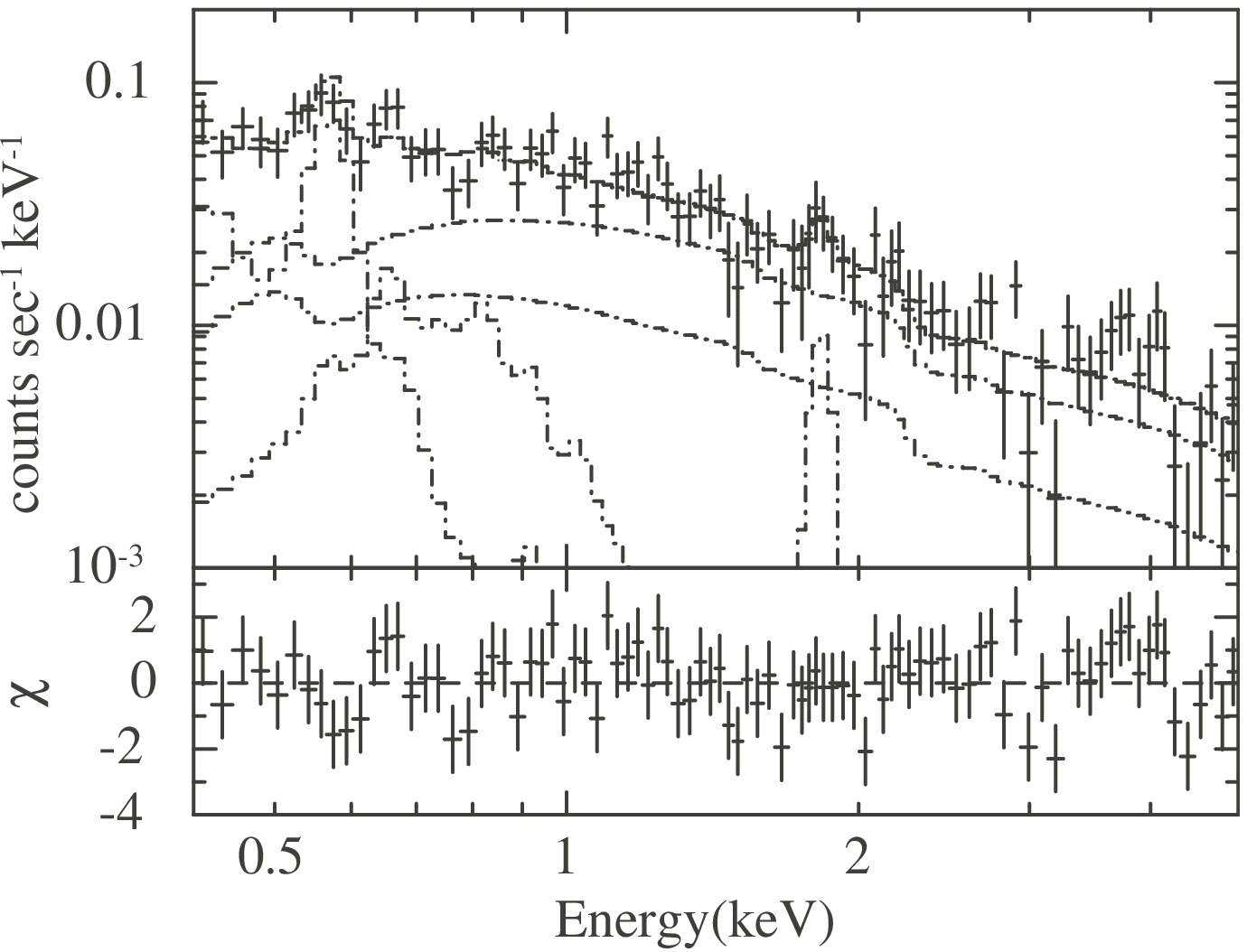}
\end{minipage}
\\
\begin{minipage}{\figw\textwidth}
({3}) Lockman hole 2\vspace{-0.055\textwidth}\\
\FigureFile(0.99\textwidth, ){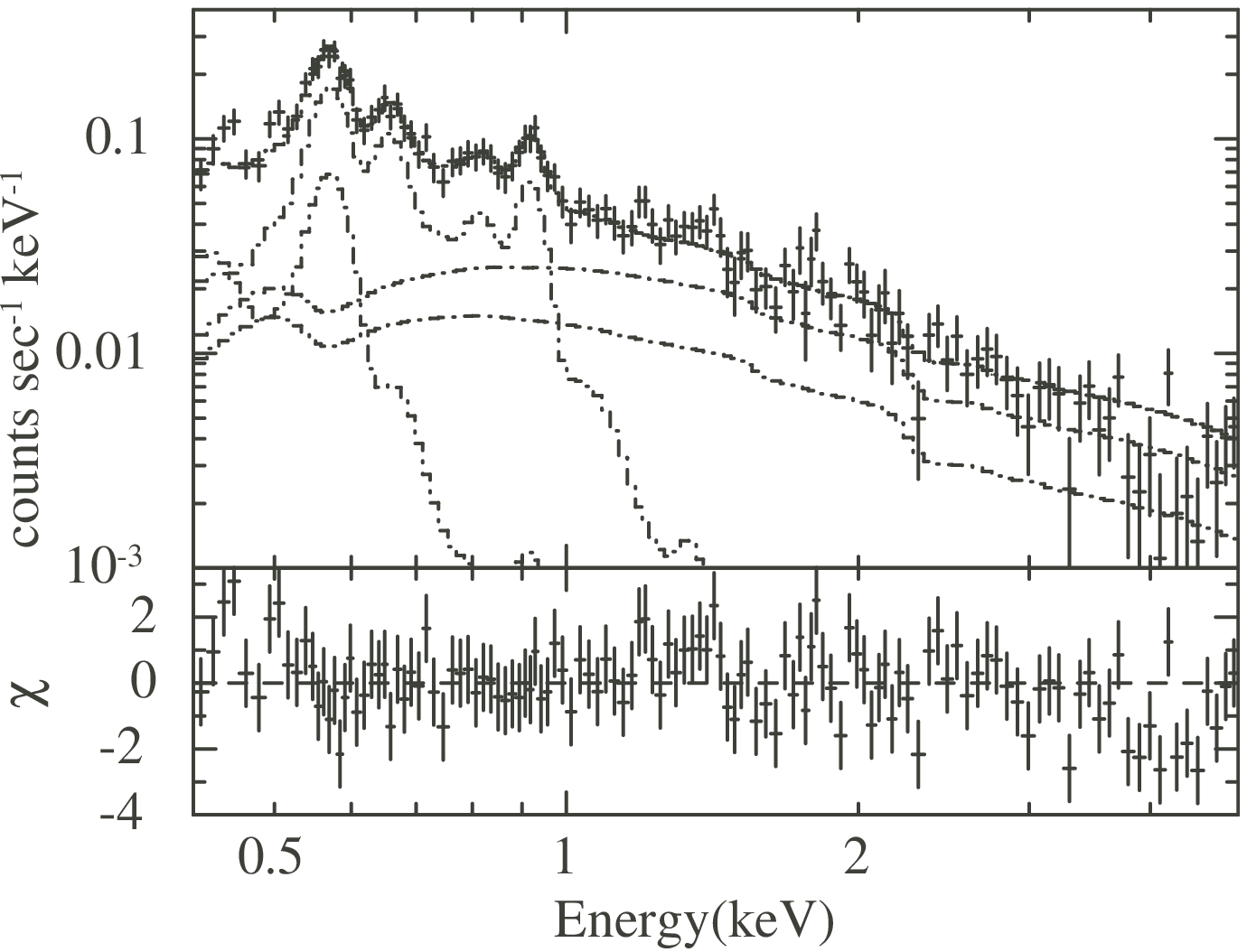}
\end{minipage}
&
\begin{minipage}{\figw\textwidth}
({4})  Lockman hole 1\vspace{-0.055\textwidth}\\
\FigureFile(0.99\textwidth, ){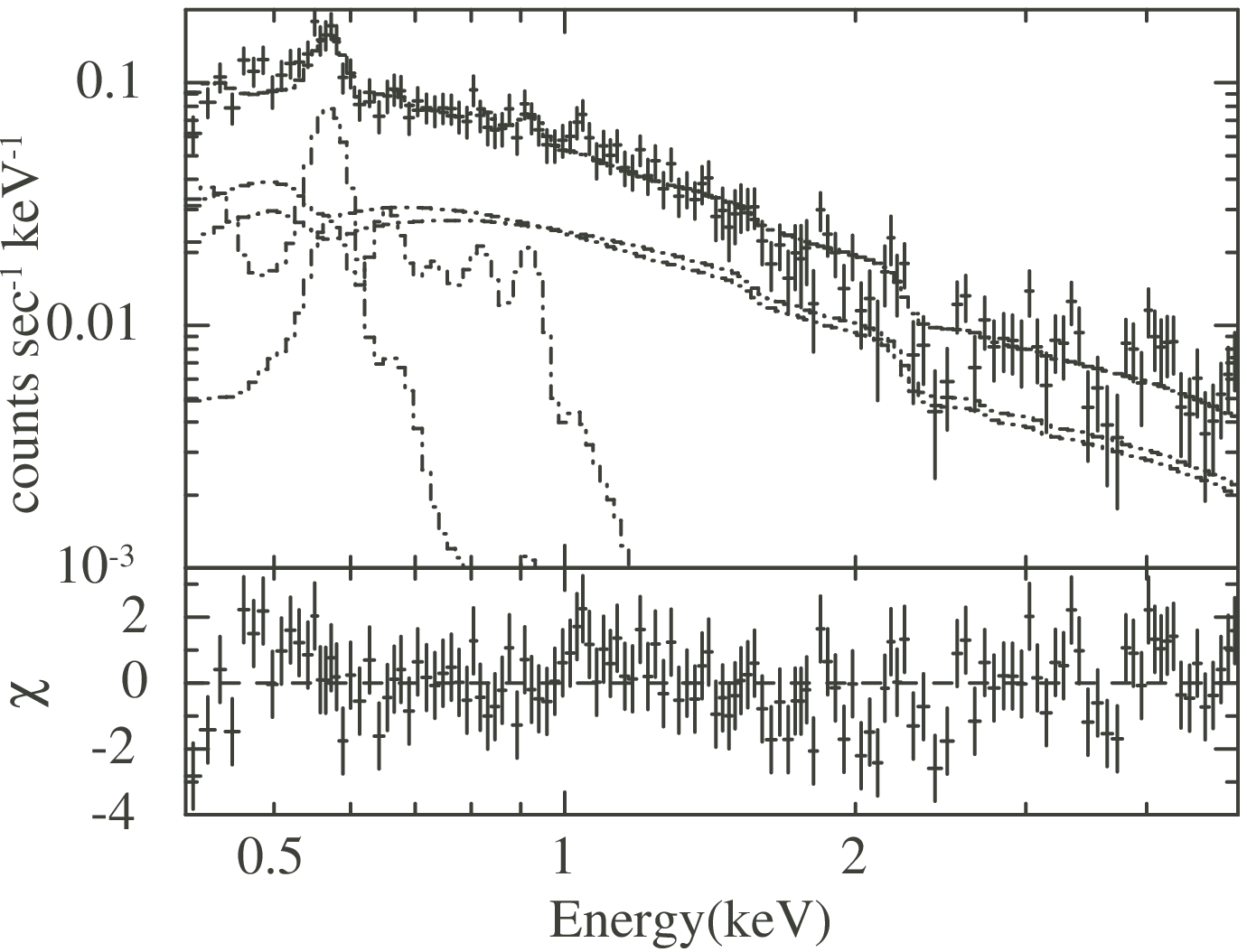}
\end{minipage}
\\
\begin{minipage}{\figw\textwidth}
({5}) Off Filament\vspace{-0.055\textwidth}\\
\FigureFile(0.99\textwidth, ){fig5_5.eps}
\end{minipage}
&
\begin{minipage}{\figw\textwidth}
({6}) On Filament\vspace{-0.055\textwidth}\\
\FigureFile(0.99\textwidth, ){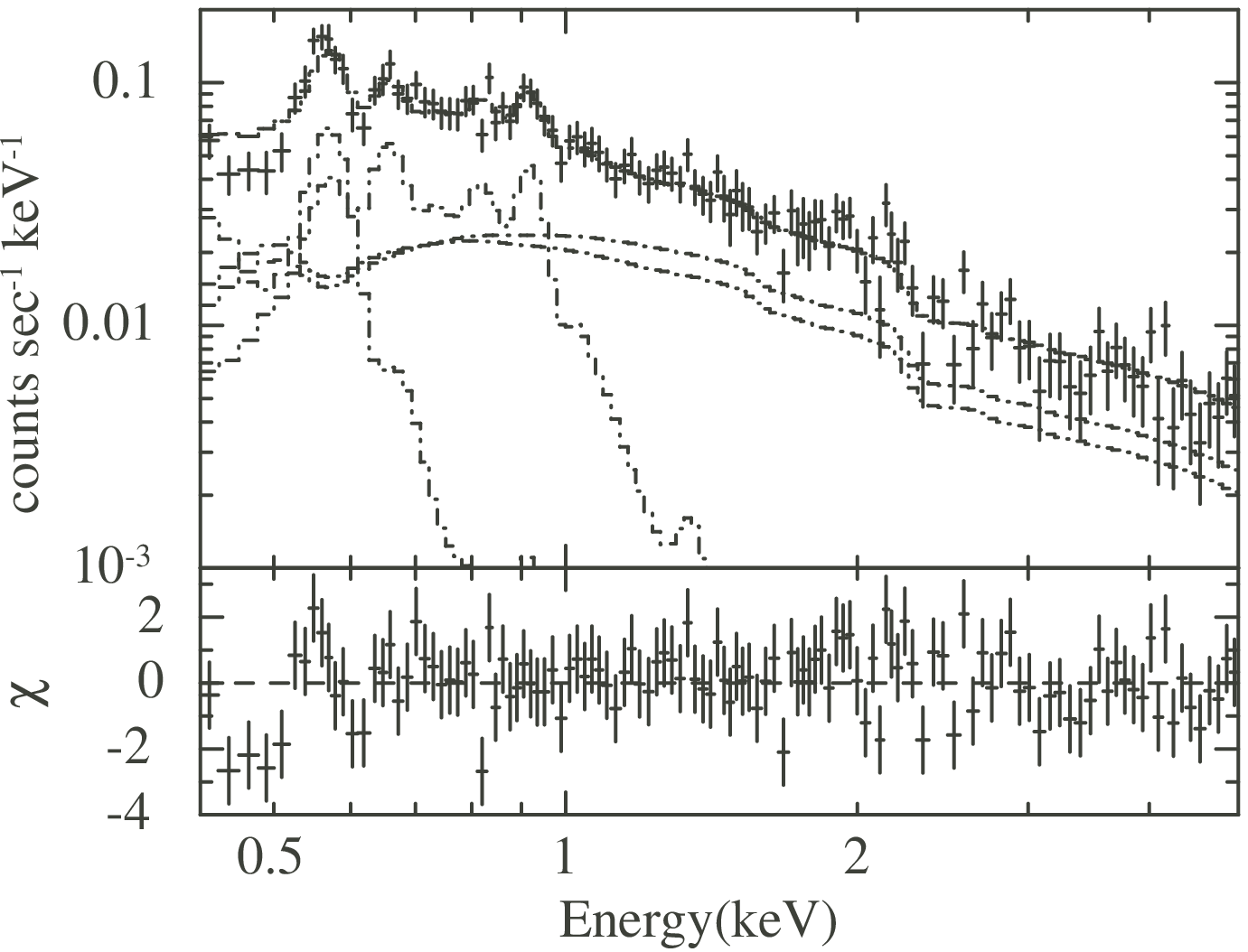}
\end{minipage}
\end{tabular}
\end{scriptsize}
\caption{
Results of spectral fits with SWCX+LHB component fixed (model 2) except for {MP235} (last panel). 
Observed spectra (crosses), best-fit model and its components (step functions)   convolved with the instrument response function and residuals of the fit (bottom panels) are shown. The vertical error bars of data points correspond to the  $1~\sigma$ statistical errors.
The model parameters of the SWCX+LHB component was taken from the midplane observation, {MP235}
in Table~\ref{tbl:specfit_pow}
\label{fig:spectrum_fixedfit} 
}
\end{figure*}

\addtocounter{figure}{-1}

\begin{figure*}
\begin{scriptsize}
\begin{tabular}{lll}
\begin{minipage}{\figw\textwidth}
({7}) High latitude A\vspace{-0.055\textwidth}\\
\FigureFile(0.99\textwidth, ){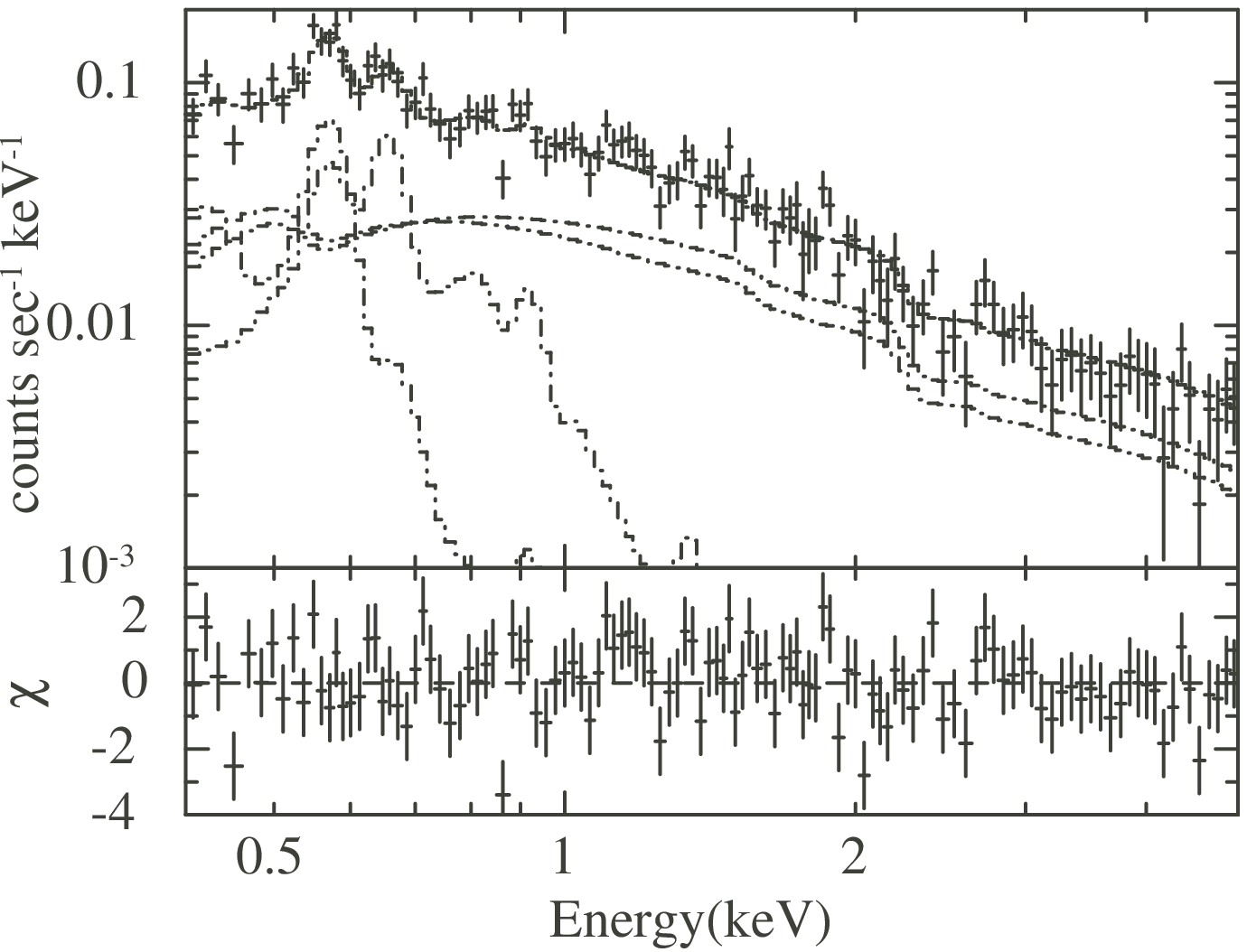}
\end{minipage}
&
\begin{minipage}{\figw\textwidth}
({8}) MBM12 off cloud\vspace{-0.055\textwidth}\\
\FigureFile(0.99\textwidth, ){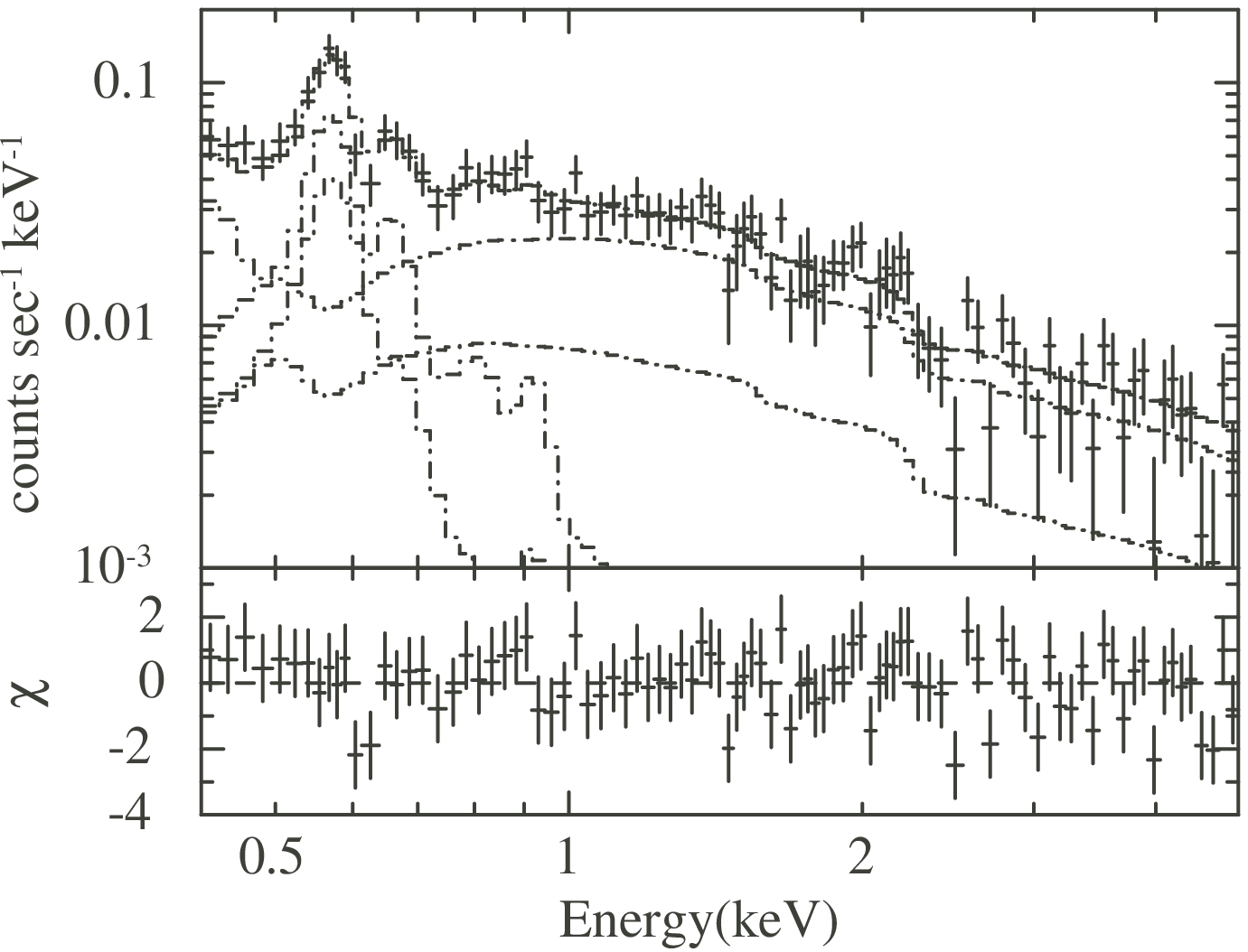}
\end{minipage}
\\
\begin{minipage}{\figw\textwidth}
({9}) LMC X-3 Vicinity\vspace{-0.055\textwidth}\\
\FigureFile(0.99\textwidth, ){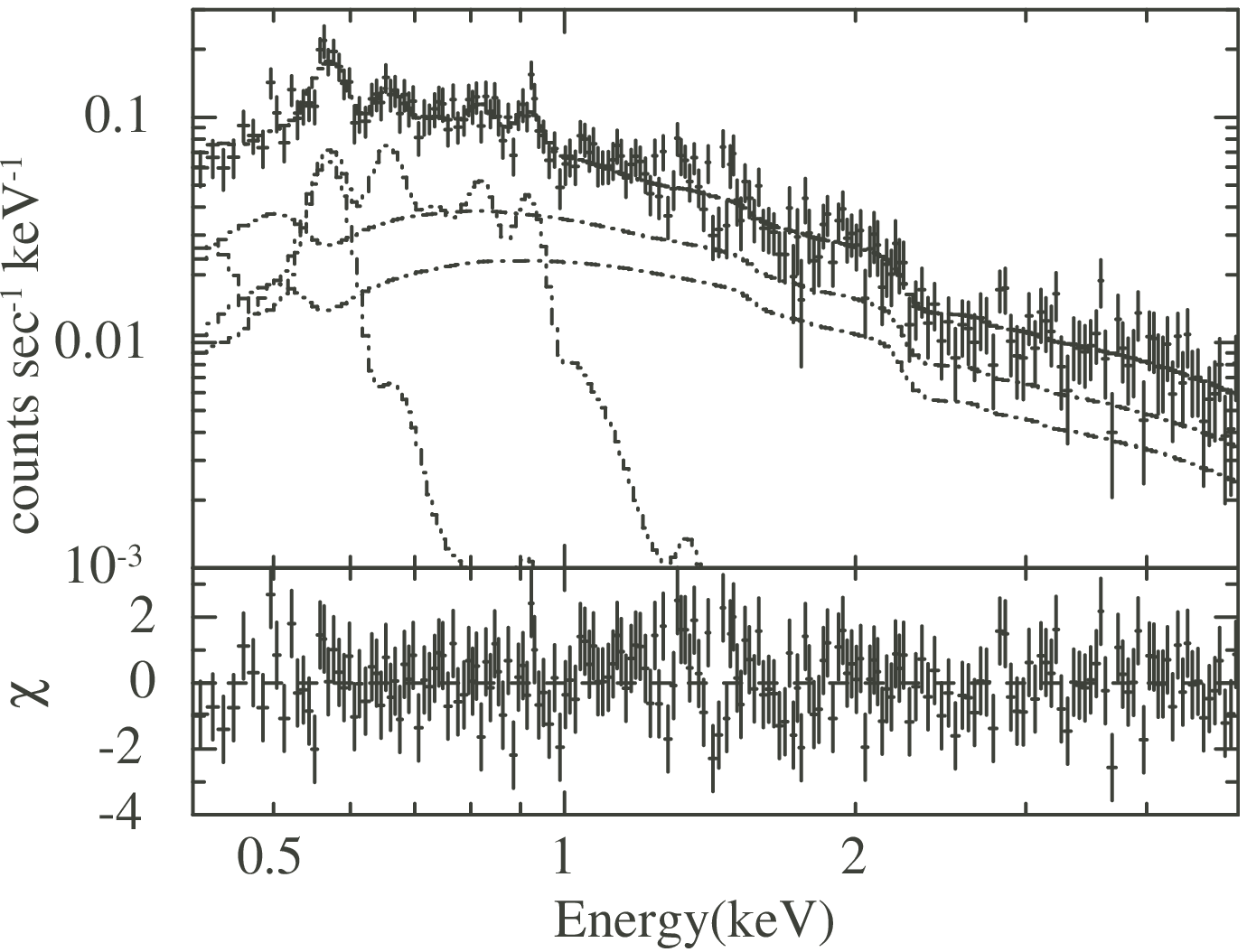}
\end{minipage}
&
\begin{minipage}{\figw\textwidth}
({10})  North Ecliptic Pole\vspace{-0.055\textwidth}\\
\FigureFile(0.99\textwidth, ){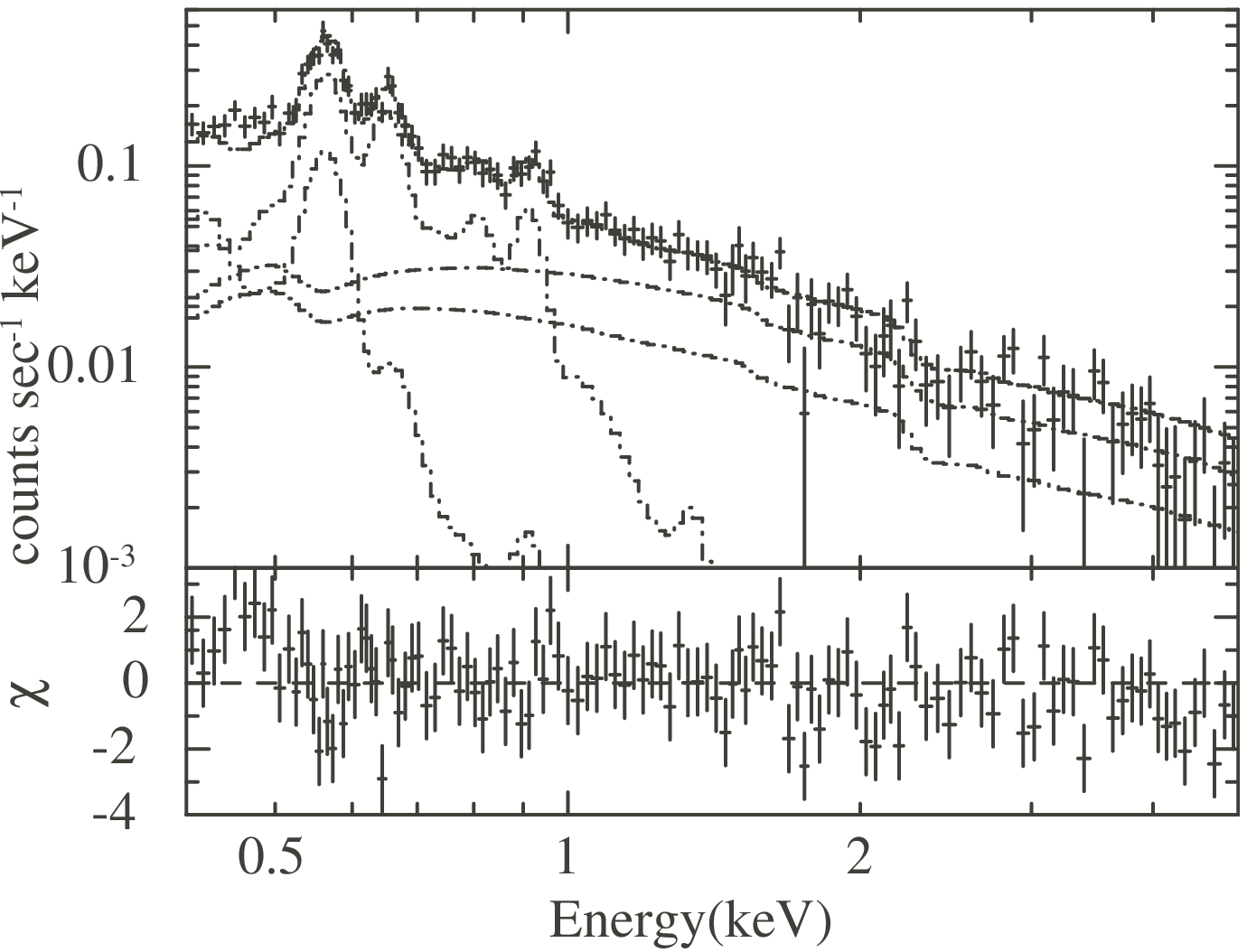}
\end{minipage}
\\
\begin{minipage}{\figw\textwidth}
({11})  North Ecliptic Pole 2\vspace{-0.055\textwidth}\\
\FigureFile(0.99\textwidth, ){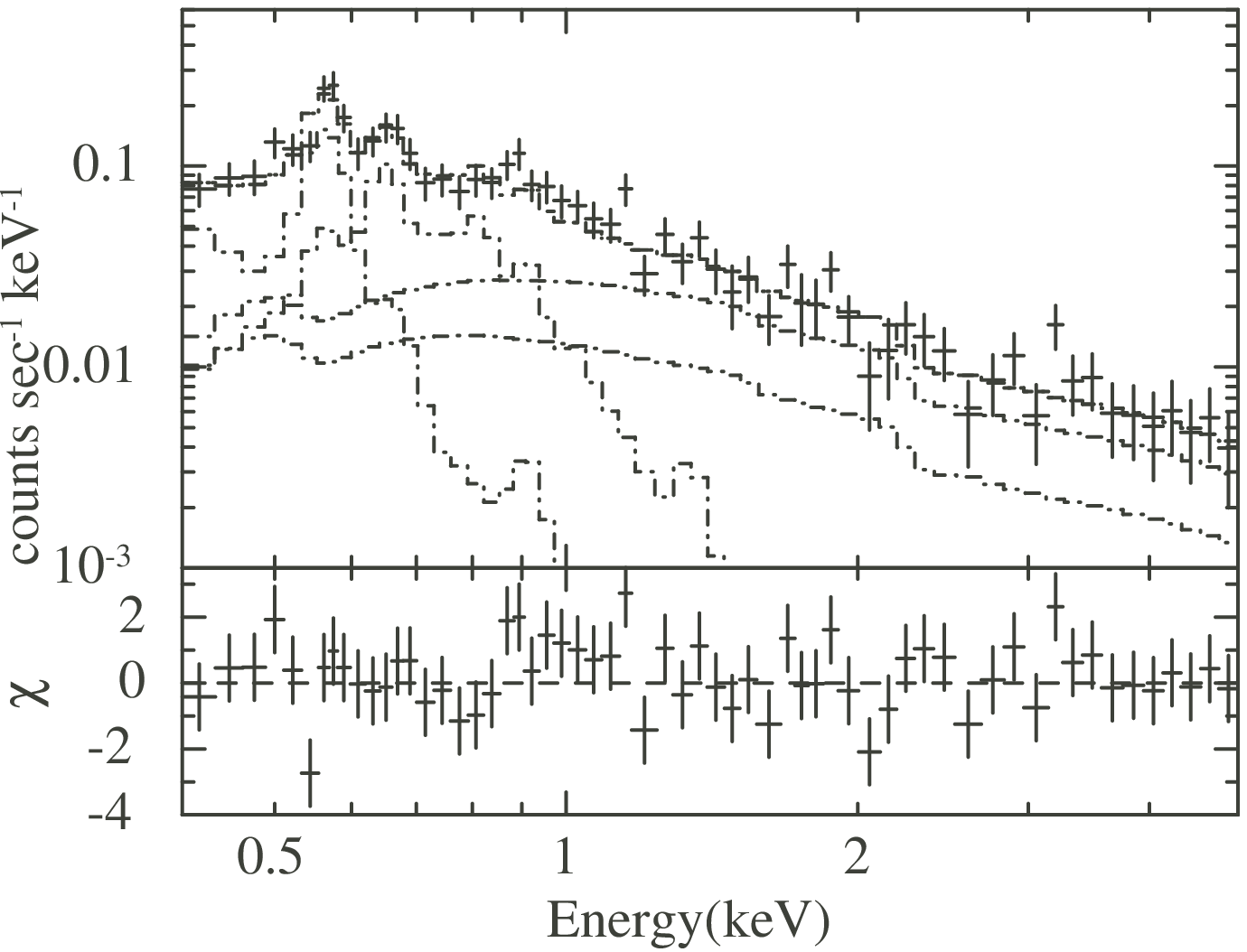}
\end{minipage}
&
\begin{minipage}{\figw\textwidth}
({12}) Low latitude 86-21\vspace{-0.055\textwidth}\\
\FigureFile(0.99\textwidth, ){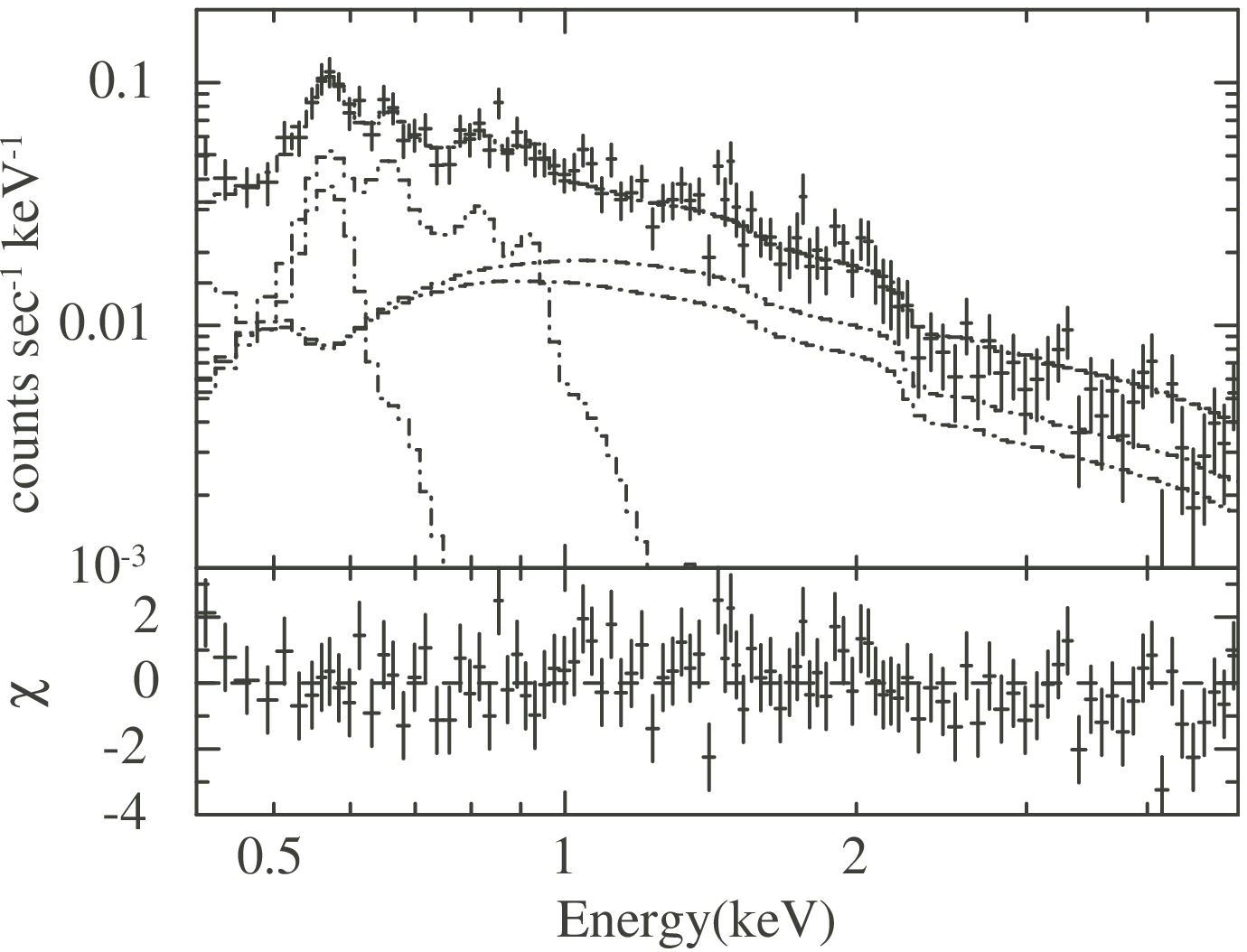}
\end{minipage}
\\
\begin{minipage}{\figw\textwidth}
({13}) Low latitude 97+10\vspace{-0.055\textwidth}\\
\FigureFile(0.99\textwidth, ){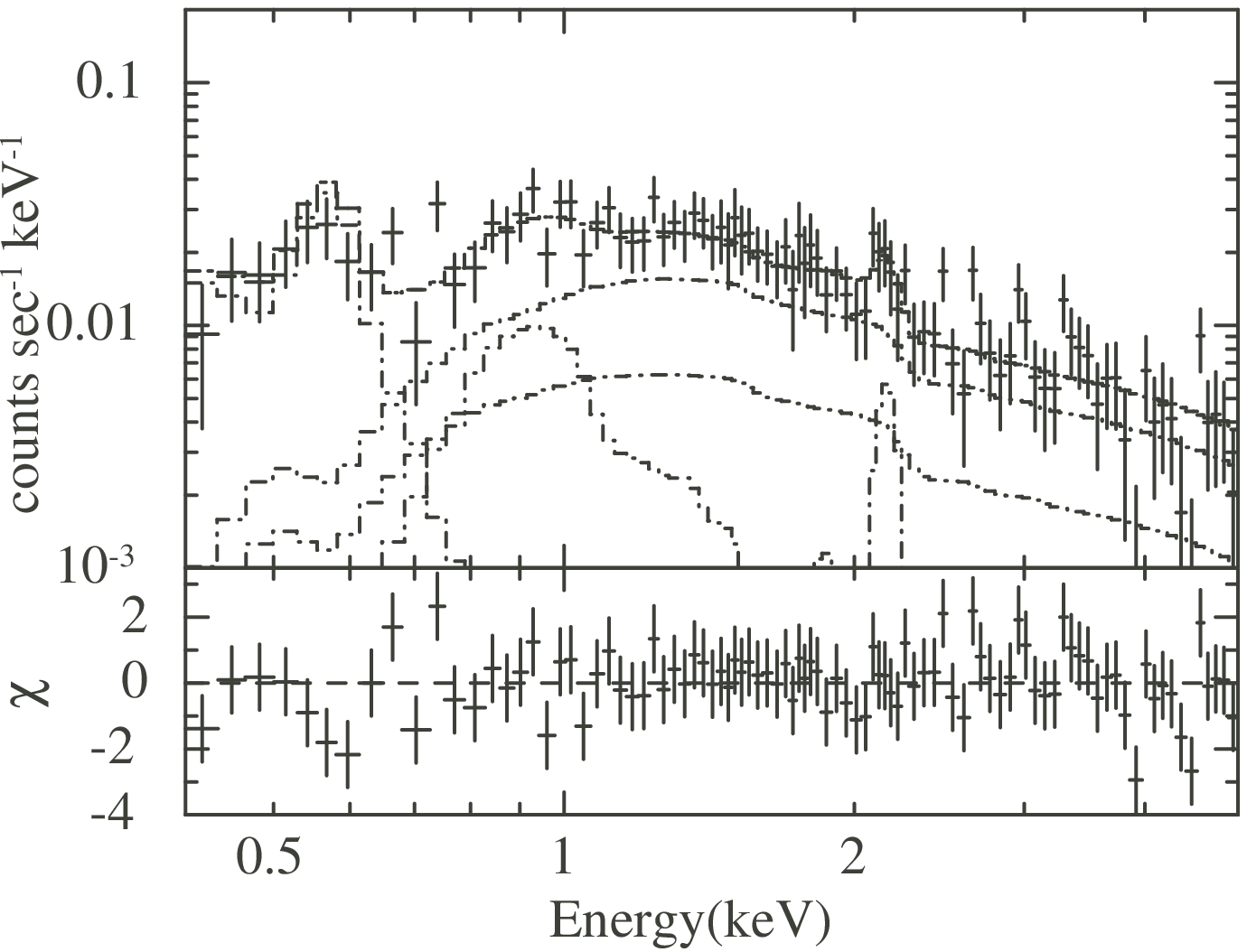}
\end{minipage}
&
\begin{minipage}{\figw\textwidth}
({R2}) Midplane 235\vspace{-0.055\textwidth}\\
\FigureFile(0.99\textwidth, ){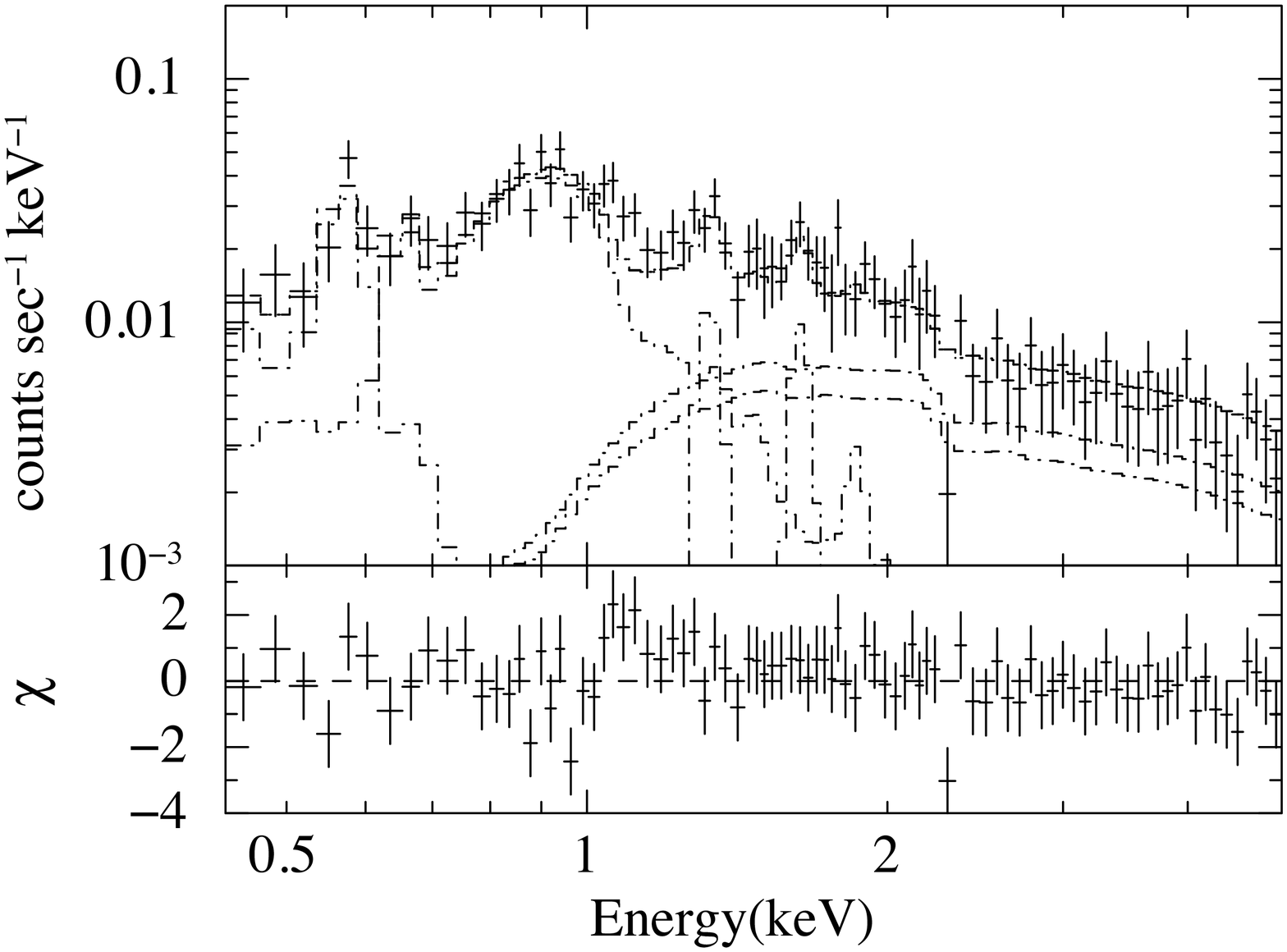}
\end{minipage}
\end{tabular}
\end{scriptsize}
\caption{
Continued.
}
\end{figure*}

In Table~\ref{tbl:specfit_fix_bknpow} and Figure~\ref{fig:spectrum_fixedfit} 
we show the results of the fits.
The $\chi^2$ values are generally good 
with the reduced $\chi^2$ values in the range of 0.95 -- 1.28 for the
degrees of freedom of 91 -- 196, although most of them are
larger than those for the fit with the first model in Table~\ref{tbl:specfit_bknpow}.
For five of the spectra,  the abundances were not constrained well. 
We thus fixed the Fe and Ne abundances  to the solar value. 
For  {LH-2}, the existence of the TAE was not significant.  
We estimated the upper limit of the intensity fixing the temperature to the 
average value of the other spectra.  
Excluding {HL-B} and {LL10}, the temperature of the TAE component are 
in the range of 0.18 to 0.24 keV with an average of 0.22 keV, which is significantly lower than the temperatures obtained by model 1.   

In model 2, the strong Fe-L and Ne K lines are explained by over-abundance of those elements.
In particular, $\sim 3 \times$ solar  abundance was required for Ne for four fields.   
We consider that the strong Ne and Fe emissions could be also represented
by a higher temperature emission component with solar  abundances.  
For model 3, we fixed the temperature and the abundance of the TAE component to the average of model 2 (0.222 keV) and to the solar value, respectively.  We introduce a fourth emission component with higher temperature which we denote TAE'.  In Table~\ref{tbl:specfit_taefix_bknpow}, we show the results for the four fields.  The resultant $\chi^2$ values are comparable to those for model 2.  In model 3, the excess Ne and Fe emissions are explained by emission with a temperature in the range of 0.6 to 0.9 keV, and an emission measure of $(1-2) \times 10^{14}~ {\rm cm}^{-5}~{\rm str}^{-1}$.

\begin{table*}
\caption{Results of four-component (CXB, TAE, TAE', SWCX+LHB)  spectral fit with SWCX+LHB normalization and temperature , and TAE temperature fixed,  and with double broken power law CXB (model 3)
\label{tbl:specfit_taefix_bknpow}}
\begin{center}
\begin{tabular}{lccccccc} \hline \hline
ID & $N_{\rm H} $$^{\rm a}$ & CXB$^{\rm b}$  &
 \multicolumn{2}{c}{TAE} & \multicolumn{2}{c}{TAE'} & $\chi^2$/dof\\ 
     & $10^{20}$cm$^{-2}$ &   Norm$^{\rm c}$ & kT (keV)  &
 Norm$^{\rm d}$  & kT (keV) & Norm$^{\rm d}$   &     \\  \hline
{4} ({LH-1}) & 0.56   &5.3$_{  -0.6 }^{+   0.5  }$ & 0.222 &1.9$_{  -0.6 }^{+   0.6  }$ & 0.746$_{ -0.382 }^{+ 0.326  }$ & 1.0$_{  -0.3 }^{+   0.4  }$ & 160.48/128 \\ 
{5} ({Off-FIL}) & 1.90 &2.6$_{  -0.5 }^{+   0.5  }$  & 0.222 & 10.3$_{  -0.7 }^{+   0.7  }$ & 0.861$_{ -0.115 }^{+ 0.132  }$ & 0.8$_{  -0.4 }^{+   0.4  }$ &197.26/123 \\
{6} ({On-FIL}) & 9.60  & 5.1$_{  -0.6 }^{+   0.5  }$ & 0.222 & 9.0$_{  -1.2 }^{+   1.1  }$ &0.676$_{ -0.090 }^{+ 0.094  }$ & 1.9$_{  -0.5 }^{+   0.6  }$ & 122.93/118 \\
{9} ({LX-3}) &4.67 & 8.5$_{  -0.6 }^{+   0.7  }$ & 0.222  & 7.4$_{  -1.1 }^{+   1.0  }$  & 0.559$_{ -0.193 }^{+ 0.109  }$ & 1.8$_{  -0.7 }^{+   0.4  }$  & 216.50/197 \\  
\hline\hline
\multicolumn{8}{l}{
\rlap{\parbox[t]{.8\textwidth}{
$^{\rm a}$~{The absorption column densities for the CXB, TAE, and TAE' components were fixed to the tabulated values. }
}}}\\
\multicolumn{8}{l}{
\rlap{\parbox[t]{.8\textwidth}{
$^{\rm b}$~{Two broken power-law model was adopted.  The  photon
 indexes below 1.2 keV were fixed to 1.52 and 1.96 .  The normalization of the former index component is fixed to 5.7, and only the normalization of the other component was allowed to vary.}
}}}\\
\multicolumn{8}{l}{
\rlap{\parbox[t]{.8\textwidth}{
$^{\rm c}$~{The normalization of one of the broken power-law components with  the unit of ${\rm phtons~s}^{-1}{\rm cm}^{-2}{\rm~keV}^{-1}{\rm str}^{-1}$@1keV.   Add 5.7 to obtain the total flux at 1 keV.}
}}}\\
\multicolumn{8}{l}{
\rlap{\parbox[t]{.8\textwidth}{
$^{\rm d}$~{The emission measure integrated over the line of sight, i.e. $(1/4\pi)\int n_{\rm e} n_{\rm H} ds$ in the unit of $10^{14}{\rm cm}^{-5}~{\rm str}^{-1}$.}
}}}\\
\end{tabular}
\end{center}
\end{table*}

\section{Discussion}
\label{sec:discussion}
\subsection{Heliospheric Solar wind charge exchange}
\label{sec:dis:H-SWCX}

Although we have been careful to avoid possible contamination by geocoronal SWCX through the selection procedures discussed in section 2, contributions from SWCX in interplanetary space are at best only partially removed by these methods since the long transit times through the solar system wash out the very large short-term intensity variations seen by monitor satellites located in the ecliptic plane near the Earth.  Models of this heliospheric SWCX emission are highly uncertain, but some predict that a large fraction of the Galactic {\it R45} emission is from this source \citep{Koutroumpa_etal_2006, Koutroumpa_etal_2008},
making it one of the most severe limitations in determining the true interstellar and halo contributions.

Solar activity affects the solar wind densities and ionization temperature.  
Near the solar maximum, the slow solar winds which have  high ionization temperatures and high densities are ejected from the sun resulting in stronger SWCX induced O\textsc{vii} and O\textsc{viii} emissions \citep{Koutroumpa_etal_2006} than in solar minimum.
Near the solar minimum, slow solar winds are emitted from the equator region of the sun, and high speed, low-density,  low-ionization-temperature winds are emitted from 
the high latitude region of the sun. 
Discrepancies between early XMM-Newton or Chandra observations which were also made near the solar maximum  in 2000-2002 and Suzaku observations near the solar minimum were noted by \citet{Koutroumpa_etal_2007} and \citet{Henley_Shelton_2008}.   
For example, Lockman Hole $(\ell, b) = (149.1, 53.6)$ by XMM-Newton observations taken in October 2002 show much higher O\textsc{vii} intensities  (7 to 18 LU, \citet{Koutroumpa_etal_2007}) than those of Suzaku (2.5 and 4.1 from the present {LH-1} and
{LH-2} fields).  
\citet{Henley_Shelton_2008} showed that O\textsc{vii} line intensities on and off directions of the shadowing filament ({On-FIL} and {Off-FIL} fields) were  $10.65^{+0.77}_{-0.82}$ and $13.86^{+1.58}_{-1.49}$ LU for the XMM-Newton observations in 2002 March, while they reported
that the intensities were $6.51^{+0.37}_{-0.45}$ and $10.53^{+0.68}_{-0.55}$ LU
for  the Suzaku observations (statistically consistent with the present results,  $5.2 ^{+1.0}_{-0.6}$ and $9.5 \pm 0.7$ LU).   If this difference is due to the different SWCX intensity as the authors suggest, it was brighter by about 5 LU during the XMM-Newton observations.
The O\textsc{vii}  emision intensity in the MBM12 on-cloud direction determined by Chandra ($1.79 \pm 0.55$) and Suzaku  $2.93 \pm 0.45$ were marginally consistent within the 90 \% statistical errors,  although the O\textsc{viii} emission intensity  by Chandra ($2.34 \pm 0.36$ LU) was larger than that of Suzaku ($0.30 \pm 0.20$ LU).

The ROSAT sky survey was able to remove SWCX contributions with time variations on scales of a day or less, which should provide efficiencies in removing geocoronal SWCS similar to the procedures used here.  
The steadier part of the heliospheric contribution is still present, however, and we might expect it to be considerably larger, since the ROSAT survey was conducted at solar maximum and the Suzaku observations near solar minimum.
Both Suzaku and ROSAT are in low Earth orbit, and observing directions of the both satellites
are almost 
perpendicular to the Sun-Earth-line.

We calculated the {\it R45} band counting rate expected for the best-fit model parameters of model 1 ( Table~\ref{tbl:specfit_bknpow}) and model 2 ( Table~\ref{tbl:specfit_fix_bknpow}) . We used the ROSAT response function, "pspcc\_gain1\_256.rsp",  in CALDB at NASA/GSFC\footnote{available from  ftp://legacy.gsfc.nasa.gov/rosat/calib\_data/pspc/cpf/matrices/pspcc}.   
Two models gave consistent results within the $1 \sigma$  statistical errors. 

\begin{figure}
\begin{center}
\FigureFile(0.8\columnwidth, ){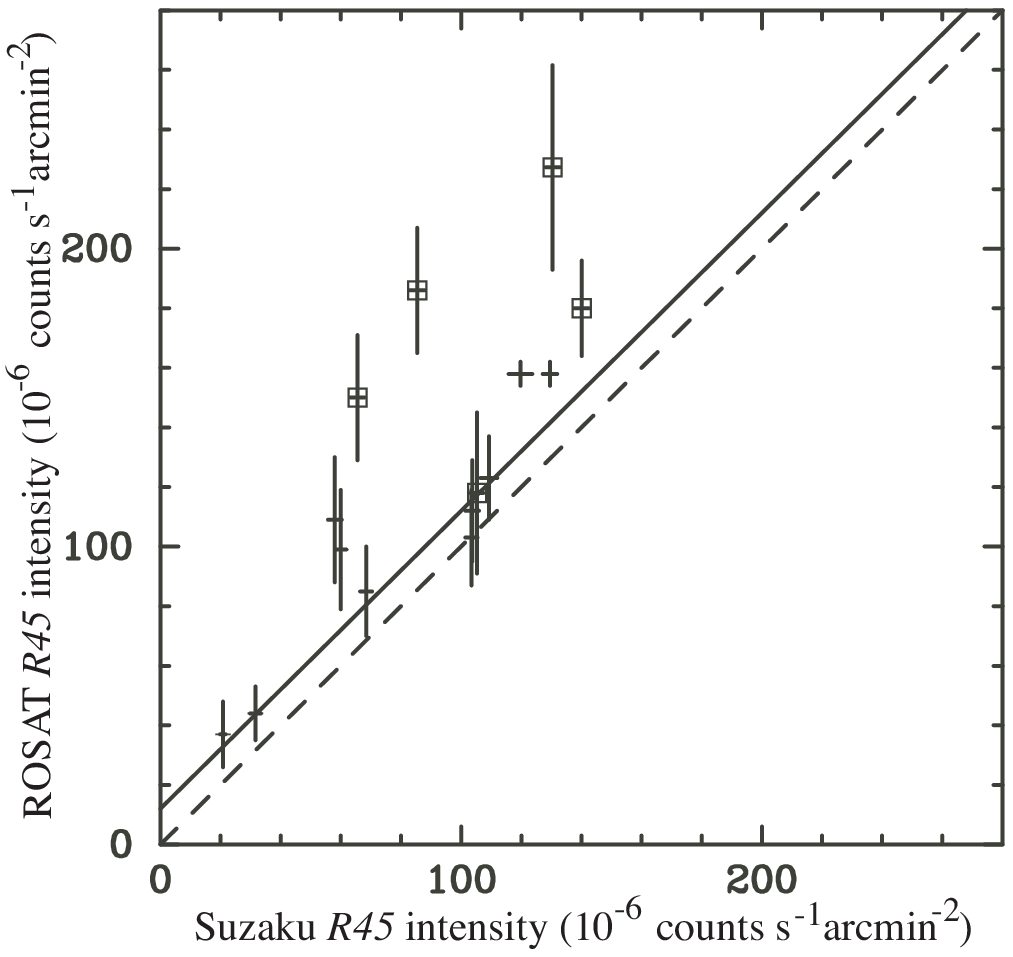}
\end{center}
\caption{
Observed ROSAT {\it R45} band counting rate v.s. counting rate expected from 
Suzaku best-fit model functions.  Points with open boxes are from possibly contaminated regions of the ROSAT survey.  The vertical error bars are the $1~\sigma$ statistical errors of the
ROSAT observation.  The horizontal error bars are $1~\sigma$ statistical errors estimated from the
spectral model fits of Suzaku data.  The diagonal solid line shows the predicted relation including the average point source removal correction, i.e. $12~\times~10^{-6}~{\rm c~s}^{-1}~{\rm amin}^{-2}$ above the 1:1 relation.   
\label{fig:rosat_vs_suzaku}}
\end{figure}

Using the database at NASA/GSFC\footnote{available from http://heasarc.gsfc.nasa.gov/cgi-bin/Tools/xraybg/xraybg.pl}, we  extracted ROSAT {\it R45} band average counting rate in a circular sky region of 36 arc minute diameter centered at the Suzaku XIS aim point. 
The size of the sky region is larger than than the Suzaku field of view.
However,  
statistical precision of survey will be inadequate on smaller fields. 
In Figure~\ref{fig:rosat_vs_suzaku}, we plot the observed ROSAT counting rate as a function of expected counting rate from the Suzaku observation.  In the figure we include {M12on} and {MP235} fields using the results from \citet{Masui_etal_2009}.
The horizontal error bars in this figure is  $1~\sigma$ statistical errors estimated from the
spectral model fits of Suzaku data: we searched for the minimum and maximum expected ROSAT counting rate for the combinations of parameters on the $\chi^2 = \chi^2_{\rm min} +1$ surface in the  5 dimensional  model parameter space.

 \begin{table}
 \caption{Comparison with ROSAT all sky survey
 \label{tbl:suzaku-rosat}}
 \begin{center}
 \begin{tabular}{lcc} \hline \hline
 ID & Suzaku$^{\rm a}$ & RASS$^{\rm b}$ {\it R45} band\\ 
      & \multicolumn{2}{c}{$10^{-6 }~{\rm c s}^{-1}~{\rm amin}^{-2}$}\\  \hline
 {1} ({GB}) & $85.4 \pm  2.6$ &  $186 \pm 21$$^{\rm c}$ \\
 {2} ({HL-B}) & $65.6 \pm 2.2$ & $150 \pm 21$$^{\rm c}$ \\
 {3}  ({LH-2}) & $68.4 \pm 1.9$ & $85 \pm 15$ \\
 {4}  ({LH-1}) & $103.4 \pm 2.0$ &  $103 \pm 16$\\
 {5} ({Off-FIL})&  $130.4 \pm 2.3$ & $227 \pm  34$$^{\rm c}$\\
 {6} ({On-FIL})& $105.3 \pm 2.1$ & $118 \pm 27$$^{\rm c}$   \\ %
 {7} ({HL-A}) & $103.7 \pm 2.2$ & $112 \pm 17$  \\
 {8}  ({M12off}) & $60.0 \pm  1.8$ & $99 \pm 20$ \\
 {9}  ({LX-3}) & $140.0 \pm 2.4$ &  $180 \pm 16$$^{\rm c}$  \\
 {10} ({NEP1}) & $129.6 \pm 2.4$ & $158 \pm 4$ \\
 {11} ({NEP2}) & $ 119.8 \pm 4.0 $ & $158 \pm 4$ \\
 {12}  ({LL21}) & $109.3 \pm 2.8$ & $123 \pm 14$ \\
 {13}  ({LL10}) &  $31.6 \pm 1.8$ &  $44 \pm 9$ \\ 
 {R1}  ({M12on})   & $20.8 \pm  0.9$ & $37 \pm 11$ \\
 {R2}  ({MP235}) & $58.0 \pm 2.3$ & $109 \pm 21$  \\
  \hline\hline
 \multicolumn{3}{l}{
 \rlap{\parbox[t]{.4\textwidth}{
 $^{\rm }$~{}{The errors in this table are all at 1-$\sigma$ significance. }
 }}}\\
 \multicolumn{3}{l}{
 \rlap{\parbox[t]{.4\textwidth}{
 $^{\rm a}$~{The intensity expected in the ROSAT {\it R45} band from the Suzaku best-fit model function. }
 }}}\\
 \multicolumn{3}{l}{
 \rlap{\parbox[t]{.4\textwidth}{
 $^{\rm b}$~{Average intensity for a circular region of 36' diamiter centered at the Suzaku-observation aim point. }
 }}}\\
 \multicolumn{3}{l}{
 \rlap{\parbox[t]{.4\textwidth}{
 $^{\rm c}$~{These directions are considered to be contaminated
  with the long-term enhancement: data points marked with open boxes in Figure~\ref{fig:rosat_vs_suzaku}.}
 }}}\\
 \end{tabular}
 \end{center}
 \end{table}

The observed and expected counting rates are well correlated.  However
the observed ROSAT counting rates are systematically larger than the rates expected from the
present Suzaku best-fit model.
The sky directions of  the five data points marked with an open rectangle 
({GB},  {HL-B}, {Off-FIL}, {On-FIL}, and {LX-3} )
 are on streaks of bright areas which follow 
the RASS scan path.  
We thus consider that they are possibly contaminated with  the long-term enhancement, i.e. with the SWCX from the geocorona, or by the solar X-ray scattering  during the ROSAT observations \citep{Snowden_etal_1994}.  
If we exclude  these data points,  the scatter of the data points from a linear relation is significantly reduced. 
We then fitted the relation of the remaining nine data points with a linear function:  (ROSAT observed rate) = $p~ \times$ (rate expected from Suzaku) $+ ~q$.   
We obtained $p = 1.08  \pm 0.12$ and  $q =(16.1^{+12.5}_{-13.6}) \times 10^{-6}~{\rm c~s}^{-1}~{\rm amin}^{-2}$.
The slope $p$ is consistent with unity within the 90\% statistical errors, while the offset $q$  is positive.

We consider that a large fraction of the positive offset can be explained by the difference in point source sensitivity of the two sets of observations.
For the ROSAT map used in the present analysis, point sources with counting rate $ > 0.020 {\rm~c~s}^{-1}$  in the {\it R45} band were removed \citep{Snowden_etal_1997}.   This threshold corresponds to $3.8 \times 10^{-13} {\rm erg~cm}^{-2}~{\rm s}^{-1}$ in 0.47 - 1.21 keV assuming a power-law spectrum of a photon index of 1.95 and an absorption column density of $5.6 \times 10^{19} {\rm cm}^{-2}$ (see below).   On the other hand, from the energy fluxes and their statistical errors of the point sources we removed from the present Suzaku analysis, we estimate the typical detection threshold for point sources to be  $1 \times 10^{-14} {\rm erg~cm}^{-2}~{\rm s}^{-1}$ in 0.47 - 1.21 keV.  Thus there is a factor of $\sim 40$ difference in the detection threshold.   We estimated the average surface brightness of point sources in the flux range of $(1 - 38) \times 10^{-14} {\rm erg~cm}^{-2}~{\rm s}^{-1}$ in 0.47 - 1.21 keV, assuming  the $\log N - \log S$ relations in 0.5 -- 2 keV band obtained by ROSAT \citep{Hasinger_etal_1993}.  To convert energy flux in 0.5 -- 2 keV to the {\it R45} band,  
we assumed a broken power-law spectrum with photon indices  of 1.96 and 1.4 respectively below and above 1.2 keV and various absorption column densities from  $5.6 \times 10^{19} {\rm cm}^{-2}$ (Lockman hole field)  to $1 \times 10^{21} {\rm cm}^{-2}$.  The result we obtained was $(14 - 10) \times 10^{-6}~{\rm c~s}^{-1}~{\rm amin}^{-2}$ in the {\it R45} band for  $N_{\rm H} = (0.056 - 1)  \times 10^{21} {\rm cm}^{-2}$.   Thus the offset $q$ is consistent with zero if we correct for contribution of point sources in the ROSAT data.   Using  the average value over the $N_{\rm H}$ range,  $12 \times 10^{-6}~{\rm c~s}^{-1}~{\rm amin}^{-2}$,  we obtain  $q = (4.1^{+12.5}_{-13.6}) \times 10^{-6}~{\rm c~s}^{-1}~{\rm amin}^{-2}$.

Although the ROSAT all sky survey was carried out  near the solar maximum in 
1990, the present results show that 
Suzaku and ROSAT {\it R45}-band intensities in at least nine directions were consistent with each other.
The intensity of the spectral component for SWCX+LHB in model 2, which contains O\textsc{vii} emission of 2.1 LU produces a ROSAT {\it R45} counting rate of $14  \times 10^{-6}~{\rm c~s}^{-1}~{\rm amin}^{-2}$.  Since this is  comparable to the 90 \% confidence upper limit of the offset $q$ ($= 17  \times 10^{-6}~{\rm c~s}^{-1}~{\rm amin}^{-2}$), it suggests that  on average, the maximum increase in heliospheric O\textsc{vii} emission intensity between the ROSAT and Suzaku is at most $\sim 2$ LU, or a factor of two.
Since the solar maximum increase should include lines from higher ionizations states such as 
O\textsc{viii} and Ne\textsc{ix} that are more efficient in producing {\it R45} counts, the increase in O\textsc{vii} is probably much less than this.

In section \ref{subsec:remov_geocorona}, we tried to remove as much as possible 
the time intervals in which the X-ray spectrum was  contaminated by the SWCX from the geocorona.  Consistency between the
ROSAT and Suzaku data also indicates successful removal.

\subsection{Intensity variations and temperature of the TAE}
\label{sec:dis:TAE}

Although the solar wind has a broadly simple structure, with the slow
wind confined to low ecliptic latitudes during solar minimum,
the spatially non-uniform distribution of neutral H and He around the
Sun leads to a complex pattern of SWCX emissivity.  The distribution on
the sky of {\em observed} SWCX emission is also a function of
the observer's particular viewing geometry, as noted for example by
\citet{Lallement_2004} and Robertson \& Cravens (2003),
requiring detailed modeling which is beyond the scope of this paper.
The general level of SWCX emission in the {\it R45} band, however, is
expected to be too small to explain the spatial variations observed
in our work, consistent with our interpretation in model 2 that the
spatially dependent components of O\textsc{vii} emission are associated primarily
with the TAE component and not with SWCX.
 
\citet{Yao_etal_2009} constructed a thick hot disk model extending above the Galactic disk in order to simultaneously explain the absorption and emission lines observed in the energy spectra of LMC X-3, the X-ray binary in the LMC,  with Chandra,  and in the energy spectra of the X-ray diffuse emission in the two directions about 30' away from LMC X-3 observed with Suzaku.  They assumed density and temperature distributions exponentially decreasing in the direction perpendicular to the Galactic disk.   They obtained as the best-fit parameters the scale heights of  $h_{\rm T} \xi = 1.4 (0.2, 5.2) $ kpc and $h_{\rm n} \xi = 2.8(1.0, 6.4)$ kpc for temperature and density respectively, and  the midplane  gas temperature and H ion density of  $T_0 = 3.6 (2.9, 4.7) \times 10^6$ K and $n_0 = 1.4 (0.3, 3.4) \times 10^{-3} {\rm cm}^{-3}$, where $\xi$ is the filling factor of the hot gas.
The emission-measure weighted average temperature along a line of sight is  $T_0/(1+h_{\rm n}/(2 h_{\rm T})) \sim 0.16$ keV.  This is lower than the best-fit temperatures of the TAE component with model 2 Table~\ref{tbl:specfit_fix_bknpow}, although the discrepancy is not very large taking the the errors in $T_0$ and $h_{\rm n}/h_{\rm T}$ into account. 

\begin{figure}
\begin{center}
\FigureFile(0.8\columnwidth, ){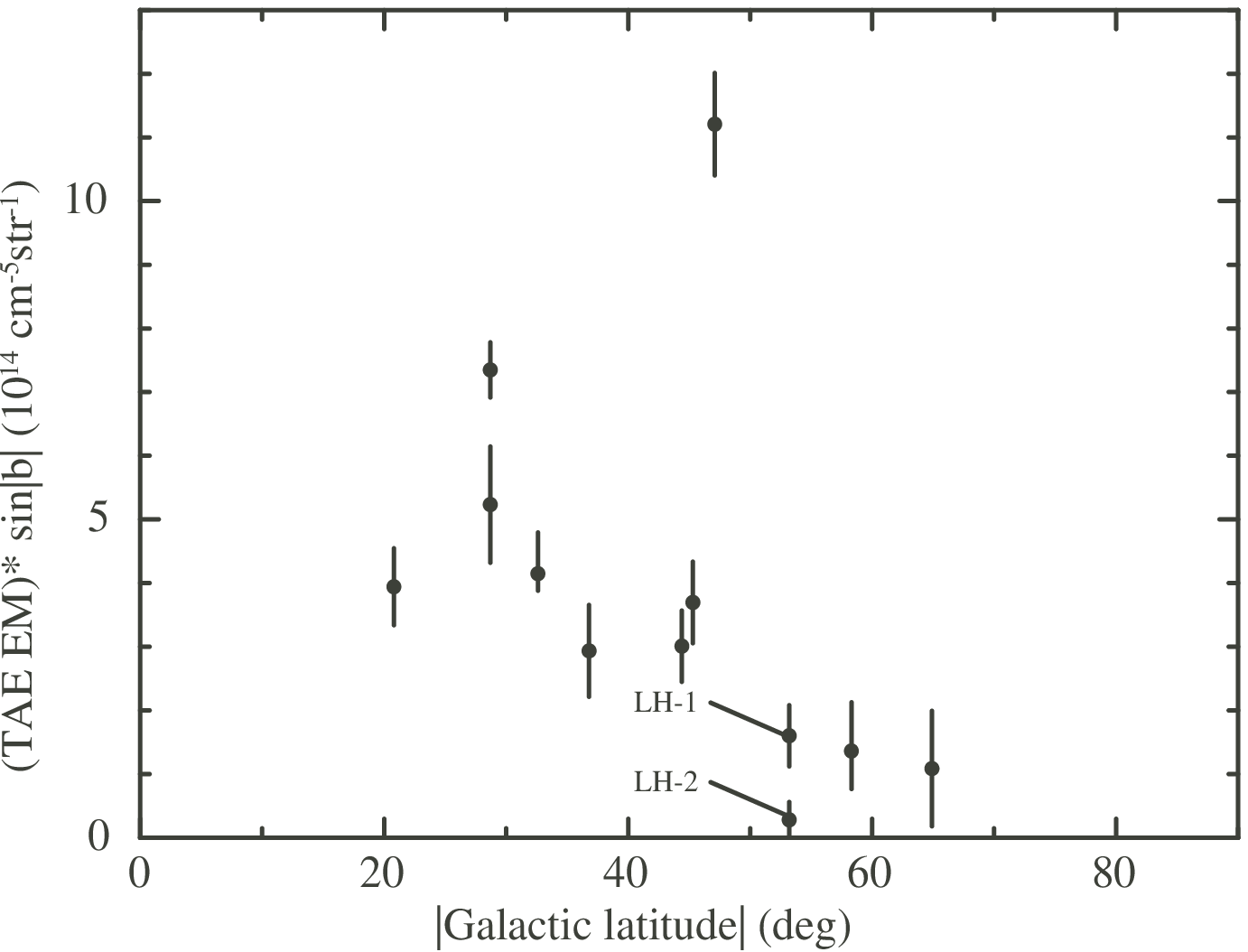}
\end{center}
\caption{
The emission measure of the TAE component multiplied by $\sin |b|$ as a function of $|b|$.
The TAE normalization factors are taken from model 2 (Table~\ref{tbl:specfit_fix_bknpow}). 
The data point  {LL10}  is not plotted because the higher temperature component of this field is like to have a different origin from other fields. 
 \label{fig:tae_norm}}
\end{figure}

For a plane-parallel configuration,  we expect the intensity of the emission to increase from high to low latitude as $\propto \sin^{-1} |b|$,with a rapid decrease at low latitudes, $b \lesssim 10^\circ$, due to Galactic absorption with a much smaller scale height than the emission.   
In Figure~\ref{fig:tae_norm},  instead of  O\textsc{vii} emission intensity, we show the emission measure of the TAE component for model 2 multiplied by $\sin |b|$ ($EM \sin |b|$).  
$EM \sin |b|$ shows a large direction-to-direction fluctuation from a constant, which suggests
that the hot gas is patchy and consists of number of blobs.   
The short angular scale spatial variation between the two Lockmann hole fields, {LH-1} and {LH-2}, is puzzling.   These are separated by only $0.42^\circ$, but the TAE component and O\textsc{vii} emission are significantly stronger for {LH-1}   (Table~\ref{tbl:specfit_taefix_bknpow}, \ref{tbl:specfit_lines_bknpow}). The ROSAT map also show a difference in  {\it R45} intensities of the two fields  (Table~\ref{tbl:suzaku-rosat}), although the field of view of the  two ROSAT fields are overlapped.
Since the distance between two lines of sight is only 10 pc at 1 kpc away from the sun, 
the density contrast of hot plasma must be high between outside and inside the blobs.
Figure~\ref{fig:tae_norm} may  also indicate that $EM \sin|b|$ is systematically larger for $|b| \lesssim 50^\circ$, than for $|b| \gtrsim 50^\circ$.  However, this could be a chance effect, and  we need more data points, in particular for  $|b| \gtrsim 50^\circ$.

We estimate the parameters of hot gas by assuming an isotropic temperature of $kT = 0.222$ keV for simplicity.  The total luminosity of the emission is estimated from the average value of $EM \sin |b|$, assuming  a plane parallel density distribution as
\[
L = 8  \pi^2 \Lambda(T)  \overline{EM \sin |b|}  R^2, 
\]
where  $\Lambda(T)$  and $R$ are, respectively, the emissivity per emission measure at a temperature $T$,  and the outer cylindrical radius of the emission region.  
Using the values of $\Lambda(T)$ for bolometric flux
\citep{Sutherland_Dopita_1993} and 0.3-2 keV band (APEC), we obtain,
\begin{eqnarray*}
L_{\rm bol} &=& 3.8 \times 10^{39} ~{\rm erg~s}^{-1}\\
 & & \times \left( \frac{\overline{EM \sin |b|}}{3.6 \times 10^{14} ~{\rm cm}^{-5}~{\rm str}^{-1}} \right)~
\left( \frac{R}{\rm 15~kpc} \right)^2,
\end{eqnarray*}
\begin{eqnarray*}
L_{\rm 0.3-2keV} &=& 1.1 \times 10^{39} ~{\rm erg~s}^{-1}\\
 & & \times \left( \frac{\overline{EM \sin |b|}}{3.6 \times 10^{14} ~{\rm cm}^{-5}~{\rm str}^{-1}} \right)~
~ \left( \frac{R}{\rm 15~kpc} \right)^2,
\end{eqnarray*}

The midplane density and the total mass of the hot gas, $M_{\rm tot} $, can be estimated by further assuming the scale height. 
\begin{eqnarray*}
n_0 &=& 1.3 \times 10^{-3}~{\rm cm}^{-3}\\
 & & \times \left( \frac{\overline{EM \sin |b|}}{3.6 \times 10^{14} ~{\rm cm}^{-5}~{\rm str}^{-1}} \right)^{1/2}~
~\left(\frac{h \xi} {1.4 {\rm kpc}}\right)^{-1/2}, 
\end{eqnarray*}
and
\begin{eqnarray*}
M_{\rm tot} &=& 6.5 \times 10^{7} M_\odot~
  \left( \frac{\overline{EM \sin |b|}}{3.6 \times 10^{14} ~{\rm cm}^{-5}~{\rm str}^{-1}} \right)^{1/2}\\
 & & \times \left( \frac{R}{\rm 15~kpc} \right)^2 \left(\frac{h \xi} {1.4 {\rm kpc}}\right)^{1/2}.
\end{eqnarray*}
As the scale height, $h$, of hot gas we assumed the temperature scale height from 
\citet{Yao_etal_2009}, since it is smaller than the density scale height.

\citet{Yao_etal_2009} concluded that the plasma responsible for the 
 O\textsc{vii} and O\textsc{viii} absorption and emission towards LMC X-3 cannot be
 isothermal.  However, the temperatures averaged over line of sight for various
 directions determined by O\textsc{vii} to O\textsc{viii} 
ratio was remarkably constant (Figure~\ref{fig:oviii-ovii_Bkn}).
A possible origin of the hot gas is supernovae.  
In fact both the total mass and the total thermal energy of the hot plasma can
be supplied at a supernova rate of $10^{-2}~{\rm y}^{-1}$ within the 
radiative cooling time
of 3 Gy. 
However, there is no particular reason
why the gas should prefer  $kT = 0.2$ keV.
The spectra of the halo emission 
of nearby star-forming galaxies and of some of normal galaxies are 
described by two-temperature models with
$kT$'s in the range of 0.1 to 0.8 keV \citep{Strickland_etal_2004, Tullman_etal_2006, Yamasaki_etal_2009}.   
Thus we consider that $kT = 0.2$ keV is specific for our Galaxy.  Then it is likely related to  
the virial temperature  ($kT = 0.2$ keV) for the rotation velocity of 200 km~s$^{-1}$
(e.g. \citet{Keres_etal_2005}).  
A possible explanation for  this coincidence is that the hot gas was formed 
by cosmological accretion \citep{Toft_etal_2002,Rasmussen_etal_2009}.  
The total luminosity of our Galaxy estimated above is consistent with the simulations.  
However, the radiative cooling time of the hot gas is 3Gy, 
and  the hot gas needs be supplied at least on this time scale.
Another interpretation is also  possible that higher temperature gas 
has escaped from the potential and that the gas at the temperature near the virial 
temperature remains, 
since the escape time scale is by a factor of five  shorter than the radiative cooling time scale.

From the spectral fits with model 2, we obtained 2 to 3 solar for [Ne/O] and [Fe/O] abundance 
ratios for four of the spectra.  
However, in model 3 (Table~\ref{tbl:specfit_taefix_bknpow}), we showed that  these Fe-L and Ne "excess" emission can be explained by a higher temperature components of $kT = 0.5 - 0.9 $ keV. 
The {HL-B} field shows a high  O emission temperature 
(0.30 keV, see Figure~\ref{fig:oviii-ovii_Bkn} and Table~\ref{tbl:specfit_fix_bknpow}).
Since all these fields are in high ecliptic latitudes, where only low-ionization-temperature winds are emitted from the sun during solar minimum, the Heliospheric SWCX is not likely origin
of the excess Ne and Fe emission.
These lines of sight may contain blobs of high ($>0.22 $ keV) temperature hot gas which may be on the way to  escape from our Galaxy.

\subsection{The spectrum of the $b =10^\circ$ sample}

\begin{figure}
\begin{center}
\FigureFile(0.8\columnwidth, ){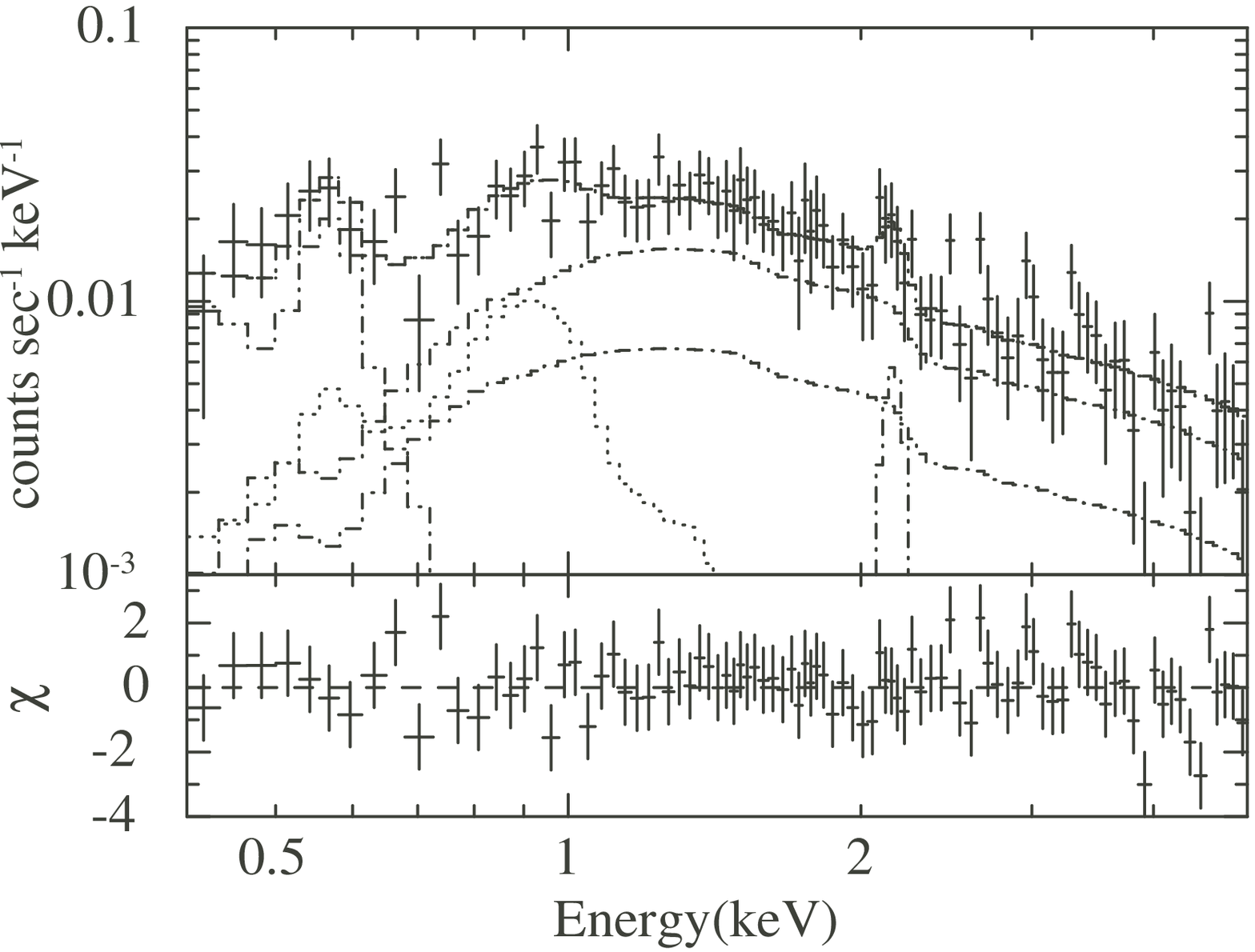}
\end{center}
\caption{
Fit result  of the Low Latitude 97+10 (data ID {13}) spectrum using the faint dM star model by \citet{Masui_etal_2009} instead of the TAE component.  
 \label{fig:b10_dMstar}}
\end{figure}

We obtained significantly higher temperature (0.75 keV) for the TAE component for Low Latitude 97+10 ({LL10}) than that of other samples (average = 0.2 keV).   
This  direction  has a high Galactic absorption of $2.78 \times 10^{21} {\rm cm}^{-2}$,
and  the transmissions for O\textsc{vii} and O\textsc{viii} are respectively only about 10 and 20 \%.  Thus the emission from the thick disk of temperature $\sim 0.2$ keV will be significantly absorbed and hard to detect, if it exists in this direction.   
\citet{Masui_etal_2009}  detected an emission  component of the similar temperature in the energy spectrum of the midplane direction, {MP235}.  
They suggested that the component is a sum of emission from 
unresolved faint dM stars existing between the bulk of Galactic absorption and the Earth, 
and constructed model spectra 
assuming an average dM star spectrum with two-temperature thermal emission, 
the stellar X-ray luminosity distribution functions, and the spatial densities of dM stars in the literature.  
The model spectrum could consistently explain the observed spectrum not only in spectral shape
but also in   absolute intensity within 30 \%.  
We fitted the {LL10} spectrum using their dM star model spectrum constructed for $(\ell, b)= (90^\circ, 10^\circ)$, instead of the TAE component in model 1.
As shown in Figure~\ref{fig:b10_dMstar}, this model reproduces the observed spectrum well ($\chi^2$ = 89.2 for 90 degrees of freedom).   However,  it was necessary to increase the intensity of
the emission by a factor of about 5 from the model.
This suggests that the emission from dM stars does not decrease so rapidly with increasing $b$ as the model predicts, or that the there are large spatial fluctuations at $b \sim 10^\circ$.  We need more observations at low Galactic latitude in order to solve this problem.

\section{Summary}

We presented spectra of soft diffuse X-ray emission in twelve  fields observed with Suzaku together with the spectra of two other fields analyzed by \citet{Masui_etal_2009}.   
In the data reduction, we carefully removed  the contributions of the solar-wind charge-exchange (SWCX) induced X-ray emission from the 
geocorona.  However, $\sim 1.5$ LU uncertainty remains in O\textsc{vii} intensity.
The discrepancy between the two NEP observations (1.9 LU) can be partly
due to yet incomplete removal of geocoronal SWCX.

To determine O emission intensities, we first fitted the spectra with a  model consisting of a broken power-law component for the CXB, and two thin thermal emission components of the solar abundance, one of which are subject to Galactic absorption  (model 1).
The  O\textsc{vii} and O\textsc{viii} emission intensities were determined using the model by setting O abundances of two thermal emission components to zero and adding two Gaussian functions to represent 
O\textsc{vii} and O\textsc{viii} .
The O\textsc{vii} and O\textsc{viii} intensities are strongly correlated and suggest the  existence of an intensity floor for 
O\textsc{vii} emission  at  
$\sim 2~{\rm photons~s}^{-1}~{\rm cm}^{-2}~{\rm str}^{-1}$ (LU).  
The O\textsc{viii} emission intensity of nine high-latitude fields show a tight correlation
with the excess of the O\textsc{vii} intensity above the floor.  
The relation is approximated as (O\textsc{vii} intensity) = 0.5~$\times$~[(O\textsc{vii} intensity)~--~2~LU)].
These suggest that the  O\textsc{vii} emission arises from two origins: approximately uniform 
emission of about 2 LU, and spatially variable (0-7 LU) emission from hot plasma of a temperature of  $\sim$ 2 keV.  The former is likely to arise from Heliospheric Solar Wind Charge Exchange plus the local hot bubble (SWCX+LHB), and the latter from hot gas in more distant parts of the Galaxy, i.e. the transabsorption emission (TAE).  It is remarkable that for most of the fields, TAE average emission temperatures are confined in a narrow range ($\sim \pm 0.2$ keV) around 0.2 keV.  
This temperature may be related to the virial temperature of the Galaxy, which locally corresponds to the rotation velocity of $\sim~200$~km~s$^{-1}$.

The O emission intensities of the two thermal components of model 1 do not reproduce the above characteristics of O emission.  This is because the abundance is fixed to the solar, while   strong Ne and Fe-L emissions exist in some of the spectra.  The observed spectra can be fitted with a model in which  the intensity and temperature of the non-absorbed SWCX+LHB component are fixed to nominal values if we set the  Ne and Fe abundances of absorbed TAE component free (model 2), or if we include an additional higher temperature component (model 3).  We found that four spectra  required  Ne to O abundances as large as 3 solar with model 2.  Alternatively, these spectra  can be fitted by model 3 with  a higher temperature (0.5 - 0.9 keV) emission component with solar abundances.   The temperatures of the TAE component obtained with these two  models were consistent with the values expected from the O\textsc{vii} to O\textsc{viii} ratio.

The surface brightnesses estimated from the present best-fit model function were statistically consistent with the ROSAT All Sky Survey (RASS) map, even though the present observations were performed during solar minimum, while the RASS was in solar maximum.   The  upper limit  for the difference was estimated to be $17 \times 10^{-6}~{\rm c~s}^{-1}~{\rm amin}^{-2}$ {\it R45}-band counting rate after corrected for the difference in point source removal threshold between the present Suzaku observations and the RASS.  

The origin of the TAE component was discussed in the context of the thick hot disk constructed by 
\citet{Yao_etal_2009}.  The emission measure determined by model 2 shows a large deviation from the 
$\sin^{-1}~|b|$ 
dependence expected from a plane parallel configuration, which we consider to suggest short spatial scale structure of the hot gas.   The total luminosity, midplane density, and the total mass of the hot gas were estimated assuming simple cylindrical geometry and a uniform temperature.  The canonical temperature of the TAE component, $kT \sim 0.2$ keV,
may be related to the virial temperature of our Galaxy.  

The lowest latitude sample of the present analysis,  $b = 10^\circ$,  contains emission with $kT$ = 0.75 keV instead of 0.2 keV.  The spectral shape of this component can be represented by the faint dM star model by \citet{Masui_etal_2009}.  However the intensity must be increased by a factor of about five from the model.

\vspace{1em}
The authors are grateful to the anonymous referee for his  comprehensive comments and useful suggestions, which improved the paper very much. They also would like to thank the Suzaku team for their effort for the operation of the spacecraft, the calibration of the instruments, and the data processing.

\end{document}